\documentclass[12pt]{article}
\usepackage{a4wide}
\usepackage{amssymb}
\usepackage{amsmath}
\usepackage{graphicx}
\usepackage{cancel}
\usepackage{array}
\usepackage{tikz}
\usetikzlibrary{decorations}
\usetikzlibrary{decorations.pathmorphing}
\usetikzlibrary{decorations.markings}
\usetikzlibrary{patterns}
\usetikzlibrary{arrows}
\usepgflibrary{patterns}
\usepackage{slashed}
\usepackage{hyperref}
\newcolumntype{C}{>{\centering\arraybackslash}m{4cm}}
\newcommand{\beq}{\begin{equation}}
\newcommand{\eeq}{\end{equation}}
\newcommand{\beqa}{\begin{eqnarray}}
\newcommand{\eeqa}{\end{eqnarray}}
\newcommand{\nn}{\nonumber}

\newcommand{\slt}{\mathfrak{sl}(2)}
\newcommand{\ub}{\tilde{u}}
\newcommand{\N}{\mathcal{N}}
\newcommand{\Z}{\mathcal{Z}}
\newcommand{\D}{\mathcal{D}}
\newcommand{\Tr}{\textrm{Tr}}
\newcommand{\K}{\mathcal{K}}
\newcommand{\ft}[2]{{\textstyle\frac{#1}{#2}}}

 \def\Xint#1{\mathchoice
 {\XXint\displaystyle\textstyle{#1}}%
 {\XXint\textstyle\scriptstyle{#1}}%
 {\XXint\scriptstyle\scriptscriptstyle{#1}}%
 {\XXint\scriptscriptstyle\scriptscriptstyle{#1}}%
 \!\int}
 \def\XXint#1#2#3{{\setbox0=\hbox{$#1{#2#3}{\int}$}
 \vcenter{\hbox{$#2#3$}}\kern-.5\wd0}}
 
 \def\dashint{\Xint-}

\begin{document}
\begin{center}
{\Large{\bf Bethe Ans\"atze for GKP strings}}
\vspace{15mm}

{\sc Benjamin Basso$^a$, Adam Rej$^{b,c}$} \\[5mm]

{\it $^a$ Perimeter Institute for Theoretical Physics, Waterloo, Ontario N2L 2Y5, Canada}\\[5mm]

{\it $^b$  School of Natural Sciences, Institute for Advanced Study, Princeton, NJ 08540, USA}\\[2mm]

{\it $^c$  Marie Curie Fellow}\\[5mm]

\texttt{\big\{bbasso $\arrowvert$ arej\big\} @ $\left\{\frac{\texttt{\normalsize perimeterinstitute.ca}}{\texttt{\normalsize ias.edu}} \right\}$}
\\[15mm]

\textbf{Abstract}\\[2mm]
\end{center}

\noindent{Studying the scattering of excitations around a dynamical background has a long history in the context of integrable models. The Gubser-Klebanov-Polyakov string solution provides such a background for the string/gauge correspondence. Taking the conjectured all-loop asymptotic equations for the $\textrm{AdS}_4/ \textrm{CFT}_3$ correspondence as the starting point, we derive the S-matrix and a set of spectral equations for the lowest-lying excitations. We find that these equations resemble closely the analogous equations for $\textrm{AdS}_5 / \textrm{CFT}_4$, which are also discussed in this paper. At large values of the coupling constant we show that they reproduce  the Bethe equations proposed to describe the spectrum of the low-energy limit of the $AdS_4\times CP^3$ sigma model.}

\newpage
\tableofcontents
\newpage

\section{Introduction}
The vacuum states lies at the heart of every quantum theory. It provides the necessary reference state on which the Hilbert space of the theory may be constructed. Once the vacuum is identified, one can proceed with studying the excitations and their dynamics. This is not to say that the vacuum itself is not of interest, as often its properties are of pivotal importance for the theory.  

When studying highly excited states, for which the number of excitations $M \to \infty$, the description based on the underlying vacuum may not be the most efficient one, however. Instead, another state may exist and provide a much more adequate state of reference. An exemplar is the antiferromagnetic state of the one-dimensional Heisenberg spin chain (see~\cite{Faddeev:1996iy} for instance). It is very excited from the point of view of the vacuum state, where all spins are aligned in one direction, and is given by an intricate superposition of magnons. On the other hand, the corresponding energy and density of roots may be explicitly found. What is even more striking is that scattering of  excitations around the antiferromagnetic state, termed spinons, is governed by an effective S-matrix that may be directly computed. Consequently,  states close to the antiferromagnetic vacuum are more easily studied by mapping them to excitations on top of the antiferromagnetic state rather than resolving the initial spin chain problem.  

In the context of AdS/CFT correspondence simple classical string solutions provide such reference states. Even though they are identified with high-in-the-spectrum eigenstates of the dilatation operator of the dual gauge theory, they have been playing a prominent role in understanding the spectral problem in the planar limit. A particular example has proven very fruitful in matching both sides of the correspondence. It was introduced by Gubser, Klebanov and Polyakov~\cite{Gubser:2002tv} and is conventionally referred to as spinning string or the GKP string. This string and its excitations are dual to large spin operators on the gauge theory side. For the $\N=4$ SYM theory one finds their simplest representatives in the so-called $\slt$  sector spanned by single-trace operators of the type
\beq \label{N4ops}
\Tr(\D\dots\D \Z\ \D\Z \dots)\,.
\eeq
The corresponding operator has $S$ covariant derivatives $\D$ and $L$ complex scalar fields $\Z$. The twist, customarily defined as the bare scaling dimension $\Delta$ minus the Lorentz spin $S$, is equal $L$. In the spin chain description the vacuum state corresponds to the $\tfrac{1}{2}$-BPS operator $\Tr(\Z^L)$ and the derivatives $\mathcal{D}$ are identified with the magnons. This logic is reversed if one adopts the GKP string as a reference state. From this perspective the scalar fields $\Z$ are seen as excitations on top of the large spin background of $\D\ldots \D$. For the $\N=6$ Chern-Simons-Matter theory there is no closed subsector with derivatives and scalar fields only. The simplest set of operators dual to the spinning string solution belongs to the $\mathfrak{osp}(2|2)$ sector,
\beq \label{N6ops}
\Tr(\D\ldots  \D Y^1\D\ldots \D \psi_{4+}^{\dagger} \D\psi^{1}_{+} \D Y^{\dagger}_4 \dots)\,,
\eeq
which is built out of bi-fundamental matter fields $(Y^1, \psi^{1}_{+}), (\psi^{\dagger}_{4+}, Y^{\dagger}_4)$ and covariant derivatives $\D$. The vacuum state of the alternating spin chain~\cite{Minahan:2008hf} corresponds to the protected operator $\Tr\, (Y^1 Y^{\dagger}_4)^L$ and excited states are made out of $K$ magnons and $\bar{K}$ anti-magnons. They are \textit{either} the Fermi fields $(\psi^{1}_{+}, \psi_{4+}^{\dagger})$ put on the vacuum sites $(Y^{1}, Y_{4})$ \textit{or} pair up to derivatives $\D$ acting on these fields. The twist of the resulting operator is $L$ and its Lorentz spin $S=\tfrac{1}{2}(K+\bar{K})$. Here, we can again turn things around and think of the matter fields as propagating through a sea of derivatives.

It is striking that both gauge theories appear to be integrable and their integrable structures are so similar, see~\cite{Beisert:2010jr} for a recent review. Besides its conceptual beauty integrability delivers powerful tools for perturbative and non-perturbative computations. In particular, the limit $S \to \infty$ may be efficiently studied using Bethe equations~\cite{Korchemsky:1995be, Belitsky:2006en, Eden:2006rx, Beisert:2006ez}. It corresponds to fixing the number of local fields and allowing the number of derivatives to approach infinity. For generic $L$ and $S$ there are several states, but the state with the lowest possible scaling dimension is unique. Furthermore, its energy grows with $L$ implying that minimal states correspond to minimal physical values of $L$. These are: $L=2$ or twist $2$ for $\N=4$ super-Yang-Mills and $L=1$ or twist $1$ for $\N=6$ ABJM theory. From now on, we will refer to these minimal states, which are dual to the GKP strings, as large spin vacuum states.

In the $\mathcal{N}=4$ theory it is known that the scaling dimensions of twist-two operators exhibit the universal behaviour~\cite{Belitsky:2006en}
\beq
\Delta^{\N=4}_{\textrm{vacuum}}-S=2\Gamma_{\textrm{cusp}}(g)(\log S+\gamma_E)+B_2(g)+o(S^0)\,,
\eeq
for large values of spin $S$. The constant $\gamma_{E}$ stands for the Euler-Mascheroni constant,  $g=\sqrt{\lambda}/4\pi$ and $\lambda$ is the 't Hooft coupling. The functions $\Gamma_{\textrm{cusp}}$ and $B_2$ go by the names of cusp anomalous dimension and virtual scaling function and may be found by solving certain integral equations~\cite{Eden:2006rx, Beisert:2006ez,Freyhult:2009my,Fioravanti:2009xt}. Quite remarkably, the very same functions control the scaling dimensions of the twist-one operators in the Chern-Simons-Matter theory~\cite{Gromov:2008qe,Beccaria:2009wb,Beccaria:2009ny}
\beq\label{deltavac6}
\Delta^{\N=6}_{\textrm{vacuum}}-S= \Gamma_{\textrm{cusp}}(g)(\log(2S)+\gamma_E)+\ft{1}{2}B_2(g)+o(S^0)\,,
\eeq
where now $g=h(\lambda)$ is some interpolating function of the 't Hooft coupling.

An interesting problem is to analyse the spectrum of energies ($E=\Delta-S$) around these minimal solutions or, equivalently, to study excitations on top of these vacua. For non-minimal values of $L$ the operators \eqref{N4ops}-\eqref{N6ops} provide examples of such excitations, but by no means do they exhaust all possibilities. A comprehensive study of the possible excitations in case of $\N=4$ may be found in~\cite{Basso:2010in}, see also~\cite{Gaiotto:2010fk}. In this paper we will carry out a similar analysis for the case of $\N=6$ ABJM theory. We will see that the lowest-lying excitations are the twist-$1/2$ matter fields which transform in $\mathbf{4}$ and $\mathbf{\bar{4}}$ of $\mathfrak{su}(4)$. This should be juxtaposed with the $\mathbf{6}$ rep of $\mathfrak{su}(4)$ which corresponds to twist-one scalar excitations for $\N=4$ theory. Exactly the opposite happens for the twist-one fermions, which for $\N=6$ are in the $\mathbf{6}$ representation of $\mathfrak{su}(4)$, while they are found in the $\mathbf{4}$ and $\mathbf{\bar{4}}$ for $\N=4$ super Yang-Mills. The remaining modes are neutral under $\mathfrak{su}(4)$ and form an infinite tower of twist-$\ell$, $\ell \geq 1$, excitations corresponding to transverse components of the gauge field $D^{\ell-1}_{\perp}F_{+ \perp}$. This is again in contrast with $AdS_5 \times S^5$ case, where one has twice as many gauge field excitations because of two possible transverse directions. What appears quite remarkable and will be discussed in this paper is that the dispersion relations for all these excitations are essentially the same in both theories.

Among the above-listed excitations the lowest-lying ones are of particular interest because they become massless at strong coupling, whereas fermions and gauge excitations retain finite masses. This is in line with the semiclassical quantisation around the GKP background performed for the $AdS_5 \times S^5$ case in~\cite{Frolov:2002av}. This observation led the authors of~\cite{Alday:2007mf} to propose a decoupling limit for the low-energy excitations
\beq \label{dlimit}
p \sim E \sim m(\lambda)\,, \qquad \qquad \lambda \to \infty\, ,
\eeq
where $m(\lambda)\sim \lambda^{1/8} e^{-\sqrt{\lambda}/4} + \ldots$ is the mass gap of the theory at strong coupling. The sigma model considerably simplifies in this limit and becomes the non-linear $O(6)$ sigma model~\cite{Alday:2007mf}. For $AdS_4\times CP^3$ the semiclassical analysis of the GKP string have been performed in~\cite{Alday:2008ut} and the low-energy effective sigma model has been worked out in~\cite{Bykov:2010tv}. The effective theory is a $CP^3$ sigma model coupled to a massless Dirac fermion.

Why is it interesting to study these effective models? The reason is at least twofold. First, at the time of writing of this article there is no known quantisation method applicable to $AdS_5\times S^5$ or $AdS_4 \times CP^3$ sigma models. Several semiclassical computations have been performed, see for example the review series \cite{Beisert:2010jr}, and it has been conjectured that the sigma models are quantum integrable. The decoupling limit leading to effective models seems to be a consistent truncation of the complex dynamics of these models. The resulting low-energy models should then inherit the integrable structures of their ``mother'' theories. Indeed, the $O(6)$ sigma model is a well-known and well-studied integrable model.

The fermionic extension of the $CP^3$ sigma model, on the other hand, received less attention. Its integrability properties have been recently studied in \cite{Basso:2012bw}. It was found to belong to an integrable class of models and the asymptotic Bethe equations (ABA) have been postulated. The mounting evidence in favour of integrability of the effective models provides a compelling  backing of the conjecture that the $AdS_5 \times S^5$ and $AdS_4 \times CP^3$ sigma models are integrable. The second incentive to study these effective models is the fact that they provide an interesting testing ground for the spectral equations of $\textrm{AdS}_5/\textrm{CFT}_4$ and $\textrm{AdS}_4/\textrm{CFT}_3$, see \cite{Beisert:2005fw} and \cite{Gromov:2008qe}. These sets of equations are postulated to describe both ends of the duality, see for example \cite{Beisert:2010jr}, as well as the system at intermediate values of the coupling. It is the aim of this article to verify if the spectral equations of the effective models may be carved out from the all-loop asymptotic Bethe equations. Our strategy will be to constructing first  the equations for the low-lying excitations of both $\mathcal{N}=4$ and $\mathcal{N}=6$ theory at finite coupling and then let the coupling goes to infinity. Please observe that in spite of the coupling constant $\lambda \to \infty$, the decoupling limit \eqref{dlimit} is of non-perturbative character. The derivation of the spectral equations for the effective models from the all-loop Bethe equations is thus a non-perturbative check on the veracity of the latter. In particular, it will be sensitive to the intricate dressing factor shared between both strong/weak coupling dualities.  

Finally, let us mention the interesting developments in gluon scattering amplitudes, where the results of this paper already found some application. The main object of our analysis is the S-matrix for excitations on top of the GKP string. Remarkably, this object is also an important ingredient in the OPE program~\cite{Basso:2013vsa, BSV,Alday:2010ku}, where it enters directly into the computation of  gluon scattering amplitudes in the so-called collinear limit. In this paper we will derive a compact expression for this S-matrix at any coupling and will study its general properties. Although we only study the lowest-lying excitations, our analysis should be amenable to a generalisation to  other excitations of the theory. An interesting question is whether our results for the $\mathcal{N}=6$ theory may be applied to scattering amplitudes in this theory. 

The paper is organised as follows. In the following section we will identify the low-lying excitations to leading order at weak coupling in the $\mathcal{N}=6$ theory and explain how to derive the corresponding asymptotic spectral equations. In Section \ref{sec:BYfc} we will lift these findings to all orders in perturbation theory. We will discuss the structure of the all-loop Bethe-Yang equations for both $\mathcal{N}=4$ and $\mathcal{N}=6$ theories and elaborate on the physical properties of their associated S-matrices. Section \ref{sec:rellimit} will be devoted to the decoupling limit \eqref{dlimit}. We will show that in this limit the Bethe-Yang equations indeed reduce to the asymptotic spectral equations of the effective string theory models. We defer several technical computations to appendices.

\section{Weak coupling analysis}\label{sec:wca}

In this section we will derive the asymptotic spectral equations for the low-energy excitations around the GKP background to leading order at weak coupling. We will begin by considering a subsector where particles and anti-particles have the same polarisation, i.e., when the isotopic degrees of freedom are left unexcited. We shall then explain how to restore the $SU(4)$ symmetry.

Even though the analysis below is not a necessary step before proceeding to the all-loop case, it has the benefit of being less technical and more intuitive, especially because closed formula for all relevant quantities may be obtained. It is also an appropriate place to remind the reader how holes appear in the realm of Bethe ansatz equations. The analysis performed below parallels the one done for $SL(2)$ operators of the $\mathcal{N}=4$ SYM theory. We refer the reader to the literature~\cite{Korchemsky:1995be,Belitsky:2006en,Eden:2006rx,Freyhult:2007pz,Basso:2010in,Bombardelli:2007ed,Bombardelli:2008ah} for related studies.

\subsection{Holes and anti-holes} \label{sec:hah}

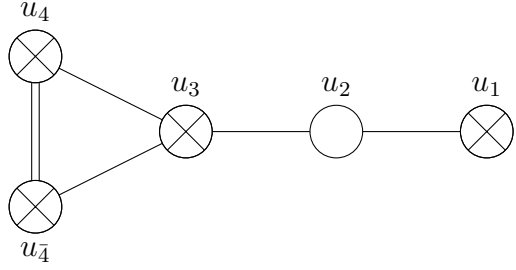
\begin{figure}
\begin{center}
\begin{tikzpicture}[cross/.style={path picture={ 
  \draw[black]
(path picture bounding box.south east) -- (path picture bounding box.north west) (path picture bounding box.south west) -- (path picture bounding box.north east);
}}]
\draw(-0.05,1)--(-0.05,-1);
\draw (0.05,1)--(0.05,-1);
\draw (0,1)--(2,0);
\draw (0,-1)--(2,0);
\draw (2,0)--(4,0);
\draw (4,0)--(6,0);
\draw[fill=white] (0,1) circle (0.35);
\draw[fill=white] (0,-1) circle (0.35);
\draw[cross] (0,1) circle (0.35);
\draw[cross] (0,-1) circle (0.35);
\draw[fill=white] (2,0) circle (0.35);
\draw[cross] (2,0) circle (0.35);
\draw[fill=white] (4,0) circle (0.35);
\draw[fill=white] (6,0) circle (0.35);
\draw[cross] (6,0) circle (0.35);
\node at (0,1.6) {$u_4$};
\node at (0,-1.6) {$u_{\bar{4}}$};
\node at (2,0.6) {$u_3$};
\node at (4,0.6) {$u_2$};
\node at (6,0.6) {$u_1$};
\end{tikzpicture}
\end{center}
\caption{Dynkin diagram in the non-compact grading. The momentum carriers of the alternating spin chain are the roots associated with the nodes $4$ and $\bar{4}$.}\label{Dynkin}
\end{figure}

The Bethe equations at weak coupling may be obtained directly from~\cite{Minahan:2008hf} after dualising the fermonic nodes or, equivalently, by taking the $g \to 0$ limit of the all-loop ansatz in the $\slt$ grading conjectured in~\cite{Gromov:2008qe}. The single-polarisation sector is obtained keeping the equations associated to the momentum carrying nodes, that is the nodes $4$ and $\bar{4}$ of the Dynkin diagram depicted in Figure~\ref{Dynkin}. One is then left with the following equations
\beqa \label{ABJM1}
&&\left(\frac{u_k+\tfrac{i}{2}}{u_k-\tfrac{i}{2}}\right)^L =\prod^{\bar{K}}_{j=1} \frac{u_k-\bar{u}_j-i}{u_k-\bar{u}_j+i}\,,\\
&&\left(\frac{\bar{u}_k+\tfrac{i}{2}}{\bar{u}_k-\tfrac{i}{2}}\right)^L =\prod^{K}_{j=1} \frac{\bar{u}_k-u_j-i}{\bar{u}_k-u_j+i}\,,
\eeqa
where $u_k = u_{4, k}$ and $\bar{u}_k = u_{\bar{4}, k}$ are the rapidities of the $K=K_{4}$ magnons and $\bar{K}=K_{\bar{4}}$ anti-magnons of the alternating spin chain of total length $2L$, see \cite{Minahan:2008hf}. They can be equivalently written as
\beq\label{1plusY}
1+Y(u_{k}) = 0\, , \qquad 1+\bar{Y}(\bar{u}_{k}) = 0\, , 
\eeq
where the function $Y(u)$ is given by
\beq\label{cf-Y}
Y(u) = -\left(\frac{u-\tfrac{i}{2}}{u+\tfrac{i}{2}}\right)^L\prod^{\bar{K}}_{j=1} \frac{u-\bar{u}_j-i}{u-\bar{u}_j+i}\, ,
\eeq
and $\bar{Y}(u)$ is obtained by exchanging of magnons and anti-magnons. These functions are closely related to the counting function $Z(u)$, a well-known object in the literature on integrable models. Explicitly, 
\beq\label{logY}
Z(u) \equiv \frac{1}{2i\pi}\log{(-1)^{L+\bar{K}-1}Y(u)} = \frac{L}{\pi}\, \textrm{arctan}{(2u)} + \frac{1}{\pi}\sum_{j=1}^{\bar{K}}\textrm{arctan}{(u-\bar{u}_{j})}\, ,
\eeq
with a similar expression for $\bar{Z}(u)$. This counting function defines a smooth and monotone function of the rapidity $u$, which interpolates between $Z(\mp \infty) = \mp \ft{1}{2} (L+\bar{K})$.

When  evaluated on any one of the magnon rapidities, the counting function \eqref{logY}, in view of the Bethe equations \eqref{1plusY},  takes values in the set of fermionic mode numbers
\beq
\mathfrak{S} = \left\{ -\frac{L+\bar{K}-2}{2} , -\frac{L+\bar{K}-4}{2}, \ldots \, , \frac{L+\bar{K}-4}{2}, \frac{L+\bar{K}-2}{2}\right\} \,.
\eeq
This set has dimension $L+\bar{K}-1$, with all elements being integers or half-integers according to the parity of $L+\bar{K}$. A similar subset $\bar{\mathfrak{S}}$ exists for the anti-magnon counting function.

The lattices of mode numbers provide us with a classification of the solutions to the Bethe ansatz equations. So far the leading characters were magnons and anti-magnons. But these ones only occupy $K$ and $\bar{K}$ points in lattices $\mathfrak{S}$ and $\bar{\mathfrak{S}}$. The remaining points correspond to \textit{holes} and \textit{anti-holes}. There are exactly $K_{h} = L+\bar{K}-K-1$ and $\bar{K}_{h} = L+K-\bar{K}-1$ such excitations. Equivalently,
\beq \label{KhpKth}
K_h+\bar{K}_h=2(L-1)\, , \qquad K_h-\bar{K}_h=2(\bar{K}-K)\, .
\eeq
They can be given rapidities $u_{h, j}$ by inverting the functions $Z(u)$ and $\bar{Z}(u)$. These rapidities solve the same set of equations as magnons and anti-magnons, namely~(\ref{1plusY}). The holes and anti-holes are the objects of interest in this article. It will become clear later that they should be identified with the low-lying excitations on top of the GKP string.

We stress that in the special case $L=1$ there are no holes nor anti-holes, $K_{h}=\bar{K}_{h}=0$, because of the first relation in \eqref{KhpKth}. The $K$ magnons and $\bar{K}=K$ anti-magnons thus provide a natural physical vacuum at large spin. The corresponding solution to Bethe equations has a symmetric distribution of roots, which is the same for magnons and anti-magnons~\cite{Gromov:2008qe}.

We notice finally that the sum and difference in~(\ref{KhpKth}) are  \textit{even} numbers. This can be understood as a selection rule on the space of solutions. This is a novel feature when compared with $\mathcal{N}=4$ SYM theory, where the number of holes was any  non-negative integer. Another type of selection rule originates from the zero-momentum condition
\beq\label{eiP}
e^{iP} = \prod_{j=1}^{K}\frac{u_{j}+\ft{i}{2}}{u_{j}-\ft{i}{2}}\prod_{j=1}^{\bar{K}}\frac{\bar{u}_{j}+\ft{i}{2}}{\bar{u}_{j}-\ft{i}{2}}=1\,.
\eeq
The zero-momentum condition is also imposed for the $\mathcal{N}=4$ SYM magnons. It should be however noted that this constraint is less potent for ABJM, since the product extends to magnons and anti-magnons. As a result, the spin of twist-one operators, for instance, may assume both even and odd values. This is not the case for $\mathcal{N}=4$, for which the counterparts are the twist-two operators that are labeled by even values of spin only. We shall come back to this ``doubling'' of solutions in Section~\ref{sec:twistq} and later when concluding the findings of this article.

\subsection{Densities of Bethe roots} \label{sec:densofB}

The simplest way of characterising the distribution of the excitations at large spin is through their densities. They can be introduced as derivatives of the counting functions
\beq
\rho(u) =\partial_{u}Z(u)\, , \qquad \bar{\rho}(u) =\partial_{u}\bar{Z}(u)\, .
\eeq
Note that in these expressions magnons and holes / anti-holes are on equal footing. The densities are smooth and positive functions of the rapidity $u$. Using the well-known Euler-Maclaurin summation formula,
\beq \label{EMformula}
\sum_{j=1}^{K}f(u_j) = \int_{-a}^{a} du\,  \rho(u)f(u) -\sum_{j=1}^{K_{h}}f(u_{h, j}) +\ldots\, .
\eeq
we can approximate sums over the roots by continuous integrals weighted by the density. The dots here stand for boundary contributions that are expected to be small at large spin. Differentiating equation \eqref{logY}  with respect to $u$ and using the above summation prescription, one can easily derive integral equation for the densities $\rho(u)$ and $\bar{\rho}(u)$. It is convenient to introduce 
\beq\label{rhopm}
\rho_{\pm}(u) = \rho(u) \pm \bar{\rho}(u)\,,
\eeq
because to leading order the supports of both functions coincide. Indeed, the common support is a symmetric interval of length $2a$, i.e. $(-a,a)$. Numerical analysis reveals $a \sim S$ for large values of $S$. The combinations~(\ref{rhopm}) allow us to decouple the system of integral equations for $\rho$ and $\bar{\rho}$. It can now be written as
\beq\label{rhop}
\begin{aligned}
&2\pi \rho_{-}(u) + 2 \int^{a}_{-a} \frac{\rho_{-}(v)}{1+(u-v)^2}dv= I_{-}(u)\, , \\
&2\pi \rho_{+}(u) -2 \int^{a}_{-a} \frac{\rho_{+}(v)}{1+(u-v)^2}dv= I_{+}(u)\,,
\end{aligned}
\eeq
where
\beq\label{sources}
\begin{aligned}
&I_{-}(u) = \sum^{K_h}_{j=1} I_{-}(u, u_{h, j})-\sum^{\bar{K}_h}_{j=1} I_{-}(u, \bar{u}_{h, j})\, ,\\
&I_{+}(u) =I_{\textrm{vacuum}}(u)+\sum^{K_h}_{j=1} I_{+}(u, u_{h, j})+\sum^{\bar{K}_h}_{j=1} I_{+}(u, \bar{u}_{h, j})\, ,
\end{aligned}
\eeq
and
\beq\label{sources-details}
I_{\textrm{vacuum}}(u) = \frac{2}{\tfrac{1}{4}+u^2}\, , \qquad I_{-}(u, v) = \frac{2}{1+(u-v)^2}\, , \qquad I_{+}(u, v) = \frac{1}{\ft{1}{4}+u^2}-\frac{2}{1+(u-v)^2}\, .
\eeq
The kernel of the first equation is identical to the one encountered in the thermodynamical limit of the antiferromagnetic $SU(2)$ XXX spin chain. As we shall realise later this equation is already exact, in other words, it receives no higher-loop corrections. Up to minor modifications the second integral equation coincides with its $\N=4$ counterpart, that is to say, with the integral equation for density of $SL(2)$ magnons. These striking  observations will continue to be valid \textit{non-perturbatively}, allowing one to compute $\rho_{+}$ directly from the density of holes found in the $\mathcal{N}=4$ theory.

The reader should note that we are interested in constructing the densities in the $u \sim O(S^0)$ domain. The integrals in the right-hand sides of~(\ref{rhop}) can be then extended over the whole real axis, as $a\rightarrow \infty$. This makes the equations soluble by means of the Fourier transform. The solution is then easily found
\beq\label{LO-densities}
\begin{aligned}
\rho_{-}(u)&=\sum^{K_h}_{j=1}\rho_{-}(u, u_{h, j})-\sum^{\bar{K}_h}_{j=1}\rho_{-}(u, \bar{u}_{h, j})\, ,\\
\rho_{+}(u)&= \rho_{\textrm{vacuum}}(u) +\sum^{K_h}_{j=1}\rho_{+}(u, u_{h, j})+\sum^{\bar{K}_h}_{j=1}\rho_{+}(u, \bar{u}_{h, j})\, ,
\end{aligned}
\eeq
where the decompositions reflect the structure of the source terms in~(\ref{sources}). We have
\beq\label{vacrho}
2\pi\rho_{\textrm{vacuum}}(u)  = C-2i\partial_u \log{\frac{\Gamma(\tfrac{1}{2}-iu)}{\Gamma(\tfrac{1}{2}+iu)}} = C-4\psi(1)+4\int_{0}^{\infty}\frac{\cos{(ut)}e^{t/2}-1}{e^{t}-1}dt\, ,
\eeq
with some constant $C$ and
\beq
\begin{aligned}
2\pi\rho_{-}(u, v) &= -i\partial_{u}\log{\frac{\Gamma(1+\tfrac{iu-iv}{2})\Gamma(\tfrac{1}{2}-\tfrac{iu-iv}{2})}{\Gamma(1-\tfrac{iu-iv}{2})\Gamma(\tfrac{1}{2}+\tfrac{iu-iv}{2})}} = 2\int_{0}^{\infty}\frac{\cos{((u-v)t)}}{e^{t}+1}dt \, , \\
2\pi\rho_{+}(u, v) & = -i\partial_{u}\log{\frac{\Gamma(1+iu-iv)\Gamma(\tfrac{1}{2}-iu)}{\Gamma(1-iu+iv)\Gamma(\tfrac{1}{2}+iu)}} =2\int_{0}^{\infty}\frac{\cos{(ut)}e^{t/2}-\cos{((u-v)t)}}{e^{t}-1}dt\, .
\end{aligned}
\eeq
The integral representations given are convenient for testing the validity of the solution when plugged into the integral equation~(\ref{rhop}).

As we can see the solution~(\ref{LO-densities}) is unique up to an arbitrary constant $C$, which is a zero mode of the integral equation~(\ref{rhop}) when $a\rightarrow \infty$. One way of fixing it is by imposing the matching with the density in the regime $u\sim a \sim S$. This is the semi-classical regime first studied in~\cite{Korchemsky:1995be,Belitsky:2006en} in the context of the $\textrm{XXX}_{-s}$ spin chain. An important feature of this regime is that the information about the distribution of holes is lost at the leading order at large spin. This is true when the holes are of order one, i.e. $O(S^0)$, which we will assume throughout this article. The semi-classical regime is then endowed with a certain universality: it does not depend on the number of holes nor on their distribution. What is more, the density is actually independent of the coupling constant. 

The easiest way to establish the form of the density in the semi-classical regime is by solving the Baxter equation following the seminal analysis~\cite{Korchemsky:1995be,Belitsky:2006en}. This strategy, originally developed for the $\textrm{XXX}_{-s}$ spin chain, is easily adapted to our case, see Appendix \ref{sec:sub}. It is no surprise that the result is almost identical to the one found for the $SL(2)$ spin chain. It reads
\beq\label{SemiCl}
\rho_{+}(u) = \frac{1}{\pi}\log{\left(\frac{1+\sqrt{1-\tilde{u}^2}}{1-\sqrt{1-\tilde{u}^2}}\right)} + o(S^0)\, ,
\eeq
where $\tilde{u} = u/a$. Having derived this density we can now relate the parameter $a$ to the spin $S$ by requiring that the density is properly normalised. Namely the integral of $\rho_{+}$ over its supporting interval $u\in (-a,a)$ should be equal to the total number of magnons and anti-magnons, that is $2S$. Integrating~(\ref{SemiCl}) allows us to fix $a = S$. This is the only difference to the $SL(2)$ spin chain, for which $a=S/2$. 

We can now fix the constant $C$ as follows. First we notice that at small $\bar{u}$ the density~(\ref{SemiCl}) exhibits logarithmic behaviour, $\rho_{+}\sim -2\log{(\tilde{u}/2)}/\pi$. Requiring the density~(\ref{LO-densities}) to match the semi-classical density~(\ref{SemiCl}) for $1 \ll u \ll S$ fixes the constant $C$ unambiguously. Expanding~(\ref{LO-densities}) at large $u$ we find
\beq\label{constant}
C = 4\log(2S)\,.
\eeq
To derive this concise expression one has to make use of the identity~(\ref{KhpKth}). The constant $C$ depends only on spin. This is due to the universality of the large rapidity regime alluded to before. In fact the behaviour ought to hold at any value of the coupling (see Appendix~\ref{sec:sub} and~\cite{Eden:2006rx}). It can be thought as a boundary condition relevant to the low-lying spectrum of scaling dimensions at large spin. This is how the scale $2\log{S}$, inherent to the large spin background enters the analysis. In the next section we will see that \eqref{constant} is consistent with the results available in the literature.

\subsection{Energy}

Equipped with the densities it is straightforward to compute the scaling dimension at large spin. The weak-coupling expression
\beq\label{EnergyOL}
\Delta = S+L + g^2\sum_{j=1}^{K}\frac{1}{u_{j}^2+\ft{1}{4}} + g^2\sum_{j=1}^{\bar{K}}\frac{1}{\bar{u}_{j}^2+\ft{1}{4}} + O(g^4)\,
\eeq
may be evaluated using formula \eqref{EMformula}. The result can only depend on the density $\rho_{+}$ since magnons and anti-magnons enter symmetrically. We find
\beq
\Delta = S+L + g^2\int \frac{\rho_{+}(u)}{u^2+\ft{1}{4}}du - g^2\sum_{j=1}^{K_h}\frac{1}{u_{h, j}^2+\ft{1}{4}}- g^2\sum_{j=1}^{\bar{K}_h}\frac{1}{\bar{u}_{h,j}^2+\ft{1}{4}} + O(g^4)\, .
\eeq
After plugging the density~(\ref{LO-densities}) together with the constant~(\ref{constant}), we observe that the scaling dimension admits the following decomposition 
\beq
\Delta -\Delta_{\textrm{vacuum}} = \sum_{j=1}^{K_h}E(u_{h, j})+\sum_{j=1}^{\bar{K}_h}E(\bar{u}_{h, j}) + o(S^0)\, ,
\eeq
where
\beq
\Delta_{\textrm{vacuum}} = S+4g^2(\log{(2S)}+\gamma_E) + 1+ o(S^0)\, ,
\eeq
is the twist-one scaling dimension at large spin~\cite{Gromov:2008qe,Beccaria:2009wb,Beccaria:2009ny}. We verify that it coincides with~(\ref{deltavac6}) at the relevant order since $\Gamma_{\textrm{cusp}}(g) = 4g^2+O(g^4)$ and $B_{2}(g) = 2+O(g^4)$. This confirms the validity of~\eqref{constant}. Having subtracted out the vacuum contribution, we may now turn to the expression for the energy of a hole and anti-hole. We find that it is given by 
\beq
E(u)= \frac{1}{2}+g^2\left(\psi(\tfrac{1}{2}+i u)+\psi(\tfrac{1}{2}-i u)-2\psi(1)\right) + O(g^4)\, ,
\eeq
which is precisely half of the energy for the holes in $\mathcal{N}=4$ SYM theory~\cite{Belitsky:2006en,Basso:2010in, Gaiotto:2010fk},
\beq\label{E1/2}
E^{\mathcal{N}=6}_{\textrm{hole}}(u) = \frac{1}{2}E^{\mathcal{N}=4}_{\textrm{hole}}(u)\, .
\eeq
Later, we will find this relation to be true at any coupling.

\subsection{Bethe-Yang equations for holes}\label{sec:BYol}

The densities \eqref{LO-densities} will allow us to derive the spectral equations for holes at weak coupling. To do this we have to express \eqref{logY} in terms of holes and anti-holes. This will be the goal of this subsection. Using \eqref{EMformula} one more time, we end up with the expression
\beq\label{Ztricky}
2\pi Z(u) = 2L\, \textrm{arctan}{(2u)} + 2\int_{-a}^{a} dv\bar{\rho}(v)\textrm{arctan}{(u-v)} -2\sum_{j=1}^{\bar{K}_h}\textrm{arctan}{(u-\bar{u}_{h, j})}\, .
\eeq
Naively, this step seems straightforward since it only amounts to integrating the density $\bar{\rho}(v)$ that we already worked out. There is a catch, however. The integral that we have to perform is not convergent when extended over the full rapidity axis. The reason is that the odd part of the density $\rho_{+}(v)$ does not decay fast enough at large rapidity. It scales as $1/v$ at large $v$,
\beq \label{odddensity}
\frac{\pi v}{2}(\rho_{+}(-v)-\rho_{+}(v)) = \sum_{j=1}^{K_{h}}u_{h, j}+\sum_{j=1}^{\bar{K}_{h}}\bar{u}_{h, j}+ O(1/v^2)\, .
\eeq
This is unfortunately insufficient to make integrals of the type
\beq
\int_{-a}^{a} dv\, \rho_{+}(v)\textrm{arctan}{(v)} \sim \int_{-a}^{a} dv\, \rho_{+}(v)\textrm{sign}{(v)}
\eeq
convergent if the parameter $a=S$ is not kept finite. To properly integrate we should then keep track of the boundaries of the support and, in view of \eqref{odddensity}, work out the leading contribution to the density for $u\sim S$. This construction is feasible and performed in Appendix \ref{sec:sub}. It is however possible to take a detour and avoid
treating the density in the semiclassical regime. This is how we shall proceed below.

The important observation is that in the end we are only interested in computing the counting function for states that fulfil the zero-momentum condition. It is then convenient to introduce a new function $y(u)$, for which the dependence on the spin-chain total momentum has been removed. We write
\beq\label{Yqy}
Y(u) \equiv q \,y(u)\, ,
\eeq
where factor $q$ is defined as
\beq \label{qfactor}
q= (-1)^{K}\sqrt{(-1)^{K+\bar{K}}e^{i P}}\, ,
\eeq
with $P$ the spin-chain total momentum, see~(\ref{eiP}). We will see later that the properties of $y(u)$ will impart a direct physical interpretation to $q$. For the time being we ask the reader to take it on faith that this is a convenient redefinition.  The function $y(u)$ may be written as
\beq
y(u)=(-1)^{K_h} \left(\frac{\ft{1}{2}+iu}{\ft{1}{2}-iu}\right)^L\prod^{\bar{K}}_{j=1} \frac{1+iu-i\bar{u}_j}{1-iu+i\bar{u}_j} \sqrt{\frac{\ft{1}{2}+i\bar{u}_j}{\ft{1}{2}-i\bar{u}_j}}\prod^K_{j=1} \sqrt{\frac{\ft{1}{2}+iu_j}{\ft{1}{2}-iu_j}} \,.
\eeq
To evaluate $y(u)$ we introduce the corresponding counting function
\beq
z(u) = \frac{1}{2\pi i}\log{(-1)^{K_h}y(u)}\,. 
\eeq
This definition provides us with a regularisation of the large rapidity divergence mentioned above. Note that this regularisation procedure does not affect the density which is given by derivative of $z(u)$. We thus have to compute
\beq\label{Zeasy}
\begin{aligned}
2\pi& z(u) = 2L\, \textrm{arctan}{(2u)}-\sum_{j=1}^{K_{h}}\textrm{arctan}{(2u_{h, j})}-\sum_{j=1}^{\bar{K}_{h}}\textrm{arctan}{(2\bar{u}_{h, j})}\\
&+ \int_{-a}^{a} dv\bigg[2\textrm{arctan}{(u-v)}\bar{\rho}(v)+\textrm{arctan}(2v)\rho_{+}(v)\bigg]  -2\sum_{j=1}^{\bar{K}_h}\textrm{arctan}{(u-\bar{u}_{h, j})}\, ,
\end{aligned}
\eeq
instead of~(\ref{Ztricky}). The integral on the right-hand side is now  finite when $a\sim S\rightarrow \infty$ and we may proceed using our previous expressions for the large spin densities found in the regime $u = O(S^0)$. The remaining steps are purely algebraic. First, we need to compute the integral
\beq
\mathcal{I}_{+}(u) = \dashint_{-\infty}^{\infty}dv\rho_{+}(v)\bigg[\textrm{arctan}{(u-v)}+\textrm{arctan}(2v)\bigg]\, ,
\eeq
where the principal value refers to the integration at infinity. Using the integral equation~(\ref{rhop}) it can be cast as
\beq
\mathcal{I}_{+}(u) = \pi\int_{0}^{u}dv\rho_{+}(v)-\frac{1}{2}\int_{0}^{u}dv I_{+}(v) + \dashint_{-\infty}^{\infty}dv\rho_{+}(v)\bigg[\textrm{arctan}{(2v)}-\textrm{arctan}{(v)}\bigg]\, .
\eeq
Now, using our expression for the density $\rho_{+}$, we get
\beq
\mathcal{I}_{+}(u) = \mathcal{I}_{\textrm{vacuum}}(u) + \sum_{j=1}^{K_{h}}\mathcal{I}_{+}(u, u_{h, j}) + \sum_{j=1}^{\bar{K}_{h}}\mathcal{I}_{+}(u, \bar{u}_{h, j}) \, ,
\eeq
where the vacuum contribution reads
\beq\label{Ivacuum}
\mathcal{I}_{\textrm{vacuum}}(u)  =  2u\log{(2S)}-i\log{\frac{\Gamma(\tfrac{3}{2}-iu)}{\Gamma(\tfrac{3}{2}+iu)}}\, ,
\eeq
while the hole with rapidity $v$ furnishes
\beq\label{Iplus}
\mathcal{I}_{+}(u, v) = -\frac{i}{2}\log{\frac{\Gamma(2+iu-iv)\Gamma(\ft{3}{2}-iu)\Gamma(\ft{3}{2}+iv)}{\Gamma(2-iu+iv)\Gamma(\ft{3}{2}+iu)\Gamma(\ft{3}{2}-iv)}} \, .
\eeq
The second relevant integral is
\beq
\mathcal{I}_{-}(u) = \int_{-\infty}^{\infty} dv\rho_{-}(v)\textrm{arctan}{(u-v)} = -\pi\int_{-\infty}^{u}dv\rho_{-}(v)+\frac{1}{2}\int_{-\infty}^{u}dv I_{-}(v) -\frac{\pi}{2}\int dv \rho_{-}(v)\, ,
\eeq
which may be evaluated to
\beq
\mathcal{I}_{-}(u) = \sum_{j=1}^{K_{h}}\mathcal{I}_{-}(u, u_{h, j})-\sum_{j=1}^{\bar{K}_{h}}\mathcal{I}_{-}(u, \bar{u}_{h, j})\, ,
\eeq
with
\beq\label{Iminus}
\mathcal{I}_{-}(u, v) = \frac{i}{2}\log{\frac{\Gamma(1+\ft{iu-iv}{2})\Gamma(\ft{3}{2}-\ft{iu-iv}{2})}{\Gamma(1-\ft{iu-iv}{2})\Gamma(\ft{3}{2}+\ft{iu-iv}{2})}} \, .
\eeq
We are now in a position to combine all the pieces together. Equations \eqref{Zeasy}, \eqref{Ivacuum}-\eqref{Iplus} and~\eqref{Iminus} lead to the sought-after expression for $y(u)$ given solely in terms of hole rapidities
\beq\label{yu}
y(u) = (2S)^{2iu}\frac{\Gamma(\ft{1}{2}-iu)}{\Gamma(\ft{1}{2}+iu)}\prod_{j=1}^{K_{h}}S(u, u_{h, j})\prod_{j=1}^{\bar{K}_{h}}\bar{S}(u, \bar{u}_{h, j})\,.
\eeq
We defined the S-matrix elements as
\beqa
&&S(u,v) =-\left[\frac{\Gamma(\tfrac{1}{2}-iu)}{\Gamma(\tfrac{1}{2}+iu)}\frac{\Gamma(\tfrac{1}{2}+iv)}{\Gamma(\tfrac{1}{2}-iv)}\right]^{1/2} 2^{i(u-v)} \frac{\Gamma \left(1+\tfrac{iu-iv}{2}\right)}{\Gamma \left(1-\tfrac{iu-iv}{2}\right)}\, ,\\
&&\bar{S}(u,v)=\, \, \, \, \, \, \left[\frac{\Gamma(\tfrac{1}{2}-iu)}{\Gamma(\tfrac{1}{2}+iu)} \frac{\Gamma(\tfrac{1}{2}+iv)}{\Gamma(\tfrac{1}{2}-iv)} \right]^{1/2}2^{i(u-v)} \frac{\Gamma \left(\tfrac{1}{2}+\tfrac{iu-iv}{2}\right)}{\Gamma \left(\tfrac{1}{2}-\tfrac{iu-iv}{2}\right)}\, .
\eeqa
A similar result may be produced for $\bar{y}(u)$ corresponding to $\bar{Y}(u)$ by swapping excitations and anti-excitations.

As an immediate application of the expression~(\ref{yu}) we derive the expression for the momentum $p(u)$ of a hole (or anti-hole) to leading order at weak coupling. The latter controls the large spin behaviour
\beq
\log{Y(u)} = 2ip(u)\log{(2S)} + \ldots\, ,
\eeq
from which we conclude that $p(u) = u$ at leading order in coupling constant. Here again we observe that it is half the result found for hole in the $\mathcal{N}=4$ SYM theory~\cite{Basso:2010in, Gaiotto:2010fk}
\beq
p_{\mathcal{N}=6}(u) = \frac{1}{2}p_{\mathcal{N}=4}(u)\, .
\eeq
As we shall see later, even though at higher orders of perturbation theory the momentum-to-rapidity maps receive corrections, the above relation remains valid. 

Having determined $Y(u)$ in terms of the hole rapidities, we may write down the Bethe-Yang equations for the holes and anti-holes. As discussed in Section \ref{sec:hah}, their rapidities should solve the equations~(\ref{1plusY}) which now read
\beq\label{BYeqol}
(2S)^{-2iu_{h, k}} = q\, \frac{\Gamma(\ft{1}{2}-iu_{h, k})}{\Gamma(\ft{1}{2}+iu_{h, k})}\prod_{j \neq k}^{K_{h}}S(u_{h, k}, u_{h, j})\prod_{j=1}^{\bar{K}_{h}}\bar{S}(u_{h, k}, \bar{u}_{h, j})\, ,
\eeq
and similarly for anti-holes. The solutions to these equations determine the spectrum of large spin operators around the minimal trajectory. From them we infer that $S(u, v)$ and $\bar{S}(u, v)$ play the role of S-matrices for scattering of holes and anti-holes, on which we will comment further in Section~\ref{sec:u4Sm}.

\subsection{Comments on the twist $q$} \label{sec:twistq}

The equations~(\ref{BYeqol}) are structurally identical to the one found for holes in $\mathcal{N}=4$ SYM theory. The only subtle difference between the two theories comes from the twists $q, \bar{q}$.

For the $\mathcal{N}=4$ theory the zero-momentum condition enforces $q=1$, so that the boundary conditions for holes are periodic. For the $\mathcal{N}=6$ theory, on the other hand, we find the following relations,
\beq\label{super-rules}
q\,\bar{q} = e^{iP}\,, \qquad q/\bar{q} = (-1)^F\,, 
\eeq
where $\bar{q}$ is obtained by replacing $(\bar{K}, K_h)$ by $(K, \bar{K}_h)$ in \eqref{qfactor} and where we introduced $F\equiv \ft{1}{2}(K_h-\bar{K}_h)$. The latter quantity is an integer according to the selection rule \eqref{KhpKth}. This number also appeared in the study of the Bykov model in~\cite{Basso:2012bw}. In the sigma model context it counts the number of fermions in vertex operators corresponding to states of the theory in finite volume. We will come back to this interpretation in Section~\ref{sec:Bykov}. Here we simply observe that even after imposing the condition that $e^{iP} = 1$ we are left with two possible values for $q = 1/\bar{q}$, associated to the two solutions to $q^2 = (-1)^F$. The possible values for the twist are then $q=\pm 1$ or $q=\pm i$ subject to the parity of $F$.

Even if the presence of the twists in the Bethe equations may seem insignificant at first, it is of pivotal importance to understand the complete spectrum of the theory and its discrete symmetries. Since \eqref{super-rules} have always \textit{two} solutions, there are typically twice as many states for the ABJM theory than there are for $\N=4$ SYM theory. This seems to be related to the fact that we have to deal with an alternating spin chain with two different reference fields at odd and even sites. The presence of the twists  reveals a $\mathbb{Z}_2$ symmetry present only at the level of the spectral equations, in other words, when the holes are put on finite cylinder. It is not part of the asymptotic or on-shell data of the theory in infinite volume ($2\log{(2S)}=\infty$). To specify the global quantum numbers of a state at large but finite spin one has to assign the number of excitations, $K_h$ and $\bar{K}_h$, fixing at the same time the value of the twist $q$ to one of the two possible solutions of~\eqref{super-rules}. Even for the large spin vacuum itself we have $q=\pm 1$ corresponding to even and odd spin trajectory! These are distinct states of the theory in finite volume and we shall comment on their energy difference in the concluding section of this paper. 

\subsection{The emergence of the $SU(4)$ symmetry}\label{su4sym}

So far we have only discussed the dynamics of the holes and anti-holes in the symmetric channel where the polarisations are homogeneous, which corresponds to the subsector  with $\mathfrak{su}(4)$ weights $[K_{h}, 0, \bar{K}_{h}]$ on top of the twist-one vacuum weights. The GKP string is expected however to have the full $SU(4)$ symmetry under which the holes and anti-holes ought to transform in fundamental and antifundamental rep, respectively. This is not directly visible at the spin-chain level in the grading used in this paper. The reason is that the ABA equations are built on top of the so-called Berenstein-Maldacena-Nastase (BMN) vacuum~\cite{Berenstein:2002jq} that breaks this symmetry down to an $SU(2)$ subgroup. The way the $SU(4)$ symmetry is recovered at large spin is by a particular arrangement of roots called stacks. In this section we will discuss these special root configurations.

\begin{figure}
\begin{center}
\begin{tikzpicture}[cross/.style={path picture={ 
  \draw[black]
(path picture bounding box.south east) -- (path picture bounding box.north west) (path picture bounding box.south west) -- (path picture bounding box.north east);
}}]
\draw (2,0)--(0,0);
\draw (0,0)--(-2,0);
\node at (-2,0.7) {$u_a$};
\node  at (0,0.7) {$u_b$};
\node  at (2,0.7) {$u_c$};

\node at (3,3.6) {\fontsize{8}{8} $u^+_{\bar{4}}$};
\node  at (3.0, 3.2) {\fontsize{8}{8} $u_{3}$};
\node  at (3.0, 2.8) {\fontsize{8}{8} $\, u^-_{\bar{4}}$};
\node  at (0,3.6) {\fontsize{8}{8} $u_2$};
\node  at (-3,3.6) {\fontsize{8}{8} $u^+_{4}$};
\node  at (-3.1,3.2) {\fontsize{8}{8} $\,\,u_3$};
\node  at (-3,2.8) {\fontsize{8}{8} $u^-_{4}$};
\draw[<-,thick] (2,0.7+0.5)--(2.5,2+0.5);
\draw[<-, thick] (0,0.7+0.5)--(0,2+0.5);
\draw[<-, thick] (-2,0.7+0.5)--(-2.5,2+0.5);
\draw[cross,fill=white,draw=white] (2.4,2.7) rectangle (2.6, 2.9);
\draw[cross,fill=white,draw=white] (2.4,3.1) rectangle (2.6,3.3);
\draw[cross,fill=white,draw=white] (2.4,3.5) rectangle (2.6,3.7);
\draw[cross,fill=white,draw=white] (-0.1, 3.1) rectangle (0.1,3.3);
\draw[cross,fill=white,draw=white] (-2.6, 2.7) rectangle (-2.4,2.9);
\draw[cross,fill=white,draw=white] (-2.6, 3.1) rectangle (-2.4,3.3);
\draw[cross,fill=white,draw=white] (-2.6,3.5) rectangle (-2.4,3.7);
\draw[fill=white,draw=black] (2,0) circle (0.35);
\draw[fill=white,draw=black] (0,0) circle (0.35);
\draw[fill=white,draw=black] (-2,0) circle (0.35);
\end{tikzpicture}
\end{center}
\caption{The $SU(4)$ Dynkin diagram for the stacks. The $SU(2)$ subgroup associated to the middle node is directly inherited from the spin chain. The $SU(2)\times SU(2)$ group associated to the left and right nodes are restored through the formation of stacks, whose contents are depicted on top of them.} \label{fig:su4}
\end{figure}
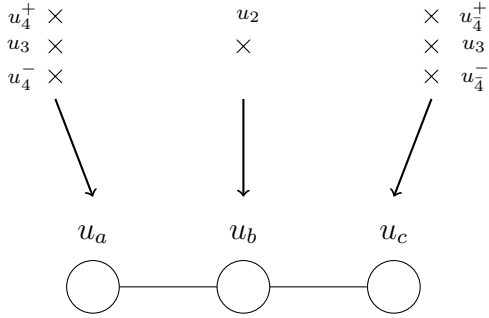

The stacks relevant for the restoration of the $SU(4)$ symmetry are particularly simple and quite similarly to those found in the context of the $\mathcal{N}=4$ SYM theory~\cite{Basso:2012bw}. One difference as compared to the $\mathcal{N}=4$ theory is that we have two types of isotopic stacks, one for each $SU(2)$ subgroup broken by the spin-chain vacuum. We shall denote these by $a$ and $c$. The stack of type $a$ is formed by \textit{two} $u_4$ roots and \textit{one} $u_3$ root,
\beq
u_{a} = \left\{u^+_{4}, u_{3}, u^-_{4} \right\}\,.
\eeq
The $u_4$ roots acquire imaginary parts and are centred around the $u_{3} = u_a$ root, i.e. $u^{\pm}_4 = u_a \pm \tfrac{i}{2}$. Similarly, the stack $c$ is formed by two $u_{\bar{4}}$ roots and one $u_3$ root
\beq
u_{c} = \left\{u^+_{\bar{4}}, u_3, u^-_{\bar{4}} \right\}\,,
\eeq
where the roots $u_{\bar{4}}$ are again $u_3$-centric. The last of the $SU(4)$ symmetry roots simply coincide with $u_2$
\beq
u_b = u_{2}\,.
\eeq
These tree types of roots are in a one-to-one correspondence with the three nodes of the ``emergent'' $SU(4)$ symmetry, see Figure~\ref{fig:su4}.

To verify that our interpretation of the stacks is correct, we should investigate their effects on the densities of roots at large spin. The starting point are the ABA equations for magnons and anti-magnons in the presence of $K_{a, b}$ stacks of type $a,b$. They read
\beqa \label{ABJMstack}
&&\left(\frac{u_k+\tfrac{i}{2}}{u_k-\tfrac{i}{2}}\right)^L =\prod^{\bar{K}'}_{j=1} \frac{u_k-\bar{u}_j-i}{u_k-\bar{u}_j+i}\prod^{K_{c}}_{j=1} \frac{u_k-u_{c,j}-\ft{3i}{2}}{u_k-u_{c,j}+\ft{3i}{2}}\prod^{K_{a}}_{j=1} \frac{u_k-u_{a,j}+\ft{i}{2}}{u_k-u_{a,j}-\ft{i}{2}}\,, \\
&&\left(\frac{\bar{u}_k+\tfrac{i}{2}}{\bar{u}_k-\tfrac{i}{2}}\right)^L =\prod^{K'}_{j=1} \frac{\bar{u}_k-u_j-i}{\bar{u}_k-u_j+i}\prod^{K_{a}}_{j=1} \frac{u_k-u_{a,j}-\ft{3i}{2}}{u_k-u_{a,j}+\ft{3i}{2}}\prod^{K_{c}}_{j=1} \frac{u_k-u_{c,j}+\ft{i}{2}}{u_k-u_{c,j}-\ft{i}{2}}\,,
\eeqa
where $K' = K-2K_{a}$ and $\bar{K}' = K-2K_{c}$ are the numbers of \textit{real} magnons and anti-magnons, respectively. These are precisely the ones that form the large spin background. The magnons that compound stacks of type $a,c$ modify the relation between the number of holes and number of magnons. The reason is that the total number of sites in the lattice of mode numbers $\mathfrak{S}$ and $\bar{\mathfrak{S}}$ are now given by $L+\bar{K}'+K_{c}-K_{a}-1$ and $L+K'+K_{a}-K_{c}-1$, respectively. This is because of the additional phases in the right-hand sides of~(\ref{ABJMstack}). From these expressions we conclude that the number of holes and anti-holes read $K_{h} = L+\bar{K}'-K' + K_{c}-K_{a}-1$ and $\bar{K}_{h} = L+K'-\bar{K}' + K_{a}-K_{c}-1$, or equivalently
\beq \label{sumanddiff}
K_{h}+\bar{K}_{h} = 2L-2\, , \qquad K_{h}-\bar{K}_{h} = 2\bar{K}'-2K'+2K_{c}-2K_{a}\, .
\eeq
The total number of both holes and anti-holes is the same as before. More importantly, the difference $F = \ft{1}{2}(K_{h}-\bar{K}_{h})$, which we called fermion number in the previous subsection, is still an integer so the same superselection rule applies. This selection rule is actually observed for any states in this theory, even after allowing for higher-twist excitations. This has a simple explanation if we think in terms of the string world-sheet description. In this context, the selection rule is understood as resulting from the $U(1)$ gauge invariance of the physical states of the theory~\cite{Basso:2012bw} (see also Section~\ref{sec:Bykov}). Among all the asymptotic excitations that can appear in the composition of such state, only those corresponding to the holes and anti-holes are charged under this symmetry. This is why the selection rule is oblivious to the presence of heavier excitations, which are neutral under the gauge symmetry, or to the presence of stacks studied here, since they only implement the $SU(4)$ symmetry that commutes with the gauge symmetry. Finally, we notice that magnons and anti-magnons become interchangeable in~(\ref{ABJMstack}) if we exchange the roles of roots of type $a$ and $c$ simultaneously.

The next step is rather straightforward. We want to compute the correction to the densities induced by the stacks. Since the equations are linear we can subtract what we have learned before and focus on the densities sourced directly by the stacks. A simple algebra reveals that we should add to~(\ref{sources})
\beq
\begin{aligned}
\delta I_{+} &=& \sum_{j=1}^{K_{a}}\bigg[\frac{3}{\ft{9}{4}+(u-u_{a, j})^2} - \frac{1}{\ft{1}{4}+(u-u_{a, j})^2}\bigg] + a \rightarrow c\, ,\\
\delta I_{-} &=& \sum_{j=1}^{K_{c}}\bigg[\frac{3}{\ft{9}{4}+(u-u_{c, j})^2} + \frac{1}{\ft{1}{4}+(u-u_{c, j})^2} \bigg] - c \rightarrow a\, .\\
\end{aligned}
\eeq
The associated contribution to the densities are
\beq\label{drho}
2\pi\delta\rho_{\pm}(u) = -\sum_{j=1}^{K_{a}}\frac{1}{\ft{1}{4}+(u-u_{a, j})^2}\mp \sum_{j=1}^{K_{c}}\frac{1}{\ft{1}{4}+(u-u_{c, j})^2}\, ,
\eeq
or equivalently
\beq
2\pi\delta\rho(u) =  -\sum_{j=1}^{K_{a}}\frac{1}{\ft{1}{4}+(u-u_{a, j})^2}\, , \qquad 2\pi\delta\bar{\rho}(u) =  -\sum_{j=1}^{K_{c}}\frac{1}{\ft{1}{4}+(u-u_{c, j})^2}\, .
\eeq
We immediately notice that stacks of type $a$ only contribute to the density of magnons and those of type $c$ to the density of anti-magnons. This was somewhat expected since stacks of type $a$ implements an $SU(2)$ symmetry under which only holes are charged. In the same vein, the $c$ stacks only couple to anti-holes. This interpretation will become more manifest below.

With the help of the densities~(\ref{drho}) we can compute the large spin energy of these stacks. Their overall contribution to the anomalous dimension is
\beq
\delta \Delta = g^2\int du \frac{\delta\rho_{+}(u)}{u^2+\ft{1}{4}} + \sum_{j=1}^{K_{a}}\frac{2g^2}{u_{a, j}^2+1}+\sum_{j=1}^{K_{c}}\frac{2g^2}{u_{c, j}^2+1} \, ,
\eeq
and it is easily seen to vanish when~(\ref{drho}) is substituted. Thus, the energy induced by stacks and stored in the continuum of \textit{real} roots exactly cancelled against the bare energy of the stacks which consist of one-strings of magnons or anti-magnons. This is the same mechanism as the one observed for the isotopic $SU(2)$ roots in the $\mathcal{N}=4$ theory~\cite{Basso:2010in}.

This observation substantiates our interpretation of stacks as isotopic roots. The presence of the stacks influences the scattering among the holes and the anti-holes as they allow scattering to take place in different $SU(4)$ channels, but at the end only the holes and anti-holes are carriers of the energy of the state. To make this point more precise we shall now consider the effect of the stacks on the scattering of holes by computing the counting functions in their background.

The analysis is very similar to the one performed in the previous subsection. We compute the function $Y(u)$ using~(\ref{Yqy}) with the twist factor modified to account for the presence of stacks
\beq\label{qiso}
q = (-1)^{K'+K_{a}}\sqrt{(-1)^{K'+\bar{K}'+K_{a}+K_{c}}\,e^{iP}}\,.
\eeq
The little counting function, $y(u)$, is
\beq
\begin{aligned}
y(u) = (-1)^{K_h}&\left(\frac{\ft{1}{2}+iu}{\ft{1}{2}-iu}\right)^L\prod_{j=1}^{\bar{K}'}\frac{1+iu-i\bar{u}_{j}}{1-iu+i\bar{u}_{j}}\prod_{j=1}^{K_{c}}\frac{\ft{3}{2}+iu-iu_{c, j}}{\ft{3}{2}-iu+iu_{c, j}}\prod_{j=1}^{K_{a}}\frac{u-u_{a, j}+\ft{i}{2}}{u-u_{a, j}-\ft{i}{2}} \\
&\times \sqrt{\prod_{j=1}^{K'}\frac{\ft{1}{2}+iu_{j}}{\ft{1}{2}-iu_{j}}\prod_{j=1}^{\bar{K}'}\frac{\ft{1}{2}+i\bar{u}_{j}}{\ft{1}{2}-i\bar{u}_{j}}\prod_{j=1}^{K_{a}}\frac{1+iu_{a, j}}{1-iu_{a,j}}\prod_{j=1}^{K_{c}}\frac{1+iu_{c, j}}{1-iu_{c,j}}}\, ,
\end{aligned}
\eeq
where again the factor under the square root in the second line is introduced to regularise the infinite product over the roots at large spin. Taking into account the additions to the densities~(\ref{drho}) and repeating the analysis performed in Section~\ref{sec:BYol}, one easily observes that the dependence on the roots of type $c$ drops out, leaving the simple product
\beq\label{yuiso}
y(u) = y(u)_{\textrm{before}}\prod_{j=1}^{K_{a}}\frac{u-u_{a, j}+\ft{i}{2}}{u-u_{a, j}-\ft{i}{2}}\,.
\eeq
The function $y(u)_{\text{before}}$ is the one in~(\ref{yu}). The expression for $\bar{y}(u)$ would of course be obtained by replacing $a$-roots by $c$-roots in the formula above. The result~(\ref{yuiso}) is telling us, as expected, that the holes are charged only with respect to \textit{one} $SU(2)$ group. They belong therefore to the $\bf{4}$ of $SU(4)$ while the anti-holes are in the $\bar{\textbf{4}}$. We notice that despite the fact that the definitions of the twists $q$ and $\bar{q}$, see (\ref{qiso}), have changed, they still satisfy the relations~(\ref{super-rules}).

Now that we explained how stacks couple to the momentum-carrying degrees of freedom, we would like to comment on the equations for the stacks. These equations are also of the type $1+Y_{a}(u_{a, k}) = 0$ with
\beq\label{Ya}
Y_{a}(u) =  -\left(\frac{u-i}{u+i}\right)^{L}\prod_{j=1}^{K'}\frac{u-u_{j}+\ft{i}{2}}{u-u_{j}-\ft{i}{2}}\prod_{j=1}^{\bar{K}'}\frac{u-\bar{u}_{j}-\ft{3i}{2}}{u-\bar{u}_{j}+\ft{3i}{2}}\prod_{j=1}^{K_{a}}\left(\frac{u-u_{a, j}+i}{u-u_{a, j}-i}\right)^2\prod_{j=1}^{K_{c}}\frac{u-u_{c, j}-2i}{u-u_{c, j}+2i}\, .
\eeq
The counting function for stacks is derived by observing that the stacks are compounds of more fundamental (spin-chain) roots. One thus simply have to fuse the scattering phases for these roots~\cite{Minahan:2008hf,Gromov:2008qe}.

To simplify~(\ref{Ya}) at large spin we essentially need to compute the following two integrals
\beq
\mathcal{J}_{\pm} = \int dv \rho_{\pm}(v)\big[\textrm{arctan}{(\ft{2}{3}(u-v))}\mp\textrm{arctan}{(2(u-v))}\big]\, .
\eeq
We can rewrite \eqref{Ya} as
\beq
\begin{aligned}\label{YalargeS}
Y_{a}(u) =  \left(\frac{1+iu}{1-iu}\right)^{L}&e^{i\mathcal{J}_{+}-i\mathcal{J}_{-}}\prod_{j=1}^{\bar{K}_{h}}\frac{\ft{3}{2}-iu+i\bar{u}_{h, j}}{\ft{3}{2}+iu-i\bar{u}_{h, j}}\prod_{j=1}^{K_{a}}\frac{1-iu+iu_{a, j}}{1+iu-iu_{a, j}}\prod_{j=1}^{K_{c}}\frac{2+iu-iu_{c, j}}{2-iu+iu_{c, j}} \\
&\times \prod_{j=1}^{K_{h}}\frac{u-u_{h, j}-\ft{i}{2}}{u-u_{h, j}+\ft{i}{2}}\prod_{j=1}^{K_{a}}\frac{u-u_{a, j}+i}{u-u_{a, j}-i}\, .
\end{aligned}
\eeq
The first line of this product greatly simplifies after using the expressions for the densities. This follows from%
\footnote{The contribution to $\mathcal{J}_{+}$ coming from the vacuum density~(\ref{vacrho}) is actually ill-defined. To give it a proper meaning we interpret it as originating from a pair hole-anti-hole carrying opposite rapidities $u_{h, \textrm{vac}} = -\bar{u}_{h, \textrm{vac}} = 2S$. Computing $\mathcal{J}_{+}$ this way and taking then $S\rightarrow \infty$ yields~(\ref{Jpm}). }
\beq\label{Jpm}
\begin{aligned}
\left(\frac{1+iu}{1-iu}\right)^{L\pm L}e^{2i\mathcal{J}_{\pm}} =  \, \, &\prod_{j=1}^{K_{h}}\frac{\ft{3}{2}+iu-iu_{h, j}}{\ft{3}{2}-iu+iu_{h, j}}\prod_{j=1}^{K_{a}}\left[\frac{1+ iu - iu_{a, j}}{1- iu+iu_{a, j}}\right]^{\pm 1}\frac{2-iu+iu_{a, j}}{2+iu-iu_{a, j}}\\
&\times (\textrm{conjugate})^{\pm 1}\, ,
\end{aligned}
\eeq
where ``conjugate''  refers to particle-antiparticle transformation, that is $\{\textrm{hole, root } a\} \rightarrow \{\textrm{anti-hole, root } c \}$. This leaves us with the simple expression
\beq
Y_{a}(u) = \prod_{j=1}^{K_{h}}\frac{u-u_{h, j}-\ft{i}{2}}{u-u_{h, j}+\ft{i}{2}}\prod_{j=1}^{K_{a}}\frac{u-u_{a, j}+i}{u-u_{a, j}-i}\, .
\eeq
This formula implies that the roots of type $a$ will satisfy Bethe equations of a $SU(2)$ Heisenberg spin chain with inhomogeneities given by the holes rapidities. This corroborates the role played by roots $a$ advocated before. We should stress that these equations are valid when no roots of type $b$ are included. The latter are however easily incorporated since they couple to roots $a$ in the same way as the roots $u_{3}$ and $u_{2}$ couple to one another in the ABA equations. The complete set of $SU(4)$ equations is then easily derived and will be given in the next section, \textit{cf.}~(\ref{ABAN=6}).

Our discussion so far was restricted to leading order in perturbation theory. Our results, however, happen to be valid at any coupling. In particular, the densities~(\ref{drho}) are exact, their energy is always zero and the scattering phases involving the isotopic roots are the same at any coupling. In other words the above analysis is in fact already exact. This was demonstrated in the context of the $\mathcal{N}=4$ theory in~\cite{Basso:2010in}. We leave it as an exercise to  incredulous readers to check that the same thing happens in the $\mathcal{N}=6$ theory.

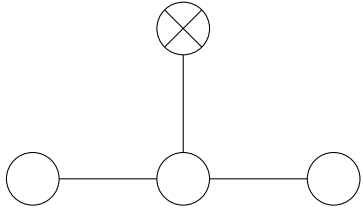
\begin{figure}
\begin{center}
\begin{tikzpicture}[cross/.style={path picture={ 
  \draw[black]
(path picture bounding box.south east) -- (path picture bounding box.north west) (path picture bounding box.south west) -- (path picture bounding box.north east);
}}]
\draw (2,0)--(0,0);
\draw (0,0)--(-2,0);
\draw (0,0)--(0,2);
\draw[fill=white] (0,0) circle (0.35);
\draw[fill=white] (2,0) circle (0.35);
\draw[fill=white] (-2,0) circle (0.35);
\draw[cross, fill=white] (0,2) circle (0.35);
\end{tikzpicture}
\end{center}
\caption{Symmetry enhancement at weak coupling. The three bosonic nodes of the $\mathfrak{su}(4)$ symmetry algebra are enhanced by a fermionic node to leading order at weak coupling. The resulting Dynkin diagram is the one of the $\mathfrak{osp}(1,1|4)$ Lie super-algebra.} \label{fig:symenh}
\end{figure}

Finally, it is interesting to observe that to leading order at weak coupling the symmetry is actually larger than $SU(4)$. This is because the fermionic $u_1$ roots, that lie at the far end of the $\mathfrak{osp}(2,2|6)$ Dynkin diagram in Figure~\ref{fig:su4}, interact with the $u_2$ roots only. It is not apparent at this order that $u_1$ roots will eventually carry energy and momentum at higher loops and will parametrize fermionic excitations on top of the GKP string. These fermions, being in the small momentum domain in the terminology of~\cite{Basso:2010in}, may be interpreted as having zero momentum at this order. In line with the arguments of~\cite{Alday:2007mf}, they behave as supersymmetry generators. The symmetry algebra $\mathfrak{su}(4)$ is then lifted to $\mathfrak{osp}(1,1|4)$,  as depicted in Figure \ref{fig:symenh}, and consequently the spectrum of excitations is classified according to this enhanced symmetry algebra. The extra excitations, obtained by including $u_1$ roots in the analysis, are descendants with respect to the $\mathfrak{osp}(1,1|4)$ algebra of the genuine all-loop ones. Up to a canonical shift they carry the same energy as their primaries. At weak coupling they are stable, but start to decay as soon as the coupling constant is set back on, because their fermionic constituents become dynamical. This is the same mechanism as the one at work for $\mathcal{N}=4$ SYM theory~\cite{Basso:2010in}, where the $\mathfrak{su}(4)$ symmetry group was found to be enhanced to $\mathfrak{su}(1,1|4)$ at one-loop order. The latter algebra includes in particular an $\mathfrak{sl}(2)$ subalgebra that was found~\cite{Gaiotto:2010fk} to play an important role in classifying the excitations emerging in the OPE decomposition of null polygonal Wilson loops at weak coupling.

\section{Bethe-Yang equations at finite coupling} \label{sec:BYfc}

In this section we discuss the general structure of the Bethe-Yang equations for the low-energy excitations of the GKP string in both $\mathcal{N}=4$ SYM and $\mathcal{N}=6$ ABJM theory. We shall first perform the analysis in the $\mathcal{N}=4$ SYM theory, for which partial results are already available in the literature, and then turn to $\mathcal{N}=6$ Chern-Simons-Matter theory. We start with an overview of the excitations around the GKP string for ABJM.

\subsection{The excitations and their dispersion relations}

We observed in the previous section that the dispersion relations for holes for $\mathcal{N}=4$ and $\mathcal{N}=6$ theories are closely related
\beq\label{EpN=46}
E^{\mathcal{N}=6}_h(u)=\frac{1}{2} E^{\N=4}_h (u)\,,\qquad p^{\mathcal{N}=6}_h(u) = \frac{1}{2} p^{\N=4}_h (u)\,.
\eeq
This has a very simple origin. We recall that energy and momentum are determined by $\rho_{+}(u) = \rho(u)+\bar{\rho}(u)$. We noticed before that this combination of densities was directly expressible in terms of the density of magnons found in the $\mathcal{N}=4$ theory hence explaining~(\ref{EpN=46}). This relationship holds true at all loops and will be established in Appendix~\ref{BYapp}. Explicitly,
\beq\label{rho46}
\rho_{+}(u) = \rho_{\mathcal{N}=4}(u)\,, \qquad \qquad \textrm{for any } g\,,
\eeq
where on the RHS the spin and the length of the spin chain in $\N=4$ are \textit{doubled}. The set of $2L-2$ holes in $\N=4$ is comprised by holes and anti-holes of $\N=6$ in accordance with \eqref{sumanddiff}\footnote{Notice that the relation~(\ref{rho46}) would have to be modified in the presence of isotopic roots.}.
Furthermore, the density $\rho_{-}(u) = \rho(u)-\bar{\rho}(u)$ does not receive higher-loop corrections! Please see Appendix~\ref{BYapp} for more details.

Using these special properties of $\rho_{\pm}(u)$ it is straightforward to observe that
\beq
\delta\Delta_{\mathcal{N}=6} = g^2\int du\, \rho_{+}(u)\bigg(\frac{i}{x^{+}(u)}-\frac{i}{x^{-}(u)}\bigg) -\textrm{holes}_{\N=6}\, - \textrm{anti-holes}_{\N=6}\, ,
\eeq
where $x^{\pm}(u) = x(u\pm \ft{i}{2})$ and $x(u)=\ft{1}{2}(u+\sqrt{u^2-4g^2})$ is the Zhukovsky map, becomes
\beq\label{deltadelta}
\delta\Delta_{\mathcal{N}=6} = g^2\int du\,  \rho_{\mathcal{N}=4}(u)\bigg(\frac{i}{x^{+}(u)}-\frac{i}{x^{-}(u)}\bigg) -\textrm{holes}_{\mathcal{N}=4} =\frac{1}{2}\delta\Delta_{\mathcal{N}=4}\, .
\eeq
Adding up the tree-level contributions on both sides, i.e. the twist, which is $1$ for each hole in the $\mathcal{N}=4$ theory and $1/2$ for each hole and anti-hole in the $\mathcal{N}=6$ theory, we arrive at the first relation in~(\ref{EpN=46}). A similar analysis leads to the second equality in~(\ref{EpN=46}) for the momenta, as shown in Appendix~\ref{BYapp}.

An important consequence of the identities~(\ref{EpN=46}) is that at strong coupling, $g = h(\lambda)\gg 1$, and for fixed $u$ we enter the relativistic regime 
\beq\label{Enpreg}
E^{\mathcal{N}=6}_h(u) = \frac{1}{2}m_{\mathcal{N}=4}\cosh{\left(\frac{\pi u}{2}\right)}\, , \qquad p^{\mathcal{N}=6}_h(u) = \frac{1}{2}m_{\mathcal{N}=4}\sinh{\left(\frac{\pi u}{2}\right)}\, ,
\eeq
similar to the one found for the $\mathcal{N}=4$ SYM theory~\cite{Alday:2007mf,Basso:2008tx,Basso:2010in}. It is in this regime that we expect the Bethe ansatz equations for the holes and anti-holes of the $\mathcal{N}=6$ spin chain to match the ones in \cite{Basso:2012bw} conjectured to encode the spectrum of the Bykov model~\cite{Bykov:2010tv}. We will come back to this issue in  Section~\ref{sec:rellimit}.

Finally, let us briefly comment on the other types of excitations. We already mentioned in the introduction that there are twist-one fermions in the $\textbf{6}$ of $\mathfrak{su}(4)$ and a tower of gauge-field excitations, with twists $1,2,3 \dots$, that are neutral under this symmetry. All these excitations have their counterparts in the $\mathcal{N}=4$ theory and their corresponding dispersion relations coincide
\beq\label{Enhighertwist}
E^{\mathcal{N}=6}_\star(p)= E^{\N=4}_\star (p)\, , \qquad \textrm{twist}_{\star} \geq 1\, .
\eeq
The reason for this equality to hold is that these excitations are embedded in the spin chain by means of roots of types $u_{1,2,3}$ situated in the ``tail'' of the Dynkin diagram in Figure~\ref{Dynkin}. For instance, fermions are associated to the fermonic node 3 or 1, while the twist-one gauge field appears as a stack of two $u_{3}$ and one $u_2$ roots. The immediate consequence is that these excitations couple symmetrically to the 4 and $\bar{4}$ nodes. They are therefore only coupled to the symmetric density $\rho_{+} = \rho+\bar{\rho}$, which satisfies an integral equation whose kernel is the same as in the $\mathcal{N}=4$ theory. The only difference is that the inhomogeneous terms come with an extra factor of $2$. This is because the roots on the wing couple to roots $4$ as they would do in the $\mathcal{N}=4$ theory. The inhomogeneous term for $\rho$, or equivalently $\bar{\rho}$, is then identical to the one for the $\mathcal{N}=4$ theory and is consequently half of the term corresponding to $\rho_{+}$. This effect compensates for the ``missing'' factor of $2$ in the formula for the anomalous dimension in the $\mathcal{N}=6$ theory, see Eq.~(\ref{deltadelta}), and eventually leads to~(\ref{Enhighertwist}).%
\footnote{Strictly speaking, this argument only establishes that $E^{\mathcal{N}=6}(u) = E^{\mathcal{N}=4}(u)$. To get to~(\ref{Enhighertwist}) one also has work out the relation between the momenta in these two theories. This is easily done.} This concludes the discussion of the spectrum of asymptotic excitations on top of  the GKP string. 

\subsection{The BY equations for $\mathcal{N}=4$ SYM theory}\label{sec:BYN4}

In the planar $\mathcal{N}=4$ SYM theory, the low-lying excitations of the GKP string are identified with the six scalar fields of the gauge theory, which transform in the vector representation of the $O(6)$ R-symmetry group. Their asymptotic Bethe ansatz or Bethe-Yang equations can be cast in the form
\beq
\begin{aligned}\label{ABAN=4}
S^{-2ip_{\mathcal{N}=4}(u_{k})} &= T_{\mathcal{N}=4}(u_{k})\prod^{K_{h}}_{j\neq k}S_{\mathcal{N}=4}(u_k, u_j)\prod_{j=1}^{K_b}\frac{u_k-u_{b, j}+\ft{i}{2}}{u_{k}-u_{b, j}-\ft{i}{2}}\, , \\
1 &= \prod_{j\neq k}^{K_a}\frac{u_{a, k}-u_{a, j}+i}{u_{a, k}-u_{a, j}-i}\prod_{j=1}^{K_{b}}\frac{u_{a, k}-u_{b, j}-\ft{i}{2}}{u_{a, k}-u_{b, j}+\ft{i}{2}}\, ,\\
\prod_{j=1}^{K_h}\frac{u_{b, k}-u_{j}+\ft{i}{2}}{u_{b, k}-u_{j}-\ft{i}{2}} &= \prod_{j\neq k}^{K_{b}}\frac{u_{b, k}-u_{b, j}+i}{u_{b, k}-u_{b, j}-i}\prod_{j=1}^{K_{a}}\frac{u_{b, k}-u_{a, j}-\ft{i}{2}}{u_{b, k}-u_{a, j}+\ft{i}{2}}\prod_{j=1}^{K_{c}}\frac{u_{b, k}-u_{c, j}-\ft{i}{2}}{u_{b, k}-u_{c, j}+\ft{i}{2}}\, ,\\
1 &= \prod_{j \neq k}^{K_c}\frac{u_{c, k}-u_{c, j}+i}{u_{c, k}-u_{c, j}-i}\prod_{j=1}^{K_{b}}\frac{u_{c, k}-u_{b, j}-\ft{i}{2}}{u_{c, k}-u_{b, j}+\ft{i}{2}}\, ,\\
\end{aligned}
\eeq
where $S$ is the spin of the GKP string.

These equations are a generalisation of the quantisation conditions for the hole rapidities $u_{k}$ studied in~\cite{Belitsky:2006en,Belitsky:2008mg} at weak coupling. They are valid up to corrections suppressed with the effective length $\sim 2\log{S}$. The latter corrections are typically of the order $\sim S^{-2m}$ at large spin, where $m$ is the mass gap of the GKP string ($m \sim 1$ at weak coupling and $m \sim \exp{(-\pi g)}$ for large values of the coupling constant). Furthermore, the equations \eqref{ABAN=4} hold for states satisfying the spin chain zero-momentum condition. If the latter  were to be relaxed, one should introduce an appropriate twist to account for it. This would amount to multiplying the right-hand side of the first equations in \eqref{ABAN=4} by the momentum phase $e^{iP}$. In other words, holes in the $\mathcal{N}=4$ theory satisfy periodic boundary conditions on the GKP string background \textit{only} at zero momentum.

The structure of the equations~(\ref{ABAN=4}) is canonical and it can be decomposed into several building blocks. The first set of equations is the most interesting from the physics' point of view, as it encodes the dynamics of the scalar excitations on top of the GKP string. These are equations for the roots $u_{k}$, with $k = 1, \ldots , K_{h}$, which stand for the rapidities of a total of $K_{h}$ holes. These roots couple to the length of the GKP string $\sim 2\log{S}$ via the momentum $p_{\mathcal{N}=4}(u)$ computed in~\cite{Basso:2010in, Gaiotto:2010fk}. A scattering process between two holes carrying rapidities $u$ and $v$ produces a phase $S_{\mathcal{N}=4}(u, v)$. This is the hole S-matrix whose properties at finite coupling will be investigated below. For small values of the coupling constant it collapses to a simple ratio of Euler Gamma functions,
\beq\label{previousS}
S_{\mathcal{N}=4}(u, v) = \frac{\Gamma(iu-iv)\Gamma(\ft{1}{2}-iu)\Gamma(\ft{1}{2}+iv)}{\Gamma(iv-iu)\Gamma(\ft{1}{2}+iu)\Gamma(\ft{1}{2}-iv)} + O(g^2)\,.
\eeq
This one-loop result may be found by casting the equations of~\cite{Belitsky:2006en} in the form~(\ref{ABAN=4}) and was already reported in~\cite{Basso:2011rc,Dorey:2011gr}. One can verify that it is a valid scattering matrix for scalars in the $\mathcal{N}=4$ theory by constructing the two-body wave function for scalar insertions on a $\Pi$-shaped light-like Wilson loop~\cite{BSV} in the spirit of~\cite{Belitsky:2011nn,Sever:2012qp}. 

We also notice the presence of the factor $T_{\mathcal{N}=4}(u)$ in the equations for the holes. This one depends on a single rapidity and in the notations of \cite{Basso:2011rc} it may be written as
\beq\label{transamp}
T_{\mathcal{N}=4}(u) = e^{2i\delta p(u)}\,.
\eeq
It lends itself to the interpretation of a transmission amplitude\footnote{We are thankful to Z.~Bajnok for pointing this out to us.}, because it accounts for the scattering of the excitation $u$ through impurities identified with the tips of the GKP string. From this perspective, the end points of the string are associated with an integrable defect of the theory which seems to preserves the $SU(4)$ symmetry. The transmission amplitude could be reflectionless or transmission-less, see~\cite{Bajnok:2004jd} and references therein. From the structure of the Bethe-Yang equations spelled out above, we observe that we are dealing with the second possibility. This factor can also be written directly in terms of the S-matrix $S_{\mathcal{N}=4}(u, v)$ of the holes 
\beq\label{tss}
T_{\mathcal{N}=4}(u) = \lim\limits_{c\rightarrow \infty}c^{-2ip_{\mathcal{N}=4}(u)}S_{\mathcal{N}=4}(u, c)S_{\mathcal{N}=4}(u, -c)\, .
\eeq
The properties of the S-matrix discussed below require the transmission amplitude to satisfy the unitarity and crossing relation
\beq
T_{\mathcal{N}=4}(u)T_{\mathcal{N}=4}(-u) = 1\, , \qquad T_{\mathcal{N}=4}(u^{2\gamma})T_{\mathcal{N}=4}(u) = 1\,.
\eeq
The mapping $\gamma:u\rightarrow u^{\gamma}$ is the mirror transformation and will be defined later in this section. To leading order at weak coupling, the transmission amplitude is given by~\cite{Basso:2011rc}
\beq
T_{\mathcal{N}=4}(u) = \frac{\Gamma(\ft{1}{2}-iu)^2}{\Gamma(\ft{1}{2}+iu)^2} + O(g^2)\, .
\eeq
The other equations in~(\ref{ABAN=4}) determine the auxiliary roots  $u_{(a, b, c), k}$, with $k = 1, \ldots, K_{(a, b, c)}$. These account for the isotopic $O(6)$ degrees of freedom of holes. 
The form of the equations for auxiliary roots will allow us to deduce the full symmetry structure of the S-matrix for holes. Here we simply notice that they are of the expected type with the hole rapidities entering as inhomogeneities in the equations for the $u_{(a, b, c), k}$, as previously observed in~\cite{Basso:2010in}. This set of equations is identical to the one for an XXX spin chain with spins transforming in the vector representation of $SO(6)$.

\subsection{The $O(6)$ S-matrix for the $\mathcal{N}=4$ theory}\label{O6Smat}

The hole S-matrix is the most interesting ingredient of the Bethe-Yang equations~(\ref{ABAN=4}). In this subsection we explain how it is defined at finite coupling and study its main properties. As far as the matrix structure is concerned, our analysis below does not differ from the conventional approach and is introduced mostly for the reader's convenience. The discussion of the scalar factor of the S-matrix, which is specific to the integrable system considered, is however new.

\subsubsection*{Matrix structure}

The form of the asymptotic Bethe ansatz equations~(\ref{ABAN=4}) is consistent with an $O(6)$ invariant S-matrix for a vector multiplet.  As such, it admits a decomposition into the invariant channels appearing in the decomposition $\textbf{6}\otimes \textbf{6} = \textbf{20} \oplus \textbf{15} \oplus \textbf{1}$, where $\textbf{20}, \textbf{15},$ and $\textbf{1}$, stand for the symmetric, adjoint and singlet representation, respectively. In other words, the S-matrix for the two-to-two process
\beq
X_{i}(u)X_{j}(v) \rightarrow X_{l}(v)X_{k}(u)\, ,
\eeq
where $X_{i}(u)$, $i\in \{1, \ldots, 6\}$, denotes one of the six scalar of the theory carrying rapidity $u$, can be written as
\beq
S_{ij}^{kl}(u, v) = \sigma_{\textbf{20}}(u, v) P_{\textbf{20}\, ij}^{\, \, \, \, \, \, \,  kl} + \sigma_{\textbf{15}}(u, v) P_{\textbf{15}\, ij}^{\, \, \, \, \, \, \,  kl}+ \sigma_{\textbf{1}}(u, v) P_{\textbf{1}\, ij}^{\, \, \, \,  kl}\,.
\eeq
The orthogonal projectors $P_{\textbf{a}\, ij}^{\, \, \, \, kl}$ ($\textbf{a} = \textbf{20}, \textbf{15}, \textbf{1}$) may be expressed through Kronecker delta functions
\beq
 P_{\textbf{20}\, ij}^{\, \, \, \, \, \, \,  kl} = \frac{1}{2}\delta_{i}^{k}\delta_{j}^{l}+\frac{1}{2}\delta_{i}^{l}\delta_{j}^{k} -\frac{1}{6}\delta_{ij}\delta^{kl}\, , \qquad P_{\textbf{15}\, ij}^{\, \, \, \, \, \, \, kl} = \frac{1}{2}\delta_{i}^{k}\delta_{j}^{l}-\frac{1}{2}\delta_{i}^{l}\delta_{j}^{k}\, ,\qquad P_{\textbf{1}\, ij}^{\, \, \, \,  kl} = \frac{1}{6}\delta_{ij}\delta^{kl}\, .
\eeq
As it is well known, the three invariant amplitudes $\sigma_{\textbf{a}}(u, v)$ may be deduced directly from the ABA equations~(\ref{ABAN=4}) by solving the equations for the isotopic roots in the presence of the two holes with rapidities $u$ and $v$. There are three solutions corresponding to three possible scattering channels. Once a solution is found, the amplitude in the associated channel may be found using the formula
\beq
\sigma_{\textbf{a}}(u, v) = S_{\mathcal{N}=4}(u, v)\prod_{j=1}^{K_b}\frac{u-u^{\textbf{a}}_{b, j}+\ft{i}{2}}{u-u^{\textbf{a}}_{b, j}-\ft{i}{2}} \, ,
\eeq
that encodes the induced scattering between the two holes $u, v$ due to the presence of the isotopic roots $u^{\textbf{a}}_{b,j}$. The relevant solutions are easily found. For illustration, in the adjoint channel ($\textbf{a} = \textbf{15}$) we have no roots of type $a,c$ and a single root of type $b$ given by $u^{\textbf{15}}_{b} = (u+v)/2$. The symmetric channel involves no isotopic roots at all and the singlet channel is characterised by two roots of type $b$ and one root of type $a$ and $c$. They immediately lead to the desired expressions
\beq
\begin{aligned}
&\sigma_{\textbf{20}}(u, v) =  S_{\mathcal{N}=4}(u, v)\, , \qquad \sigma_{\textbf{15}}(u, v) = \frac{u-v+i}{u-v-i}S_{\mathcal{N}=4}(u, v)\, , \\
&\qquad \qquad \sigma_{\textbf{1}}(u, v) = \frac{u-v+2i}{u-v-2i}\frac{u-v+i}{u-v-i}S_{\mathcal{N}=4}(u, v)\, ,
\end{aligned}
\eeq
and neatly combine to
\beq\label{Smatrix-final}
S_{ij}^{kl}(u, v) = S_{\mathcal{N}=4}(u, v)\left[\frac{u-v}{u-v-i}\delta_{i}^{j}\delta_{k}^{l} - \frac{i}{u-v-i}\delta_{i}^{k}\delta_{k}^{j} + \frac{i(u-v)}{(u-v-i)(u-v-2i)}\delta_{ij}\delta^{kl}\right]\, .
\eeq
The matrix structure, in other words the term in the square brackets, is identical to the one found for the factorisable S-matrix of the non-linear $O(6)$ sigma model~\cite{Zamolodchikov:1977nu}. This was certainly expected since it is largely fixed by the Yang-Baxter equations, which are oblivious to the specific model under study as long as it is integrable. Nevertheless, the Yang-Baxter equation alone does not fix the scale of the rapidities entering the matrix structure above. This scale or normalisation is determined by the crossing relations that relate the three amplitudes. It is here where the intricate dynamics of $\mathcal{N}=4$ sediments. 

\subsubsection*{Scalar factor}

Though the S-matrix~(\ref{Smatrix-final}) is identical in form to the one of the non-linear $O(6)$ sigma model, it describes nevertheless a different scattering theory. This is because the scalar factor $S_{\mathcal{N}=4}(u, v)$ is, for generic values of the coupling $g$ and rapidities $u, v$, completely different from the expression found for the non-linear $O(6)$ model. In particular, the phase $S_{\mathcal{N}=4}(u, v)$ is generally \textit{not} a function of the difference of rapidities $(u-v)$, due to lack of boost invariance. Only at strong coupling, and for rapidities $u, v$ of order $O(g^0)$, does the scalar factor become identical to the the one for the $O(6)$ model. This will be shown in Section~\ref{O(6)limit}. At intermediate values of $g$ the best one can do is to find a representation of the scalar factor $S_{\mathcal{N}=4}(u, v)$ in terms of the large spin density.  To be more precise, we shall represent the scattering phase in terms of the two functions $\gamma^{v}(2gt), \tilde{\gamma}^{v}(2gt)$ that parameterise the correction to the density induced by a hole carrying rapidity $v$. There are two of them because it turns out to be convenient to distinguish between the odd and even parts with respect to $v$. We thus have $\gamma^{-v} = \gamma^{v}$ and $ \tilde{\gamma}^{-v} = - \tilde{\gamma}^{v}$. These are the same functions that determine the expression for the energy and momentum of a hole at generic value of the coupling constant~\cite{Basso:2010in}. They can be defined non-perturbatively as a solution to linear integral equations similar to the Beisert-Eden-Staudacher (BES) equation~\cite{Eden:2006rx,Beisert:2006ez}. We discuss these equations in Appendix~\ref{all-loop-densities}. We find that the scattering phase $S_{\mathcal{N}=4}(u, v)$ can be decomposed as
\beq\label{Sgeneral}
S_{\mathcal{N}=4}(u, v) = \hat{S}_{\mathcal{N}=4}(u, v)\exp{(-2if_{1}(u, v)+2if_{2}(u, v))}\, .
\eeq
The first factor in this equation is explicitly known and is given by the following integral
\beq\label{Shat}
\log{\hat{S}_{\mathcal{N}=4}(u, v)} = 2i\int_{0}^{\infty}\frac{dt}{t}\frac{e^{t/2}J_{0}(2gt)\sin{(ut)}-e^{t/2}J_{0}(2gt)\sin{(vt)}-e^{t}\sin{((u-v)t)}}{e^{t}-1}\, ,
\eeq
where $J_{0}(z) = 1+O(z^2)$ is the zeroth Bessel function. The two extra terms $f_{1, 2}(u, v)$ appearing exponentiated in~(\ref{Sgeneral}) are more dynamical and may be expressed through the aforementioned functions $\gamma^{v}, \tilde{\gamma}^{v}$. Explicitly, 
\beq
\begin{aligned}\label{f-functions}
&f_{1}(u, v) = \int_{0}^{\infty}\frac{dt}{t}\frac{\sin{(ut)}e^{t/2}\gamma^{v}(2gt)}{e^{t}-1}\, ,\\
&f_{2}(u, v) = \int_{0}^{\infty}\frac{dt}{t}\frac{(\cos{(ut)}e^{t/2}-J_{0}(2gt))\tilde{\gamma}^{v}(2gt)}{e^{t}-1}\, .\\
\end{aligned}
\eeq
Even though the functions $\gamma^{v}$ and $\tilde{\gamma}^{v}$ are not known in a closed form for generic coupling, their weak- and strong-coupling expansions may be found. For instance, it is easy to see that to leading order at weak coupling we have that $f_{1}(u, v) \sim f_{2}(u, v) \sim g^2$. We then immediately see that
\beq\label{loSmat}
S_{\mathcal{N}=4}(u, v) \simeq \hat{S}_{\mathcal{N}=4}(u, v) = \frac{\Gamma(iu-iv)\Gamma(\ft{1}{2}-iu)\Gamma(\ft{1}{2}+iv)}{\Gamma(iv-iu)\Gamma(\ft{1}{2}+iu)\Gamma(\ft{1}{2}-iv)} + O(g^2)\, ,
\eeq
in perfect agreement with~(\ref{previousS}). It is not difficult to expand further the general formulae~(\ref{Sgeneral})-(\ref{f-functions}). We perform the next-to-leading computation in Appendix~\ref{wce}. The true advantage of the representation \eqref{Sgeneral} lies somewhere else. As we shall now show, it allows one to explore general properties of the S-matrix like unitarity, crossing, etc., some of which being difficult to address without a non-perturbative definition like \eqref{Sgeneral}. 

Although it is not entirely manifest from their general definition \eqref{f-functions}, the functions $f_{1, 2}(u, v)$ are \textit{not} independent. As a consequence of the equations satisfied by $\gamma^{v}$ and $\tilde{\gamma}^{v}$, it is indeed possible to show that the relation
\beq\label{exchange-rel}
f_{1}(u, v) = f_{2}(v, u)\, ,
\eeq
holds true at any coupling, see Appendix~\ref{Unitarity}. It is quite easy to verify that this identity guarantees the unitarity of the S-matrix 
\beq
S_{\mathcal{N}=4}(u, v)S_{\mathcal{N}=4}(v, u) = 1\, .
\eeq
This identity follows immediately from \eqref{Sgeneral} and~(\ref{exchange-rel}), after taking into account that $\hat{S}_{\mathcal{N}=4}(u, v)\hat{S}_{\mathcal{N}=4}(v, u)=1$.

Let us now use \eqref{Sgeneral} to deduce the crossing transformation of the hole S-matrix. We start with identifying the crossing map in the rapidity plane of the hole. One may think about the crossing transformation as a sequence of two mirror rotations (or a double Wick rotation), each one of them exchanging the space and time dimensions. This mirror rotation, that we shall denote as $u\rightarrow u^{\gamma}$, was previously considered in~\cite{Basso:2011rc}. For a hole, it simply amounts to a shift of the rapidity by the imaginary unit $u^{\gamma} = u+i$. Despite appearing to be very simple, this step has to been taken with care. In order to correctly implement the mirror rotation, the shift $u^{\gamma}=u+i$ has to been understood as an analytical continuation that goes through the strip $-2g < \textrm{Re}(u) < 2g$ in the complex $u$-plane, see Figure~\ref{fig:cpstrong}. The reason is that many of the hole-related observables have cuts in this strip. This is the case for the energy, momentum and the S-matrix. The mirror transformation for all these quantities entails passing through the first cut in the strip, which lies at $\textrm{Im}(u) = i/2$. When applied to a hole with energy $E(u)$ and momentum $p(u)$, this procedure gives~\cite{Basso:2011rc}
\beq
E(u^{\gamma}) = ip(u)\, , \qquad p(u^{\gamma}) = iE(u)\, ,
\eeq
which demonstrates the mirror invariance of the GKP background~\cite{Alday:2007mf}. Getting back to our definition of crossing, we can now compose two  mirror transformations to obtain
\beq
E(u^{2\gamma}) = -E(u)\, , \qquad p(u^{2\gamma}) = -p(u)\, ,
\eeq
with $u^{2\gamma} = u+2i$. This is the expected crossing relation. It swaps the sign of energy and momentum and thus the role of particles and antiparticles. Note that the latter are one and the same thing for scalar excitations.

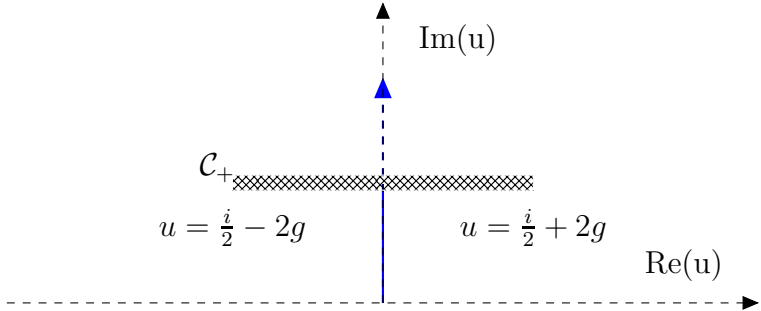
\begin{figure}
\begin{center}
\begin{tikzpicture}[cross/.style={path picture={ 
  \draw[black]
(path picture bounding box.south east) -- (path picture bounding box.north west) (path picture bounding box.south west) -- (path picture bounding box.north east);
}}]
\draw[-triangle 45,dashed] (0,0)--(10,0);
\draw[thick, blue] (5,0)--(5,1.5);
\draw[-triangle 45, thick, blue, dashed] (5,0)--(5,3);
\draw[-triangle 45,dashed] (5,0)--(5,4);
\draw[draw=white, pattern=crosshatch]  (3,1.5) rectangle (7,1.7);
\node at (3,1) {$u=\tfrac{i}{2}-2g$};
\node at (7,1) {$u=\tfrac{i}{2}+2g$};
\node at (2.8,1.8) {$\mathcal{C}_{+}$};
\node at (9,0.5) {Re(u)};
\node at (6,3.5) {Im(u)};
\end{tikzpicture}
\end{center}
\caption{The mirror path at small values of the coupling constant. It entails shifting the rapidity $u$ by the imaginary unit $i$ such that one crosses the cut $\mathcal{C}_{+}$ connecting the square-root branch points $i/2\pm 2g$. This cut closes up at weak coupling, hence performing the mirror transformation is impossible perturbatively. } \label{fig:cpweak}
\end{figure}
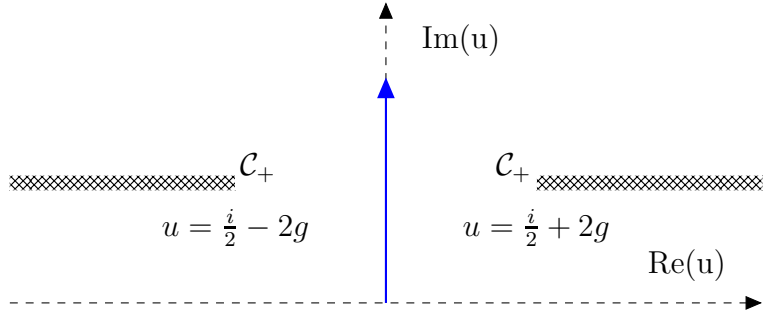
\begin{figure}
\begin{center}
\begin{tikzpicture}[cross/.style={path picture={ 
  \draw[black]
(path picture bounding box.south east) -- (path picture bounding box.north west) (path picture bounding box.south west) -- (path picture bounding box.north east);
}}]
\draw[-triangle 45,dashed] (0,0)--(10,0);
\draw[-triangle 45,dashed] (5,0)--(5,4);
\draw[-triangle 45, thick, blue] (5,0)--(5,3);
\draw[draw=white, pattern=crosshatch]  (0,1.5) rectangle (3,1.7);
\draw[draw=white, pattern=crosshatch] (7,1.5) rectangle (10,1.7);
\node at (3,1) {$u=\tfrac{i}{2}-2g$};
\node at (7,1) {$u=\tfrac{i}{2}+2g$};
\node at (3.3,1.8) {$\mathcal{C}_{+}$};
\node at (6.7,1.8) {$\mathcal{C}_{+}$};
\node at (9,0.5) {Re(u)};
\node at (6,3.5) {Im(u)};
\end{tikzpicture}
\end{center}
\caption{The mirror path at large values of the coupling constant. At strong coupling the cut $\mathcal{C}_{+}$ opens up hence revealing the simplicity of the mirror transformation.} \label{fig:cpstrong}
\end{figure}

We may now apply the very same procedure to the S-matrix. The matrix part is easily transformed, since it is just a rational function of the rapidity $u$. The transformation of the scalar factor $S(u^{2\gamma}, v)$ is more interesting. We find the following simple equation
\beq\label{cross-rel}
S_{\mathcal{N}=4}(u^{2\gamma}, v) = \frac{u-v}{u-v+2i}S_{\mathcal{N}=4}(v, u)\, .
\eeq
It matches exactly with the expected crossing relation for an $O(6)$ invariant S-matrix. We prove \eqref{cross-rel} in Appendix~\ref{Crossing}, where the expression for the ``mirror" S-matrix $S(u^{\gamma}, v)$, which describes the scattering between a mirror and a real excitation, is also derived. It is also shown there that the S-matrix is mirror invariant, i.e.,
\beq\label{mirrS}
S(u^{\gamma}, v^{\gamma}) = S(u, v)\, .
\eeq
This equality means that the scattering between two excitations in the same channel is the same in the real or mirror kinematics, in agreement with~\cite{Alday:2007mf}.

The S-matrix for holes in $\mathcal{N}=4$ SYM theory looks very similar to the one considered by Zamolodchikov. Both exhibit the same matrix structure, satisfy the same crossing equations and are mirror invariant. The only difference is that there is no relativistic invariance. The dynamics the GKP string is complex and as a consequence the scalar factor is different from the one found for $O(6)$ sigma model. They essentially differ by a proportionality factor~$\varphi(u, v)$
\beq
S_{\mathcal{N}=4}(u, v) = S_{\textrm{O(6)}}(u-v) \varphi(u, v)\, ,
\eeq
which by definition is neutral under crossing and satisfies unitarily on its own. It \textit{does not} introduce any new poles (bound states) into the S-matrix, so that the scalar factor $S_{\mathcal{N}=4}(u, v)$ is minimal. In other words it helps to implement the analytic structure proper to $\mathcal{N}=4$ S-matrix, but it does not introduce new physical states. It is instructive to spell out the leading term in the strong coupling expansion of $\varphi(u,v)$
\beq
-i\log{\varphi(u, v)} \sim c(g)\,m_{\mathcal{N}=4}\,\sinh{\frac{\pi}{2}(u-v)}\,.
\eeq
The coefficient $c(g)$ has a power-law behaviour at strong coupling. It trivially satisfies $\varphi(u^{2\gamma}, v) \equiv  \varphi(u+2i, v)= \varphi(v, u)$ and is negligible at strong coupling because it is suppressed by the mass gap $m_{\mathcal{N}=4} \sim \exp{(-\pi g)}$ of the theory. 

For conventional relativistic systems such CDD factors are typically ruled out by the analytical properties at large rapidities, which impose that the S-matrix should grow no quicker than as a certain power of the relativistic rapidity $\theta$. In our case, the argument would not apply since the large $\theta$ behaviour lies very far away from the relativistic regime. This is because at short distance the effective $O(6)$ model for the GKP excitations is very different from any relativistic model. 

Finally, we stress that the corrections to the $O(6)$ regime contained  in the CDD factor $\varphi(u, v)$ are exponentially small only for rapidities $u, v$ of order $O(1)$ at strong coupling. As we move away from this regime they re-sum and their magnitude may become comparable with that of the scalar factor for the $O(6)$ model. This is certainly the case when we reach the so called giant hole domain, which was first introduced and analysed in~\cite{Dorey:2010iy,Dorey:2010id,Dorey:2010zz,Losi:2010hr}. This is a strong-coupling regime where both rapidities are large relatively to the coupling,  i.e. $u, v \geqslant g\gg 1$. It is quite universal in the sense that all scattering phases are expected to be the same to leading order at strong coupling, no matter what type of excitations are being scattered. All excitations in this limit behave like semiclassical bumps propagating and scattering on top of the GKP string and connect with the so-called spiky-string solutions~\cite{Kruczenski:2004wg,Dorey:2008vp}. The scattering phase for giant holes bears little similarity with the one for holes in the $O(6)$ regime. Its explicit expression was worked out in~\cite{Dorey:2011gr}.

\subsection{The BY equations for $\mathcal{N}=6$ ABJM theory}

The asymptotic Bethe ansatz equations for the holes of the $\mathcal{N}=6$ Chern-Simons-Matter theory are quite similar to the ones found before. Naturally, this time there are two types of momentum-carrying roots,
\beq
\begin{aligned}\label{ABAN=6}
(2S)^{-ip_{\mathcal{N}=6}(u_{k})} &= q\, T_{\mathcal{N}=6}(u_{k})\prod^{K_{h}}_{j\neq k}S_{\mathcal{N}=6}(u_k, u_j)\prod^{\bar{K}_{h}}_{j = 1}\bar{S}_{\mathcal{N}=6}(u_k, \bar{u}_j)\prod_{j=1}^{K_a}\frac{u_k-u_{a, j}+\ft{i}{2}}{u_{k}-u_{a, j}-\ft{i}{2}}\, , \\
\prod_{j=1}^{K_h}\frac{u_{a, k}-u_{j}+\ft{i}{2}}{u_{a, k}-u_{j}-\ft{i}{2}} &= \prod_{j\neq k}^{K_a}\frac{u_{a, k}-u_{a, j}+i}{u_{a, k}-u_{a, j}-i}\prod_{j=1}^{K_{b}}\frac{u_{a, k}-u_{b, j}-\ft{i}{2}}{u_{a, k}-u_{b, j}+\ft{i}{2}}\, ,\\
1 &= \prod_{j\neq k}^{K_{b}}\frac{u_{b, k}-u_{b, j}+i}{u_{b, k}-u_{b, j}-i}\prod_{j=1}^{K_{a}}\frac{u_{b, k}-u_{a, j}-\ft{i}{2}}{u_{b, k}-u_{a, j}+\ft{i}{2}}\prod_{j=1}^{K_{c}}\frac{u_{b, k}-u_{c, j}-\ft{i}{2}}{u_{b, k}-u_{c, j}+\ft{i}{2}}\, ,\\
\prod_{j=1}^{\bar{K}_h}\frac{u_{c, k}-\bar{u}_{j}+\ft{i}{2}}{u_{c, k}-\bar{u}_{j}-\ft{i}{2}} &= \prod_{j \neq k}^{K_c}\frac{u_{c, k}-u_{c, j}+i}{u_{c, k}-u_{c, j}-i}\prod_{j=1}^{K_{b}}\frac{u_{c, k}-u_{b, j}-\ft{i}{2}}{u_{c, k}-u_{b, j}+\ft{i}{2}}\, ,\\
(2S)^{-ip_{\mathcal{N}=6}(\bar{u}_{k})} &= \bar{q}\, T_{\mathcal{N}=6}(u_{k})\prod^{\bar{K}_{h}}_{j\neq k}S_{\mathcal{N}=6}(\bar{u}_k,\bar{u}_j)\prod^{K_{h}}_{j = 1}\bar{S}_{\mathcal{N}=6}(\bar{u}_k, u_j)\prod_{j=1}^{K_c}\frac{\bar{u}_k-u_{c, j}+\ft{i}{2}}{\bar{u}_{k}-u_{c, j}-\ft{i}{2}}\,. \\
\end{aligned}
\eeq
We recall that the parameter $S$ is the spin of the GKP string. The transmission amplitude $T_{\mathcal{N}=6}(u)$ is directly related to the one of the $\mathcal{N}=4$ theory and reads
\beq\label{T4T6}
T_{\mathcal{N}=6}(u) = T_{\mathcal{N}=4}(u)^{1/2}\, .
\eeq
Exploiting the analogy with the $\mathcal{N}=4$ theory even further, it can be related to the S-matrix by the formula
\beq
T_{\mathcal{N}=6}(u) = \lim\limits_{c\rightarrow \infty}c^{-2ip_{\mathcal{N}=6}(u)}S_{\mathcal{N}=6}(u, c)\bar{S}_{\mathcal{N}=6}(u, -c)\,.
\eeq
Actually, in the above we can replace $\bar{S}_{\mathcal{N}=6}$ with $S_{\mathcal{N}=6}$ without changing the validity of this representation. At weak coupling we find
\beq
T_{\mathcal{N}=6}(u) = \frac{\Gamma(\ft{1}{2}-iu)}{\Gamma(\ft{1}{2}+iu)} + O(g^2)\, ,
\eeq
a result already familiar from Section~\ref{sec:BYol}.

We stressed several times that the most significant difference between the equations for $\mathcal{N}=4$ and $\mathcal{N}=6$ is the presence of the twist factors $q, \bar{q}$. For $\mathcal{N}=6$ ABJM theory these factors are present even after imposing the spin-chain zero momentum condition, as we carefully explained in Section~\ref{sec:twistq}. They are not independent and satisfy relations~(\ref{super-rules}). Equations \eqref{ABAN=6} are also equipped with a selection rule on the numbers of holes and anti-holes that enforces $F=\ft{1}{2}(K_{h}-\bar{K}_h)$ to be an integer. It can be interpreted as a restriction on the quadrality of the $SU(4)$ representations allowed in this theory
\beq
\left(K_{h} +3\bar{K}_h\right) \,\,\, \textrm{mod} \,\,\, 4 = \left(2F\right) \,\,\, \textrm{mod} \,\,\, 4 = \left(1-(-1)^F\right)  \,\,\, \textrm{mod} \,\,\, 4 .
\eeq

\subsection{The $U(4)$ S-matrix for the $\mathcal{N}=6$ theory}\label{sec:u4Sm}

We shall now derive the S-matrix for the low-lying excitations on top of the GKP string in the $\mathcal{N}=6$ theory. The derivation will follow closely the one encountered in Section~\ref{O6Smat}.

\subsubsection*{Matrix structures}

The form of the ABA equations~(\ref{ABAN=6}) is consistent with a reflectionless $U(4)$ S-matrix for the two sets of excitations of the $\mathcal{N}=6$ theory, namely the holes and anti-holes. The only two-to-two scattering processes consistent with transmission-only scattering are 
\beq\label{hole-antihole-processes}
Z_{i}(u)Z_{j}(v) \rightarrow Z_{l}(v)Z_{k}(u)\, , \qquad Z_{i}(u)\bar{Z}_{j}(v) \rightarrow \bar{Z}_{l}(v)Z_{k}(u)\, , 
\eeq
where $Z_{j}(u), \bar{Z}_{j}(u)$,  $j=1, \ldots , 4$, stand for a hole and anti-hole with rapidity $u$, respectively. 

For each process in~(\ref{hole-antihole-processes}) we associate the corresponding S-matrix, denoted as $S_{ij}^{kl}(u, v)$ and $\bar{S}_{ij}^{kl}(u, v)$. Thanks to the $U(4)$ symmetry, they can be decomposed over the irreducible channels in $\textbf{4}\otimes \textbf{4} = \textbf{10} \oplus \textbf{6}$ and $\textbf{4}\otimes \bar{\textbf{4}} = \textbf{15} \oplus \textbf{1}$. This way we write
\beq
\begin{aligned}
&S_{ij}^{kl}(u, v) = \sigma_{\textbf{10}}(u, v)P_{\textbf{10}\, ij}^{\, \, \, \, \, \, \,  kl}+\sigma_{\textbf{6}}(u, v)P_{\textbf{6}\, ij}^{\, \, \, \, kl}\, , \\
&\bar{S}_{ij}^{kl}(u, v) = \sigma_{\textbf{15}}(u, v)\bar{P}_{\textbf{15}\, ij}^{\, \, \, \, \, \, \,  kl}+\sigma_{\textbf{1}}(u, v)\bar{P}_{\textbf{1}\, ij}^{\, \, \, \, kl}\,. \\
\end{aligned}
\eeq
The orthogonal projectors are given by
\beq
\begin{aligned}
&P_{\textbf{10}\, ij}^{\, \, \, \, \, \, \,  kl} = \frac{1}{2}\delta_{i}^{k}\delta_{j}^{l}+\frac{1}{2}\delta_{i}^{l}\delta_{j}^{k}\, , \qquad P_{\textbf{6}\, ij}^{\, \, \, \,kl} = \frac{1}{2}\delta_{i}^{k}\delta_{j}^{l}-\frac{1}{2}\delta_{i}^{l}\delta_{j}^{k}\, , \\
&\bar{P}_{\textbf{15}\, ij}^{\, \, \, \, \, \, \,  kl} = \delta_{i}^{k}\delta_{j}^{l}-\frac{1}{4}\delta_{ij}\delta^{kl}\, , \qquad \bar{P}_{\textbf{1}\, ij}^{\, \, \, \, kl} = \frac{1}{4}\delta_{ij}\delta^{kl}\ . \\
\end{aligned}
\eeq
As in the previous case the invariant amplitudes can be found from the solutions to the Bethe ansatz equations associated to a pair of holes and a pair of hole and anti-hole, respectively. This way we find
\beq
\begin{aligned}
&\sigma_{\textbf{10}}(u, v) = S_{\mathcal{N}=6}(u, v)\, , \qquad \sigma_{\textbf{6}}(u, v) = \frac{u-v+i}{u-v-i}\, S_{\mathcal{N}=6}(u, v)\, ,\\
&\sigma_{\textbf{15}}(u, v) = \bar{S}_{\mathcal{N}=6}(u, v)\, , \qquad \sigma_{\textbf{1}}(u, v) = \frac{u-v+2i}{u-v-2i}\, \bar{S}_{\mathcal{N}=6}(u, v)\, ,\\
\end{aligned}
\eeq
or, equivalently,
\beq
\begin{aligned}
&S_{ij}^{kl}(u, v) = S_{\mathcal{N}=6}(u, v)\bigg[\frac{u-v}{u-v-i}\delta_{i}^{k}\delta_{j}^{l}-\frac{i}{u-v-i}\delta_{i}^{l}\delta_{j}^{k}\bigg]\, , \\
&\bar{S}_{ij}^{kl}(u, v) = \bar{S}_{\mathcal{N}=6}(u, v)\bigg[\delta_{i}^{k}\delta_{j}^{l}+\frac{i}{u-v-2i}\delta_{ij}\delta^{kl}\bigg]\, . \\
\end{aligned}
\eeq
These two matrix structures are identical to the ones for a factorizable and reflectionless relativistic model with $U(4)$ symmetry~\cite{Berg:1977dp}.

\subsubsection*{Scalar factors}

The scattering phases of the $\mathcal{N}=6$ theory, $S_{\mathcal{N}=6}(u, v)$ and $\bar{S}_{\mathcal{N}=6}(u, v)$, are closely related to the scalar factor of the $\mathcal{N}=4$ theory. This follows from the relation~(\ref{rho46}) between the densities of the two theories. Its direct consequence is
\beq\label{prodS}
S_{\mathcal{N}=6}(u,v)\bar{S}_{\mathcal{N}=6}(u, v) = S_{\mathcal{N}=4}(u, v)\, ,
\eeq
which is valid for \textit{any} coupling. The second interesting property found before concerns the difference $\rho_{\mathcal{N}=6}(u)-\bar{\rho}_{\mathcal{N}=6}(u)$. We found that this difference is one-loop exact. Using the results of the analysis in Section~\ref{sec:BYol} this translates to
\beq\label{ratioS}
S_{\mathcal{N}=6}(u, v)/\bar{S}_{\mathcal{N}=6}(u, v) = \frac{\Gamma(\frac{iu-iv}{2})\Gamma(\ft{1}{2}-\frac{iu-iv}{2})}{\Gamma(\frac{iv-iu}{2})\Gamma(\ft{1}{2}+\frac{iu-iv}{2})}\, .
\eeq
The above two identities completely determine the scattering phases of the $\mathcal{N}=6$ theory up to an overall sign. The latter ambiguity can always be fixed by imposing the requirement
\beq
S_{\mathcal{N}=6}(u,u) = -\bar{S}_{\mathcal{N}=6}(u,u) = -1\, .
\eeq
Equivalently, one can write
\beq
\begin{aligned}
&S_{\mathcal{N}=6}(u, v) = \hat{S}_{\mathcal{N}=6}(u, v)\exp{(-if_{1}(u, v)+if_{2}(u, v))}\, ,\\
&\bar{S}_{\mathcal{N}=6}(u, v) = \hat{\bar{S}}_{\mathcal{N}=6}(u, v)\exp{(-if_{1}(u, v)+if_{2}(u, v))}\, ,\\
\end{aligned}
\eeq
with the functions $f_{1, 2}(u, v)$ given in Eq.(\ref{f-functions}) and
\beq
\begin{aligned}
&\log{\hat{S}_{\mathcal{N}=6}(u, v)} = \frac{1}{2}\log{\hat{S}_{\mathcal{N}=4}(u, v)}+ i\int_{0}^{\infty}\frac{dt}{t}\frac{e^{t/2}\sin{(\frac{1}{2}(u-v)t)}-e^{t}\sin{(\frac{1}{2}(u-v)t)}}{e^{t}-1}\, , \\
&\log{\hat{\bar{S}}_{\mathcal{N}=6}(u, v)} = \frac{1}{2}\log{\hat{S}_{\mathcal{N}=4}(u, v)}- i\int_{0}^{\infty}\frac{dt}{t}\frac{e^{t/2}\sin{(\frac{1}{2}(u-v)t)}-e^{t}\sin{(\frac{1}{2}(u-v)t)}}{e^{t}-1}\, .
\end{aligned}
\eeq
As we discussed before the dispersion relation for the holes in the $\mathcal{N}=6$ theory is identical to the one found in $\mathcal{N}=4$ SYM after rescaling the energy and the momentum by $1/2$. It follows that the crossing map for these excitations must also be the same. Implementing this transformation for the scattering phases of the $\mathcal{N}=6$ theory one easily derives that
\beq\label{cross-rel-bis}
S_{\mathcal{N}=6}(u^{2\gamma}, v) = \frac{u-v+i}{u-v+2i}\, \bar{S}_{\mathcal{N}=6}(v, u)\, .
\eeq
This relation is in agreement with the crossing relation for a relativistic and factorizable S-matrix with $U(4)$ symmetry.

\section{The relativistic limit} \label{sec:rellimit}
In this section we take the strong coupling limit of the S-matrices presented in the previous section. As already mentioned, for rapidities of order $O(1)$ these S-matrices become relativistic at strong coupling. In the case of the $\mathcal{N}=4$ SYM the hole scattering matrix becomes identical with the S-matrix for massive excitations of the non-linear $O(6)$ sigma model. This is the prediction from the string theory~\cite{Alday:2007mf}, which was previously tested using integrability in~\cite{Basso:2008tx, Fioravanti:2008rv, Gromov:2008en}. For the $\mathcal{N} = 6$ theory the string theory prediction is that the low-energy dynamics of the GKP string is controlled by the Bykov model~\cite{Bykov:2010tv}. In this section we shall establish the relation between the strongly coupled $\mathcal{N}=6$ theory and the Bykov model. As a first step, we shall re-derive and complete the relation between the $O(6)$ sigma model and the $\mathcal{N}=4$ theory using our general expressions.

\subsection{The non-linear $O(6)$ sigma model}\label{O(6)limit}

We start with the S-matrix. To get its expression at strong coupling, one can begin with the general formula~(\ref{Sgeneral}) and proceed as follows. First,  we need the identity
\beq\label{parteq}
\int_{0}^{\infty}\frac{(1-J_{0}(2gt))\tilde{\gamma}^{v}(2gt)}{e^{t}-1} = -\int_{0}^{\infty}\frac{dt}{t}\tilde{\gamma}^{v}_{+}(2gt) -\int_{0}^{\infty}\frac{dt}{t}\frac{(1-J_{0}(2gt))e^{t/2}\sin{(vt)}}{e^{t}-1}\, ,
\eeq
where $\tilde{\gamma}^{v}_{+}$ is the even part of $\tilde{\gamma}^{v}$ with respect to $t$. As shown in Appendix~\ref{alldens}, it may be easily derived from the properties of $\tilde{\gamma}^{v}$. When applied to~(\ref{Sgeneral}) it removes certain terms involving the Bessel function $J_{0}(2gt)$. One is left with
\beq
\begin{aligned}\label{Sgeneralbis}
&-i\log{S}_{\mathcal{N}=4}(u, v) = 2\int_{0}^{\infty}\frac{dt}{t}\frac{J_{0}(2gt)e^{t/2}\sin{(ut)}-e^{t/2}\sin{(vt)}-e^{t}\sin{((u-v)t)}}{e^{t}-1}\\
& -2\int_{0}^{\infty}\frac{dt}{t}\frac{\sin{(ut)}e^{t/2}\gamma^{v}(2gt)}{e^{t}-1}+ 2\int_{0}^{\infty}\frac{dt}{t}\frac{(\cos{(ut)}e^{t/2}-1)\tilde{\gamma}^{v}(2gt)}{e^{t}-1} -2\int_{0}^{\infty}\frac{dt}{t}\tilde{\gamma}^{v}_{+}(2gt)\,.
\end{aligned}
\eeq
Despite the fact that this expression looks even less symmetric w.r.t. $u$ and $v$ than before, it is actually better suited for the strong coupling analysis performed below.

The next step is to find the strong coupling expressions for the functions $\gamma^{v},\tilde{\gamma}^{v}$ for fixed rapidity $v$. The relevant solution is given by
\beq\label{particular}
\gamma^{v}(2gt)_{\textrm{part}} = J_{0}(2gt)-\frac{e^{t/2}\cos{(vt)}}{\cosh{t}} \, , \qquad \tilde{\gamma}^{v}(2gt)_{\textrm{part}} = -\frac{e^{t/2}\sin{(vt)}}{\cosh{t}} \, . \\
\eeq
It can be deduced from previous analysis of the $O(6)$ regime~\cite{Basso:2008tx,Fioravanti:2008rv,PhD}. To verify that the above functions indeed satisfy the equations for the density, one simply needs to plug them into~(\ref{Eqs-uform}) and make use of the relation
\beq
\int^{\infty}_0 dt\frac{J_0(2gt)}{t} (\cos({ut})-1)=0\, \qquad \textrm{for} \qquad u^2 < (2g)^2\,.
\eeq
The functions \eqref{particular} provide only a particular solution to \eqref{Eqs-uform} and they \textit{do not} coincide with the physical solution. The reason is that they are too singular as functions of $t$. They have poles in the complex $t$-plane, while the physical solution has to be holomorphic everywhere except for $t=\infty$, where it has an essential singularity. Even though the particular solution~(\ref{particular}) is not exactly the one we are looking for, it is nonetheless a good starting point at strong coupling. The equations for $\gamma^{v},\tilde{\gamma}^{v}$ are linear, so it is always possible to correct the ansatz~(\ref{particular}) by adding the homogeneous solutions $\gamma^{v}_{\textrm{hom}},\tilde{\gamma}^{v}_{\textrm{hom}}$ that take care of the unwanted poles,
\beq
\gamma^{v} = \gamma^{v}_{\textrm{part}} + \gamma^{v}_{\textrm{hom}} \, , \qquad \tilde{\gamma}^{v} = \tilde{\gamma}^{v}_{\textrm{part}} + \tilde{\gamma}^{v}_{\textrm{hom}} \, .
\eeq
It is a lucky coincidence that at strong coupling these zero-mode solutions are suppressed by powers of $\exp{(\pm \pi v/2-\pi g)}$. To see how this comes about, let us look at the nearest singularities of the particular solution~(\ref{particular}), i.e.,  the ones that are closest to the origin $t=0$. These singularities are simple poles located at $t=\pm i\pi/2$. They are proportional to the factors $\exp{(\pm \pi v/2)}$ or to linear combinations thereof. To remove them one should find a zero-mode solution with poles at $t=\pm i\pi/2$, but with residues cancelling the one in~(\ref{particular}). Solutions of this type were considered in~\cite{Basso:2009gh} in a related context and were shown to be exponentially suppressed with the coupling. To be more precise, for singularities located at $t = \pm i\pi/2$, the zero-mode solution comes with a factor $\exp{(-\pi g)}$ and the exponential damping will get even stronger for zero-mode solution with singularities located further away from the origin. One thus immediately concludes that the homogeneous solution is negligible at strong coupling as long as $v^2 \ll (2g)^2$. The particular solution~(\ref{particular}) should therefore correctly describe the dynamics of the holes up to exponentially small corrections.

To give a simple illustration of the above discussion, we can look at the energy and momentum of a hole. Given the functions $\gamma^{v}$ and  $\tilde{\gamma}^{v}$, they can be obtained by using the general relations~\cite{Basso:2010in}
\beq
E_{\mathcal{N}=4}(v) = 1 + 2\lim\limits_{t\rightarrow 0} \frac{\gamma^{v}(2gt)}{t}\, , \qquad p_{\mathcal{N}=4}(v) = 2v+ 2\lim\limits_{t\rightarrow 0} \frac{\tilde{\gamma}^{v}(2gt)}{t}\, .
\eeq
Evaluating them with the the help of the particular solution~(\ref{particular}), we find
\beq
E_{\mathcal{N}=4}(v)_{\textrm{part}} = 0\, , \qquad p_{\mathcal{N}=4}(v)_{\textrm{part}} = 0\, .
\eeq
This was expected since energy and momentum are exponentially small at strong coupling. Their leading order expressions are thus captured by the homogeneous solutions, i.e., we verify that $\gamma^{v}_{\textrm{hom}}\propto m\cosh{(\pi v/2)}$ and $\tilde{\gamma}^{v}_{\textrm{hom}}\propto m\sinh{(\pi v/2)}$, with $m\sim \exp{(-\pi g)}$ being the mass gap of the theory.

It is quite straightforward to obtain the scattering phase for holes in the relevant strong coupling regime. Plugging the particular solution~(\ref{particular}) into the general expression~(\ref{Sgeneralbis}) we derive
\beq\label{So6reg}
S_{\mathcal{N}=4}(u, v)_{\textrm{part}} = S_{\textrm{O(6)}}(u-v)\,,
\eeq
where
\beq\label{So6ZZ}
S_{\textrm{O(6)}}(u) = -\frac{\Gamma(1+\frac{iu}{4})\Gamma(\frac{1}{2}-\frac{iu}{4})\Gamma(\frac{3}{4}+\frac{iu}{4})\Gamma(\frac{1}{4}-\frac{iu}{4})}{\Gamma(1-\frac{iu}{4})\Gamma(\frac{1}{2}+\frac{iu}{4})\Gamma(\frac{3}{4}-\frac{iu}{4})\Gamma(\frac{1}{4}+\frac{iu}{4})}\, ,
\eeq
is the S-matrix of the $O(6)$ model~\cite{Zamolodchikov:1977nu} written in terms of $u=2\theta/\pi$. This generalises the result of~\cite{Basso:2008tx}, where the relation with the $O(6)$ model was established at the level of the kernel of the integral equation for symmetric density of holes. In the above derivation no assumption was made about the hole rapidities except that $u,v \sim O(g^0)$ at strong coupling.

The last ingredient entering the Bethe-Yang equations for holes is the transmission amplitude $T_{\mathcal{N}=4}(v)$, see Eq.~(\ref{ABAN=4}). It can be expressed in terms of the function $\tilde{\gamma}^{v}$, as done in~(\ref{defTv})-(\ref{defTvalt}). Plugging the particular solution~(\ref{particular}) into the formula~(\ref{defTvalt}) yields
\beq
T_{\mathcal{N}=4}(v)_{\textrm{part}} = 1\, .
\eeq
This result suggests that in the $O(6)$ regime the holes do not experience any scattering when moving through the tips of the string. This is actually not completely correct since a non-trivial contribution to the transmission amplitude may still be generated by the homogeneous part $\tilde{\gamma}^{v}_{\textrm{hom}}$ of the solution. To the leading order in the mass gap $m$, the homogeneous solution is proportional to the momentum, as we have just demonstrated. It implies that $\log{T_{\mathcal{N}=4}}(u) \propto ip_{\mathcal{N}=4}(u)$ up to coupling dependent constant. This simply means that the net effect of the transmission amplitude is to modify the expression for the length  of the system $R$. More precisely, as discussed in~\cite{Basso:2011rc}, the effective length of the $O(6)$ model reads
\beq\label{renorR}
2\log S\rightarrow R = 2\log{\left(\frac{8\pi S}{\sqrt{\lambda}}\right)} + O(1/\sqrt{\lambda})\,.
\eeq
This agrees with the results obtained in a related context in~\cite{Freyhult:2011gf, Fioravanti:2011xw}.

In summary, the holes are controlled by the $O(6)$ model sigma model when the rapidities are kept to be of order $O(1)$ at strong coupling. There is no nontrivial transmission amplitude and the Bethe-Yang equations are the same as for a system with periodic boundary conditions put on a cylinder of length $R$. More generally, we expect that the Thermodynamic Bethe Ansatz (TBA) equations for  holes will coincide with the TBA equations for $O(6)$ model in the regime considered. Some evidence supporting this conjecture comes from the L\"usher formula for the vacuum energy, which was proposed in~\cite{Basso:2011rc} and shown to reduce to the one for the $O(6)$ model in the non-perturbative regime.

\subsection{The Bykov model}\label{sec:Bykov}

We now turn to the discussion of the low-energy effective model for the GKP string in $AdS_4\times CP^3$. The relevant bosonic excitations parameterise the fluctuations of the string in $CP^3$. They couple to a massless Dirac fermion that is the only remnant of the superstring coordinates in the Alday-Maldacena decoupling limit~\cite{Alday:2007mf}. The Lagrangian for these low-energy modes was constructed by Bykov in~\cite{Bykov:2010tv}. Its dynamics and S-matrix were studied in~\cite{Basso:2012bw}, building on previous studies of similar models~\cite{Witten:1978bc,D'Adda:1978kp,Koberle:1987wc,Azaria:1995wg,Campostrini:1993fr}. It was found that the spectrum is gapped and spanned by two multiplets of massive excitations in the $\mathbf{4}$ and $\mathbf{\bar{4}}$ of $SU(4)$, respectively. They were called spinons and anti-spinons in~\cite{Basso:2012bw}. Here, they are identified with holes and anti-holes of the $\mathcal{N}=6$ theory. It is interesting to note that the spinons are neither fermions nor bosons~\cite{Basso:2012bw}. They have a fractional statistics corresponding to spin $1/4$. This is somewhat in line with the gauge theory description, since in the spin chain picture the elementary mixing  $\psi^1_+\psi^{\dagger}_{+4}\sim Y^{1}\mathcal{D}Y^{\dagger}_{4}$ seems to prevent us from any clear conclusion regarding the statistics of the holes and anti-holes.

An important consequence of the string theory description of these excitations is that their mass has to be exponentially suppressed with the string tension. This is the same phenomenon as for the $\mathcal{N}=4$ theory~\cite{Alday:2007mf}. The spinons are massive and their mass is proportional to the dynamical scale $\Lambda$ of the Bykov model. The latter depends on the $CP^3$ coupling as
\beq
\Lambda \propto \kappa^{1/4} e^{-\pi\kappa/2}\,.
\eeq
This may be read off from Eq.~(10) of~\cite{Basso:2012bw} evaluated for $N=4$ and $p=1$. To prove that this scaling is correctly reproduced by the gauge theory, we should first determine the relation between the coupling constants of both theories. This can be done as follows. First, we observe that the coupling $\kappa$ of the effective theory and the string tension $\sigma$ may be related by the low-energy expansion of the Nambu-Goto action,
\beq
\mathcal{L} = \sigma \sqrt{ds^2_{\textrm{AdS}_3}+4ds^2_{CP^3}} = \sigma + 2\sigma ds^2_{CP^3} +\ldots = \sigma + \kappa ds^2_{CP^3} +\ldots\,.
\eeq
We used that $ds^2_{\textrm{AdS}_3}=1$ evaluates to $1$ on the GKP background. Hence, $\kappa$ is twice the string tension $\sigma$, which in the $\mathcal{N}=6$ theory matches with the spin-chain coupling $g$, since $g\equiv h(\lambda) = \sqrt{\lambda/2} +\ldots = \sigma + \ldots$ for $\lambda, g, \sigma \gg 1$. In short, $\kappa = 2g +\ldots$ and thus $\Lambda \propto g^{1/4}e^{-\pi g}$. We immediately observe that this is exactly the same dependence on $g$ as for the $O(6)$ model. Our prediction that $m_{\mathcal{N}=6}$ is equal to $m_{\mathcal{N}=4}$, up to an irrelevant factor of $1/2$, agrees with what one expects from the Bykov model.

A stronger evidence that the holes of the strongly coupled $\mathcal{N}=6$ theory are described by the Bykov model comes from  matching the S-matrices. The scattering of spinons and anti-spinons was analyzed in~\cite{Basso:2012bw} and argued to be controlled by the minimal reflectionless S-matrix with $U(4)$ symmetry, which was originally constructed in~\cite{Berg:1977dp}. Its global structure is therefore the same as the one found in Section~\ref{sec:u4Sm} for the scattering of holes in the $\mathcal{N}=6$ theory. Thanks to the general expressions~(\ref{prodS})-(\ref{ratioS}), we can prove that in the decoupling limit the S-matrices of the two theories are identical. After plugging~(\ref{So6reg})-(\ref{So6ZZ}) into~(\ref{prodS})-(\ref{ratioS}), we immediately derive
\beq\label{SmatU4final}
S(u, v)  = -\frac{\Gamma(1+\ft{i(u-v)}{4})\Gamma(\ft{1}{4}-\ft{i(u-v)}{4})}{\Gamma(1-\ft{i(u-v)}{4})\Gamma(\ft{1}{4}+\ft{i(u-v)}{4})}\, , \qquad \bar{S}(u, v) = \frac{\Gamma(\ft{1}{2}-\ft{i(u-v)}{4})\Gamma(\ft{3}{4}+\ft{i(u-v)}{4})}{\Gamma(\ft{1}{2}+\ft{i(u-v)}{4})\Gamma(\ft{3}{4}-\ft{i(u-v)}{4})}\, .
\eeq
We observe an agreement with the expressions for scattering phases of spinons written in~\cite{Basso:2012bw} when the map $\theta = \pi u/2$, \textit{cf.} formula~(\ref{Enpreg}), between relativistic and Bethe rapidities is utilised.

Let us now comment on the Bethe-Yang equations. They involve the transmission amplitude $T_{\mathcal{N}=6}(u)$, see Eq.(\ref{ABAN=6}). This quantity is related to its $\mathcal{N}=4$ counterpart~(\ref{T4T6}) in such a way that it gets absorbed into the  effective length of the model. This follows closely the mechanism in~(\ref{renorR}). It also follows from~(\ref{T4T6}) that, when expressed in terms of the spin-chain coupling $g$, the effective lengths for $\mathcal{N}=4$ and $\mathcal{N}=6$ theories coincide up to rescaling of the spin by a factor of $2$. Hence, we get
\beq
2\log{(2S)}\rightarrow R = 2\log{\left(\frac{4S}{g}\right)} + O(1/g) = 2\log{\left(\frac{4S}{\sigma}\right)} + O(1/\sigma)\, .
\eeq
Thanks to this doubling of the spin, this result is the same as the one found in~(\ref{renorR}) for the string in $AdS_5\times S^5$ when written in terms of the string tension $\sigma = \sqrt{\lambda}/(2\pi) = 2g$. This is not very surprising since the expression $2\log{(4S/\sigma)}$ has the meaning of the classical length of the GKP string in the long string limit $S\gg \sigma$. It is fixed by the geometry of the $AdS_3$ subspace in which the string is rotating and is thus oblivious at the classical level to the embedding in $AdS_5\times S^5$ or  $AdS_4\times CP^3$.

Finally, we already mentioned several times that the BY equations for the holes are equipped with a selection rule and with the twist factor $q$. Remarkably, both features are present in the description of the spectrum of the Bykov model in finite volume~\cite{Basso:2012bw}. The selection rule originates from the $U(1)$ gauge symmetry of the model. Under this symmetry the two multiplets of fields $z_i$ and $\bar{z}_i$, parametrising the $CP^3$ space, have charges $\pm 1$, while the massless fermion of the Bykov model carries charge $+2$. Any state of the finite-volume theory can be mapped to a vertex operator built out of the previous elementary fields. Since physical operators ought to be neutral under the gauge symmetry, the overall charge should sum up to zero. This translates to the requirement that the number of Dirac fermions $F$ is half of the difference between the number of $z_i$ and $\bar{z}_i$ fields. For a state of the theory, which is only sensitive to the number of spinons and anti-spinons, this turns into the condition $K_h-\bar{K}_h = 2F$. This is exactly what we found for the holes in the $\N=6$ theory. The origin of the twist $q$ is somewhat more subtle and is associated to a $\mathbb{Z}_2$ symmetry of the theory that is spontaneously broken in infinite volume. This explains why this symmetry is not directly visible at the level of the asymptotic data. The symmetry gets restored, however, when the system is put on finite cylinder and manifests itself by the presence of the twist which takes two possible values. We refer the reader to~\cite{Basso:2012bw} for further clarifications.

The above discussion leads us to the conclusion that in the decoupling limit the holes of the $\mathcal{N}=6$ theory and the spinons of the Bykov model are indistinguishable from one another, at least at the level of the Bethe-Yang equations. It would be interesting to see whether this is still the case once finite-size corrections are included~\cite{Gromov:2009tv}.

\section{Conclusions}

In this paper we have derived and analysed Bethe ansatz equations for the low-lying excitations of the GKP string in both $AdS_5\times S^5$ and $AdS_4\times CP^3$ string theory. The integrable structures for both theories bear much resemblance and so it comes as no surprise that the effective equations for holes we derived are similar in both models. Yet, at strong coupling, they describe spectra of \textit{two different} integrable models, the $O(6)$ sigma model and the Bykov model. Does this mean that both integrable models are closely related as well? The answer seems to be positive after recalling that the S-matrix of the $O(6)$ and Bykov model coincide in a specific subsector. This is valid at any coupling and simply reflects the fact that in this subsector  the spectrum of excitations is the same in the two gauge theories~\cite{Gromov:2008qe}. The S-matrix nevertheless operates in two different channels from the $SU(4)$ perspective, namely $[0, K_{h}, 0]$ for $\N=4$ theory and $[K_{h}, 0, K_{h}]$ for $\mathcal{N}=6$ Chern-Simons-Matter theory. It is then a bit surprising that they happen to be related to one another in such a simple way. The phenomenon is actually not coincidental  and may be lifted to higher-rank models. More precisely, the integrable $O(N+2)$ non-linear sigma model appears related to the integrable $U(1)\times SU(N)$ model studied in~\cite{Basso:2012bw}. Their S-matrices coincide in the subsector $K_h=\bar{K}_h$, when holes and anti-holes become indistinguishable. Although we expect this relation  to break down after including finite-size corrections, it is interesting to ask whether this fact may be understood directly from the world-sheet theories.

It is interesting to comment on the vacuum structure of the Bykov model. In fact, the theory has two vacua, as was argued in~\cite{Basso:2012bw} from the the world-sheet perspective. In gauge theory they correspond to the twist-one solutions associated with even and odd values of spin. These have different analytic properties.  This is manifest already at the leading order at weak coupling, where the anomalous dimension is given by $4g^2(\psi(\ft{S}{2}+\ft{1}{2})-\psi(\ft{1}{2}))$ for even spins. The expression for odd spins is obtained by replacing $S$ by $S+1$ in this formula, see~\cite{Zwiebel:2009vb,Beccaria:2009ny}. At large spin they become degenerate, with their energy bands separated by $O\left(1/S\right)$ at weak coupling. The degeneracy is lifted by finite-size corrections, or, to be more precise, by the exchange of lowest-twist particles going around the GKP string with the circumference $2\log{S}$. This effect induces the separation of vacua of the order
\beq
E_{\textrm{even}}-E_{\textrm{odd}} \sim 1/S^{2m}\,,
\eeq
where $m$ is the mass gap of the theory. At weak coupling $m=1/2+O(g^2)$, while at strong coupling $m\rightarrow 0$ and the two bands in the stringy perturbative regime are expected to be separated by
\beq\label{cformula}
E_{\textrm{even}}-E_{\textrm{odd}} = -\frac{\pi}{\log{S}} + \ldots 
\eeq
according to the Eq.(110) in~\cite{Basso:2012bw}.
More generally, it is possible to write down the first L\"uscher formula for the vacuum energy shift
\beq\label{Luescher}
E_{\textrm{even}}-E_{\textrm{odd}}  = -N_{h}(q_{+}-q_{-})\int \frac{dp(u)}{2\pi} y_{\textrm{vacuum}}(u^{\gamma})  -\bar{N}_{h}(q_{+}-q_{-})\int \frac{dp(u)}{2\pi} \bar{y}_{\textrm{vacuum}}(u^{\gamma}) \,.
\eeq
It should be valid up to contributions from higher-twist excitations and multi-particle states, which are suppressed with higher powers of $1/S$. There are $N_{h} = \bar{N}_{h} = 4$  holes and anti-holes in the mirror channel and the twist factors of the even and odd vacua are $q_{\pm } = \pm 1$. The $y_{\textrm{vacuum}} = \bar{y}_{\textrm{vacuum}}$ functions, as was argued in this paper, are directly related to the corresponding function for $\mathcal{N}=4$ theory. It was constructed in~\cite{Basso:2011rc}, so that we  immediately find
\beq
y_{\textrm{vacuum}}(u^{\gamma}) = (2S)^{-2E_{\mathcal{N}=6}(u)}T_{\mathcal{N}=6}(u^{\gamma}) = \frac{\pi g^2}{2\cosh{(\pi u)}S} + O(g^4)
\eeq
to leading order at weak coupling. Plugging this expression into \eqref{Luescher} we derive
\beq
E_{\textrm{even}}-E_{\textrm{odd}} = -\frac{4g^2}{S} + \ldots\,,
\eeq
which reproduces the two-loop $\sim g^2 \sim h(\lambda)^2\sim \lambda^2$ result. It would be interesting to find the higher-loop expansion using the L\"uscher formula proposed above. For instance, at higher orders in perturbation theory the formula~(\ref{Luescher}) should capture large spin wrapping corrections, about which little is known so far. The first L\"uscher correction on its own will not  reproduce the strong coupling prediction~(\ref{cformula}). This is because it ceases to be valid when $m\log{S}$ is no longer large enough. This calls for a better understanding of the TBA equations for the GKP string, with help of which one would be able to re-sum all L\"uscher corrections.

In \cite{Basso:2012bw} we pointed out that the massless world-sheet fermions inherited by the low-energy sigma model are subject to anti-periodic boundary conditions. This seems a puzzle from the point of view of the Green-Schwarz formulation of the $AdS_4 \times CP^3$ sigma model, where all fermionic excitations are periodic. Since our analysis of the AdS/CFT Bethe equations confirms the presence of twist factors induced by the anti-periodic fermion, this twisting of boundary conditions calls for further investigation. How exactly does the low-energy truncation interferes with the standard boundary conditions? 

Finally, let us comment on the the dressing phase present in the all-loop Bethe ansatz for $AdS_4 \times CP^3$ correspondence. It was conjectured in \cite{Gromov:2008qe} and \cite{Ahn:2008aa} to coincide with the BES dressing phase \cite{Beisert:2006ez}. While there are several tests of the dressing phase available for the $\N=4$ case, see for example the review article \cite{Vieira:2010kb}, the situation is less clear in case of $\N=6$ Chern-Simons-Matter theory. Recently, the weak-coupling computation of \cite{Mauri:2013vd} confirmed the validity of the dressing factor at the leading order. The solutions \eqref{particular} yielding the low-energy models \eqref{So6ZZ} and \eqref{SmatU4final} are very sensitive to the actual expression for the dressing phase. Since we are able to reproduce the spectral equations for low-energy sigma models from Bethe ansatz equations of \cite{Gromov:2008qe}, we feel that this result strongly supports the veracity of the conjectured dressing factor.

\section*{Acknowledgments}
We would like to thank Zolt\'an Bajnok and Juan Maldacena for discussions. Benjamin Basso is grateful to the Wigner Research Center for Physics for hospitality while this article was still in preparation. 
The research of Adam Rej was supported by a Marie Curie International Outgoing Fellowship within the 7th European Community Framework Programme, grant number PIOF-GA-2010-273854. Adam Rej also gratefully acknowledges support from the Institute for Advanced Study.

\appendix

\section{Derivation of the Bethe-Yang equations}\label{BYapp}

In this appendix we first recall the expression for the correction to the density of roots sourced by a hole in the $\mathcal{N}=4$ SYM theory~\cite{Basso:2010in}. This result is then used to construct all the relevant quantities  that enter the expression for hole counting function in both the $\mathcal{N}=4$ and $\mathcal{N}=6$ theories. 

\subsection{The all loop density}\label{all-loop-densities}\label{alldens}

The density $\rho(u)$ is defined as the logarithmic derivative of the function $Y(u)$
\beq\label{rhosig}
\rho(u) = -\frac{i}{2\pi}\partial_{u}\log Y(u)\, ,
\eeq
and it contains all the information one needs to carry out the analysis at large spin. In practice, it is convenient to separate the contribution of holes from the large spin vacuum. Because the large spin equation is linear, we may as well consider only one hole and write
\beq \label{vactohole}
\rho(u) = \rho_{\textrm{vacuum}}(u) - \sigma^{v}_{\textrm{hole}}(u) \, .
\eeq
The density $\sigma^{v}_{\textrm{hole}}(u)$ can be shown to admit the representation~\cite{Basso:2010in}
\beq\label{sigv}
\sigma^{v}_{\textrm{hole}}(u) = \frac{1}{\pi}\int_{0}^{\infty}dt \cos{(ut)}e^{t/2}\Omega^{v}(t) +\frac{1}{\pi} \int_{0}^{\infty}dt \sin{(ut)}e^{t/2}\tilde{\Omega}^{v}(t)\, ,
\eeq
where $\Omega^{-v}(t) = \Omega^{v}(t)$ and $\tilde{\Omega}^{-v}(t) = -\tilde{\Omega}^{v}(t)$ correspond to odd and even parts of the hole density. We further decompose these functions as follows
\beqa \label{FTomega}
&&\Omega^{v}(t) = \frac{\cos{(vt)}e^{-t/2}-J_{0}(2gt)}{e^{t}-1}+\frac{\gamma^{v}(2gt)}{e^{t}-1}\, ,\\ \label{FTomegat}
&&\tilde{\Omega}^{v}(t) = \frac{\sin{(vt)}e^{-t/2}}{e^{t}-1}+\frac{\tilde{\gamma}^{v}(2gt)}{e^{t}-1}\, ,
\eeqa
where $J_{0}(z)$ is the zeroth Bessel function and $\gamma^{v},\tilde{\gamma}^{v}$ are yet to be determined. The first term in the right hand side of these equations roughly accounts for the one-loop density, while the other one takes care of the higher-loop corrections.

The functions $\gamma^{v}$ and $\tilde{\gamma}^{v}$ contain all the dynamics of the hole beyond the one-loop order. As we shall see they can be used to construct expressions for the energy, momentum, transmission amplitude and the S-matrix. It was argued in~\cite{Basso:2010in} that these two functions should be holomorphic in the $t$-plane, excluding $t=\infty$ where they are expected to have an essential singularity. Their Fourier transforms are supported on the interval $(-2g, 2g)$ with square-root branch points at both ends. Both properties can be derived from the fact that $\gamma^{v}, \tilde{\gamma}^{v}$ admit absolutely convergent Neumann expansions over Bessel functions,
\beq\label{Neumann}
\gamma^{v}(2gt) = \sum_{n\geq 1}2n\gamma^{v}_{n}J_{n}(2gt)\,, \qquad \tilde{\gamma}^{v}(2gt) = \sum_{n\geq 1}2n \tilde{\gamma}^{v}_{n}J_{n}(2gt)\,.
\eeq
Note that the expansion coefficients $\gamma^{v}_{n},\tilde{\gamma}^{v}_{n}$ are only functions of the coupling constant and rapidity. After decomposing the function $\gamma^{v}$ into odd and even parts w.r.t. to $t$
\beq
\gamma^{v}(2gt) = \gamma^{v}_{+}(2gt)+\gamma^{v}_{-}(2gt)\, , \qquad \gamma^{v}_{\pm}(-2gt) = \pm \gamma^{v}_\pm (2gt)\, ,
\eeq
one can also deduce the following integral representations
\beq\label{NeumannInt}
\gamma_{2n-1}^{v} = \int_{0}^{\infty}\frac{dt}{t}J_{2n-1}(2gt)\gamma_{-}(2gt)\, , \qquad \gamma_{2n}^{v} = \int_{0}^{\infty}\frac{dt}{t}J_{2n}(2gt)\gamma_{+}(2gt)\,.
\eeq
Similar formulae may be found for the coefficients $\tilde{\gamma}_n$. They follow from the orthogonality relations for odd or even Bessel functions\footnote{We assume here that it is permissible to exchange the summation and integration operations when deriving these expressions. This can be justified by estimating the large $n$ behaviour of the Neumann coefficients that solve the system of equations~(\ref{Eqs-system}).}
\beq\label{orthoBessel}
\int_{0}^{\infty}\frac{dt}{t}J_{2n-1}(t)J_{2m-1}(t) =  \frac{\delta_{n, m}}{2(2n-1)}\, , \qquad \int_{0}^{\infty}\frac{dt}{t}J_{2n}(t)J_{2m}(t) =  \frac{\delta_{n, m}}{4n}\,.
\eeq

The first coefficient in~(\ref{Neumann}) is especially interesting as it controls the energy of a hole carrying rapidity $v$
\beq
E(v) = 1+4g\gamma^{v}_{1}\, .
\eeq
The momentum, on the other hand, is given by 
\beq \label{pholeN4}
p(v) = 2v + 4g\tilde{\gamma}^{v}_{1}\,.
\eeq
More complicated observables, like the S-matrix, $S(u, v)$, or the transmission amplitude, $T(u)$, require determining all coefficients, or equivalently the entire functions $\gamma^{v}, \tilde{\gamma}^{v}$. This can be done by solving a system of linear equations for the Neumann coefficients $\gamma^{v}_n, \tilde{\gamma}^{v}_n$, see \cite{Basso:2010in},
\beq\label{Eqs-system}
\begin{aligned}
&\gamma_{n}^{v} + \int_{0}^{\infty}\frac{dt}{t}J_{n}(2gt)\frac{\gamma^{v}_{+}(2gt)-(-1)^n\gamma^{v}_{-}(2gt)}{e^{t}-1} = \kappa^{v}_{n}\, ,\\
&\tilde{\gamma}_{n}^{v} + \int_{0}^{\infty}\frac{dt}{t}J_{n}(2gt)\frac{\tilde{\gamma}^{v}_{-}(2gt)+(-1)^n\tilde{\gamma}^{v}_{+}(2gt)}{e^{t}-1} = \tilde{\kappa}^{v}_{n}\,.\\
\end{aligned}
\eeq
In both equations $n\geqslant 1$ and the source terms are given by
\beq\label{source-terms}
\begin{aligned}
&\kappa^{v}_{n} = -\int_{0}^{\infty}\frac{dt}{t}J_{n}(2gt)\frac{\cos{(vt)}e^{t/2}-J_{0}(2gt)}{e^{t}-1}\, ,\\
&\tilde{\kappa}^{v}_{n} = -\int_{0}^{\infty}\frac{dt}{t}J_{n}(2gt)\frac{\sin{(vt)}e^{t/2}}{e^{t}-1}\, .
\end{aligned}
\eeq
Solving these equations at weak coupling is rather straightforward. For instance, to leading order we find
\beq
\gamma^{v}_{n} = \kappa^{v}_{n} = \tilde{\gamma}^{v}_{n} = \tilde{\kappa}^{v}_{n}=O(g^{n})\, .
\eeq
In particular, one easily verifies that
\beq
E(v) =  1+2g^2(\psi(\ft{1}{2}+iv)+\psi(\ft{1}{2}-iv)-2\psi(1)) + O(g^4)\, ,
\eeq
and
\beq
p(v) =  2v-2\pi g^2\textrm{tanh}(\pi v) + O(g^4)\, ,
\eeq
in agreement with the one-loop expressions for the energy and momentum of a hole~\cite{Belitsky:2006en,Basso:2010in,Gaiotto:2010fk}. Subleading corrections can be obtained by iterating~(\ref{Eqs-system}). Later on we shall use the system of equations~(\ref{Eqs-system}) to prove the unitarity of the S-matrix.

Finally, let us point out that \eqref{Eqs-system} may be repackaged into
\beqa \label{Eqs-uform}
&&\int_{0}^{\infty}\frac{dt}{t}\sin{(ut)}\bigg[\frac{\gamma^{v}_{-}(2gt)}{1-e^{-t}}+\frac{\gamma^{v}_{+}(2gt)}{e^{t}-1}+\frac{\cos{(vt)}e^{t/2}-J_{0}(2gt)}{e^{t}-1}\bigg] =0\,,\\ \nn
&&\int_{0}^{\infty}\frac{dt}{t}(\cos{(ut)}-J_{0}(2gt))\bigg[\frac{\gamma^{v}_{+}(2gt)}{1-e^{-t}}-\frac{\gamma^{v}_{-}(2gt)}{e^{t}-1}+\frac{\cos{(vt)}e^{t/2}-J_{0}(2gt)}{e^{t}-1}\bigg] =0\,,\\ \nn
&&\int_{0}^{\infty}\frac{dt}{t}\sin{(ut)}\bigg[\frac{\tilde{\gamma}^{v}_{-}(2gt)}{1-e^{-t}}-\frac{\tilde{\gamma}^{v}_{+}(2gt)}{e^{t}-1}+\frac{\sin{(vt)}e^{t/2}}{e^{t}-1}\bigg] = 0\, ,\\ \nn
&&\int_{0}^{\infty}\frac{dt}{t}(\cos{(ut)}-J_{0}(2gt))\bigg[\frac{\tilde{\gamma}^{v}_{+}(2gt)}{1-e^{-t}}+\frac{\tilde{\gamma}^{v}_{-}(2gt)}{e^{t}-1}+\frac{\sin{(vt)}e^{t/2}}{e^{t}-1}\bigg] =0\,,
\eeqa
which hold for $u^2 \leqslant (2g)^2$ only. This form will be useful for working out the crossing transformation of the S-matrix. Here we will apply these equations to derive the relation~(\ref{parteq}) which was relevant for the strong coupling discussion in Section~\ref{O(6)limit}. We easily notice that the latter relation is more or less the same as the last equation in~(\ref{Eqs-uform}) evaluated at $u=0$. The only thing missing is the following identity
\beq\label{J0tilde}
\int_{0}^{\infty}\frac{dt}{t}J_{0}(2gt)\tilde{\gamma}_{+}^{v}(2gt) = 0\, ,
\eeq
which follows from the orthogonality relations~(\ref{orthoBessel}) and leads immediately to~(\ref{parteq}). 

\subsection{The counting function of the $\mathcal{N}=4$ SYM theory}\label{app:cfN=4}

We now turn to the evaluation of the counting function. We recall that we already derived the all-loop representation for the density. The final step is to integrate it properly.

As explained in Section~\ref{sec:BYol} computing the counting function at large spin requires separating contributions from the two  regions of the rapidity: $u\sim 1$ and $u\sim S$.  Although this analysis may be performed explicitly, it is simpler to redefine the counting function so to avoid computations in the semiclassical regime $u \sim S$. This is the approach we will take here. We will not include isotopic roots in our derivation, as they were treated previously in~\cite{Basso:2010in}. At the end of this section, however, we will point out how they modify the final expression.

Our starting point is the ``little'' counting function $y(u)$. We have already introduced a similar quantity at weak coupling for the ABJM spin chain in Section \ref{sec:BYol}. Observing that for $\N=4$ there is only one momentum-carrying node, the adequate definition is
\beq\label{yN4}
y(u) = (-1)^{K_{h}}\left({-\frac{x^-}{x^+}}\right)^L \prod^{K}_{j=1} \frac{i-u+u_j}{i+u-u_j} \left(\frac{1-g^2/x^+ x^{-}_j}{1-g^2/x^- x^+_j}\right)^2\sigma^2(u,u_j)\left(-\frac{x_{j}^{-}}{x_{j}^{+}}\right)\, ,
\eeq
where $x^{\pm} = x(u\pm i/2)$, $x(u) = \ft{1}{2}(u+\sqrt{u^2-4g^2})$ is the Zhukovsky variable and $\sigma(u, v)$ is the BES dressing phase~\cite{Beisert:2006ez}. Note that in  the $\mathfrak{sl}(2)$ subsector we have  $K_{h} = L-2$ holes and $K = S$ magnons, where $L$ is the length of the spin chain and $S$ is the Lorentz spin. It is related to the usual counting function $Y(u)$ via
\beq
Y(u) = e^{iP}y(u)\, , \qquad \textrm{with} \qquad e^{iP} = \prod_{j=1}^{K}\frac{x^{+}_j}{x^{-}_j}\, ,
\eeq
such that for physical states $Y(u) = y(u)$. Note that there is $\textit{no}$ sign ambiguity related with this regularisation scheme as opposed to $\N=6$ theory, \textit{cf.} Section \ref{sec:BYol}.

The vacuum contribution corresponding to the vacuum density in \eqref{vactohole} was previously constructed in~\cite{Basso:2011rc} and reads
\beq
y_{\textrm{vacuum}}(u) = e^{2ip(u)\log{S}+2i\delta p(u)} = e^{2ip(u)\log{S}}T(u) \, ,
\eeq
where $p(u)$ is the hole momentum and $2\delta p(u) = -i\log{T(u)}$ is the transmission amplitude, see~(\ref{transamp}). A representation of $\delta p(u)$ was given in~\cite{Basso:2011rc}. Here we would like to point out that it admits an equivalent representation directly in terms of the function $\tilde{\gamma}^{v}$. The explicit expression reads
\beqa \nn
\log{T(v)} &=& -2ip(v)\psi(1)+4i\int_{0}^{\infty}\frac{dt}{t}\frac{\sin{(vt)}e^{t/2}J_{0}(2gt)-vt}{e^{t}-1}\\ \label{defTv}
&&\, +4i\int_{0}^{\infty}\frac{dt}{t}\frac{\tilde{\gamma}^{v}(2gt)J_{0}(2gt)-2g\tilde{\gamma}^{v}_{1}t}{e^{t}-1} \, ,
\eeqa
and it follows from the one given in~\cite{Basso:2011rc} by using an exchange relation similar to the one proposed in~\cite{Basso:2008tx}, see also Appendix~\ref{Unitarity}. This can also be written as
\beq\label{defTvalt}
\log{T(v)} = -2ip(v)\psi(1)+4i\int_{0}^{\infty}\frac{dt}{t}\frac{\sin{(vt)}e^{t/2}+\tilde{\gamma}^{v}(2gt)-p(v)t/2}{e^{t}-1} -4i\int_{0}^{\infty}\frac{dt}{t}\tilde{\gamma}^{v}_{+}(2gt)\, ,
\eeq
and may be deduced from~(\ref{defTv}) by using  \eqref{Eqs-uform}-\eqref{J0tilde}. In practice, once the S-matrix $S(u, v)$ has been computed, it is often easier, especially at weak coupling, to extract the transmission amplitude by means of~(\ref{tss}).

The contribution to $y(u)$ of a single hole carrying a rapidity $v$ is  
\beq\label{logy}
\log{(-y^{v}(u))} = -2i\int_{0}^{\infty}\frac{dt}{t} \sin{(ut)}e^{t/2}\Omega^{v}(t) + 2i \int_{0}^{\infty}\frac{dt}{t}(\cos{(ut)}e^{t/2}-J_{0}(2gt))\tilde{\Omega}^{v}(t)\, .
\eeq
It is such that $-\sigma^{v}_{\textrm{hole}}(u) =- \tfrac{i}{2\pi}\partial_{u}\log{y^{v}(u)}$ is the correction to the density of roots induced by a hole with rapidity $v$, see Eqs.~(\ref{rhosig}-\ref{sigv}). We point out here that the large-rapidity divergence in the counting function mentioned in Section~\ref{sec:BYol} is directly related to the small $t$ behaviour
\beq
\frac{\cos{(ut)}e^{t/2}\tilde{\Omega}^{v}(t)}{t} \sim \frac{p(v)}{2t}\, ,
\eeq
which appears when integrating the density~(\ref{sigv}). The refined counting function in~(\ref{yN4}) is rid of this problem since the second integrand in~(\ref{logy}) behaves regularly at small $t$. Adding the factor $\prod_{j}(-x_{j}^{-}/x_{j}^{+})$ in~(\ref{yN4}) produces the zeroth Bessel function, $J_{0}(2gt)$, in~(\ref{logy}) and serves as a regulator. Now, plugging the expressions~(\ref{FTomega}-\ref{FTomegat}) into~(\ref{logy}) we immediately arrive at
\beq
\begin{aligned}\label{logyf}
\log{(-y^{v}(u))} = &-2i\int_{0}^{\infty}\frac{dt}{t} \frac{\sin{(ut)}e^{t/2}\gamma^{v}(2gt)}{e^{t}-1} +2i\int_{0}^{\infty}\frac{dt}{t}\frac{(\cos{(ut)}e^{t/2}-J_{0}(2gt))\tilde{\gamma}^{v}(2gt)}{e^{t}-1} \\
&+2i\int_{0}^{\infty}\frac{dt}{t}\frac{\sin{(ut)}e^{t/2}J_{0}(2gt)-\sin{(vt)}e^{t/2}J_{0}(2gt)-\sin{((u-v)t)}}{e^{t}-1}\, .
\end{aligned}
\eeq
In line with our discussion in Section~\ref{sec:BYfc}, the final expression is of the type
\beq\label{yN=4}
y(u) = y_{\textrm{vacuum}}(u) \prod_{j=1}^{K_h}S(u, u_{h, j})\, ,
\eeq
with the S-matrix $S(u, v) = y^{v}(u)$ given by~(\ref{logyf}) or equivalently by~(\ref{Sgeneral}-\ref{f-functions}). 

Finally, following the discussion in~\cite{Basso:2010in}, adding isotopic roots $u_{b, j}$ is straightforward and leads to the following extension of the little counting function
\beq
y(u) = y_{\textrm{vacuum}}(u) \prod_{j=1}^{K_h}S(u, u_{h, j})\prod_{j=1}^{K_{b}}\frac{u-u_{b, j}+\ft{i}{2}}{u-u_{b, j}-\ft{i}{2}}\,.
\eeq
This is the result we used in~(\ref{ABAN=4}).

\subsection{The counting functions of the $\mathcal{N}=6$ ABJM theory}

In this section we extend the results of Section~\ref{sec:BYol} to all-loop order. This will allow us to present non-perturbative expressions for the S-matrix of the holes. 

The single-polarisation sector for the $\N=6$ ABJM theory is comprised by the two momentum-carrying nodes only. This way the all-loop equations \cite{Gromov:2008qe} are reduced to  
\beqa \label{BE1}
&&\left(\frac{x^+_k}{x^-_k}\right)^L=\prod^{\bar{K}}_{j=1}\frac{u_k-\bar{u}_j-i}{u_k-\bar{u}_j+i} \frac{1-g^2/x^+_k \bar{x}^-_j}{1-g^2/x^-_k \bar{x}^+_j}\sigma(u_k,\bar{u}_j)\prod^K_{j\neq k}\frac{1-g^2/x^+_k x^-_j}{1-g^2/x^-_k x^+_j}\sigma(u_k,u_j)\, , \\ \label{BE2}
&&\left(\frac{\bar{x}^+_k}{\bar{x}^-_k}\right)^L=\prod^{K}_{j=1}\frac{\bar{u}_k-u_j-i}{\bar{u}_k-u_j+i} \frac{1-g^2/\bar{x}^+_k x^-_j}{1-g^2/\bar{x}^-_k x^+_j} \sigma(\bar{u}_k,u_j)\prod^{\bar{K}}_{j\neq k} \frac{1-g^2/\bar{x}^+_k \bar{x}^-_j}{1-g^2/\bar{x}^-_k \bar{x}^+_j} \sigma(\bar{u}_k,\bar{u}_j)\,,
\eeqa
where the dressing phase $\sigma(u, v)$ is the same as in the $\mathcal{N}=4$ theory~\cite{Gromov:2008qe}.
Following the analysis in Section \ref{sec:hah}, we introduce $\rho_{\pm}(u)=\rho(u) \pm \bar{\rho}(u)$. It may be easily verified using \eqref{BE1}-\eqref{BE2} that the density $\rho_-(u)$ is already exact to leading order at weak coupling and coincides with the one given in~\eqref{LO-densities}. The density $\rho_+(u)$, on the other hand, satisfies the same integral equation as the density \eqref{vactohole} in the single-polarisation sector of $\N=4$ and may be found by solving \eqref{Eqs-uform}. We thus have all ingredients in place to reconstruct the counting function in this case as well.

The little counting functions $y(u)$ is defined similarly as in Section \ref{sec:BYol},
\beq
Y(u) = q\, y(u)\, ,
\eeq
with
\beq
q = (-1)^{K}\sqrt{(-1)^{K+\bar{K}}e^{iP}}\, , \qquad e^{iP} = \prod_{j=1}^{K}\frac{x^{+}_j}{x^{-}_{j}}\prod_{j=1}^{\bar{K}}\frac{\bar{x}^{+}_j}{\bar{x}^{-}_{j}}\, .
\eeq
The corresponding function for anti-excitations is defined analogously. We thus have to compute
\beq
\begin{aligned}
y(u) = (-1)^{K_{h}}\left({-\frac{x^-}{x^+}}\right)^L \prod^{\bar{K}}_{j=1} &\frac{1+iu-i\bar{u}_j}{1-iu+i\bar{u}_j} \frac{1-g^2/x^+ \bar{x}^{-}_j}{1-g^2/x^- \bar{x}^+_j}\sigma(u,\bar{u}_j)\sqrt{-\frac{\bar{x}_{j}^{-}}{\bar{x}_{j}^{+}}} \\
\qquad \qquad  \qquad &\times \prod^{K}_{j=1}\frac{1-g^2/x^+ x^{-}_j}{1-g^2/x^- x^+_j}\sigma(u,u_j)\sqrt{-\frac{x_{j}^{-}}{x_{j}^{+}}}\, ,
\end{aligned}
\eeq
together with $\bar{y}(u)$ obtained by swapping fundamental and anti-fundamental degrees of freedom. It is now when the relation with $\N=4$ SYM becomes so helpful! We notice that the product $y(u)\bar{y}(u)$ is nothing else but the little counting function for $\mathcal{N}=4$ theory
\beq
y(u)\bar{y}(u) = y_{\mathcal{N}=4}(u)\,.
\eeq
The quantum number in $y_{\mathcal{N}=4}$ have to be doubled, i.e. $L \to 2L$ and $S \to 2S$. Of course, this translates into equality between the density $2i\pi\rho_{+}(u) = \partial_{u}\log{(y(u)\bar{y}(u))}$ and $\rho_{\mathcal{N}=4}(u)$. We have already pointed out this relation in~(\ref{rho46}). The formula \eqref{yN=4} then yields
\beq\label{ybary}
y(u)\bar{y}(u) = e^{2ip_{\mathcal{N}=4}(u)\log{(2S)}}T_{\mathcal{N}=4}(u)\prod_{j=1}^{K_{h}}S_{\mathcal{N}=4}(u, u_{h, j})\prod_{j=1}^{\bar{K}_{h}}S_{\mathcal{N}=4}(u, \bar{u}_{h, j})\, .
\eeq 
The ratio of the $y$-functions is given exactly by its weak coupling expression because it only depends on $\rho_-(u)$. We find
\beq\label{yoverbary}
y(u)/\bar{y}(u) = (-1)^{K_{h}-\bar{K}_{h}}\prod_{j=1}^{\bar{K}}\frac{1+iu-i\bar{u}_{j}}{1-iu+i\bar{u}_{j}}\prod_{j=1}^{K}\frac{1-iu+iu_{j}}{1+iu-iu_{j}}\, .
\eeq
This ratio is directly related to the scattering phase of spinons on top of the antiferromagnetic state in Heisenberg spin chain, as briefly mentioned in Section~\ref{sec:BYol}. Exploiting this analogy, we write it as
\beq
y(u)/\bar{y}(u) = \prod_{j=1}^{K_{h}}S_{\mathfrak{su}(2)}(u, u_{h, j})\prod_{j=1}^{\bar{K}_{h}}S^{-1}_{\mathfrak{su}(2)}(u, \bar{u}_{h, j})\, ,
\eeq
with the $\mathfrak{su}(2)$ spinons S-matrix
\beq\label{su2S}
S_{\mathfrak{su}(2)}(u, v) = \frac{\Gamma \big(\tfrac{iu-iv}{2}\big)}{\Gamma \big(\tfrac{iv-iu}{2}\big)}\frac{\Gamma \big(\tfrac{1}{2}-\tfrac{iu-iv}{2}\big)}{\Gamma \big(\tfrac{1}{2}+\tfrac{iu-iv}{2}\big)}\, .
\eeq
The relations~(\ref{ybary}) and~(\ref{yoverbary}) determine the $y$-functions up to a sign that is fixed by the analysis performed in Section~\ref{sec:BYol}. This leads to
\beq
S_{\mathcal{N}=6}(u, v) = -\sqrt{S_{\mathcal{N}=4}(u, v)S_{\mathfrak{su(2)}}(u, v)}\, , \qquad \bar{S}_{\mathcal{N}=6}(u, v) = \sqrt{S_{\mathcal{N}=4}(u, v)S^{-1}_{\mathfrak{su(2)}}(u, v)}\, ,
\eeq
or equivalently to~(\ref{prodS}) and~(\ref{ratioS}). For the momentum $p_{\mathcal{N}=6}(u)$ we find the relation~(\ref{EpN=46}) and for the transmission amplitude $T_{\mathcal{N}=6}(u)$ we get the expression~(\ref{T4T6}).

Finally, we recall that the isotopic roots modify the expressions for $y(u), \bar{y}(u)$ by simple phases, as shown at weak coupling in Section~\ref{su4sym}. These phase factors, however, do not receive radiative corrections. Hence the full Bethe-Yang equations~(\ref{ABAN=6}) take a relatively simple form.

\section{Checking unitarity and crossing}\label{cucrss}

In this appendix we shall derive the unitarity and crossing equations for the S-matrix of holes.

To keep the amount of algebra to a minimum, it is convenient to introduce the special functions
\beq
\begin{aligned}\label{original-f-functions}
&f_{1}(u, v) = -\sum \limits_{n\geqslant 1}2(2n-1)\tilde{\kappa}^{u}_{2n-1}\gamma^{v}_{2n-1}-\sum \limits_{n\geqslant 1}2(2n)\tilde{\kappa}^{u}_{2n}\gamma^{v}_{2n}\, , \\
&f_{2}(u, v) = -\sum \limits_{n\geqslant 1}2(2n-1)\kappa^{u}_{2n-1}\tilde{\gamma}^{v}_{2n-1}-\sum \limits_{n\geqslant 1}2(2n)\kappa^{u}_{2n}\tilde{\gamma}^{v}_{2n}\, , \\
&f_{3}(u, v) = +\sum \limits_{n\geqslant 1}2(2n-1)\tilde{\kappa}^{u}_{2n-1}\tilde{\gamma}^{v}_{2n-1}-\sum \limits_{n\geqslant 1}2(2n)\tilde{\kappa}^{u}_{2n}\tilde{\gamma}^{v}_{2n}\, , \\
&f_{4}(u, v) = +\sum \limits_{n\geqslant 1}2(2n-1)\kappa^{u}_{2n-1}\gamma^{v}_{2n-1}-\sum \limits_{n\geqslant 1}2(2n)\kappa^{u}_{2n}\gamma^{v}_{2n}\, , \\
\end{aligned}
\eeq
which are formed out of the Neumann coefficients $\gamma_{n}, \tilde{\gamma}_{n}$ and source terms $\kappa_{n}, \tilde{\kappa}_{n}$ entering the system of equations~(\ref{Eqs-system}). The reason to consider these particular combinations lies in the fact that they are building blocks for the S-matrix of holes in both real and mirror kinematics. Taking into account the expressions for the source terms~(\ref{source-terms}) and the representations \eqref{Neumann}, one easily finds
\beq
\begin{aligned}\label{f-functions-App1}
&f_{1}(u, v) = \int_{0}^{\infty}\frac{dt}{t}\frac{\sin{(ut)}e^{t/2}\gamma^{v}(2gt)}{e^{t}-1}\, ,\\
&f_{2}(u, v) = \int_{0}^{\infty}\frac{dt}{t}\frac{(\cos{(ut)}e^{t/2}-J_{0}(2gt))\tilde{\gamma}^{v}(2gt)}{e^{t}-1}\, ,\\
&f_{3}(u, v) = \int_{0}^{\infty}\frac{dt}{t}\frac{\sin{(ut)}e^{t/2}\tilde{\gamma}^{v}(-2gt)}{e^{t}-1}\, ,\\
&f_{4}(u, v) = \int_{0}^{\infty}\frac{dt}{t}\frac{(\cos{(ut)}e^{t/2}-J_{0}(2gt))\gamma^{v}(-2gt)}{e^{t}-1}\, .\\
\end{aligned}
\eeq
We immediately recognise that the two first entries, i.e.,  $f_{1}(u, v)$ and $f_{2}(u, v)$, are the two functions that control the core part of the S-matrix for two holes carrying rapidities $u$ and $v$. We shall see later on that the two last functions play a similar role for the mirror S-matrix. 

We will now discuss some of the properties of the functions~(\ref{f-functions-App1}) and relate them to unitarity and crossing of the S-matrix.

\subsection{Unitarity}\label{Unitarity}

To prove the unitarity of the S-matrix for holes we only have to show that
\beq\label{f12}
f_{1}(u, v)  = f_{2}(v, u)\, .
\eeq
This relation is not obvious at first glance. This is because the original definition of the functions $f_{1}(u, v)$ and $f_{2}(u, v)$, see Eq.~(\ref{original-f-functions}), does not treat symmetrically the two rapidities. To derive \eqref{f12} we need to swap them. This can be done by using the exchange relations
\beq
\begin{aligned}
&\sum \limits_{n\geqslant 1}2(2n-1)\kappa^{v}_{2n-1}\gamma^{u}_{2n-1}+\sum \limits_{n\geqslant 1}2(2n)\kappa^{u}_{2n}\gamma^{v}_{2n} = \int_{0}^{\infty}\frac{dt}{t}\frac{\gamma^{u}_{-}(2gt)\gamma^{v}_{-}(2gt)+\gamma^{v}_{+}(2gt)\gamma^{u}_{+}(2gt)}{1-e^{-t}}\, , \\
&\sum \limits_{n\geqslant 1}2(2n-1)\tilde{\kappa}^{v}_{2n-1}\tilde{\gamma}^{u}_{2n-1}+\sum \limits_{n\geqslant 1}2(2n)\tilde{\kappa}^{u}_{2n}\tilde{\gamma}^{v}_{2n} = \int_{0}^{\infty}\frac{dt}{t}\frac{\tilde{\gamma}^{u}_{-}(2gt)\tilde{\gamma}^{v}_{-}(2gt)+\tilde{\gamma}^{v}_{+}(2gt)\tilde{\gamma}^{u}_{+}(2gt)}{1-e^{-t}}\, , \\
&\sum \limits_{n\geqslant 1}2(2n-1)\tilde{\kappa}^{v}_{2n-1}\gamma^{u}_{2n-1}-\sum \limits_{n\geqslant 1}2(2n)\kappa^{u}_{2n}\tilde{\gamma}^{v}_{2n} = \int_{0}^{\infty}\frac{dt}{t}\frac{\gamma^{u}_{-}(2gt)\tilde{\gamma}^{v}_{-}(2gt)-\tilde{\gamma}^{v}_{+}(2gt)\gamma^{u}_{+}(2gt)}{1-e^{-t}}\, , \\
&\sum \limits_{n\geqslant 1}2(2n-1)\kappa^{v}_{2n-1}\tilde{\gamma}^{u}_{2n-1}-\sum \limits_{n\geqslant 1}2(2n)\tilde{\kappa}^{u}_{2n}\gamma^{v}_{2n} = \int_{0}^{\infty}\frac{dt}{t}\frac{\tilde{\gamma}^{u}_{-}(2gt)\gamma^{v}_{-}(2gt)-\gamma^{v}_{+}(2gt)\tilde{\gamma}^{u}_{+}(2gt)}{1-e^{-t}}\, . \\
\end{aligned}
\eeq
They follow from the universality of the kernel of the equations \eqref{Eqs-system} and were proposed in~\cite{Basso:2008tx} in a related context. Observing that the right-hand side of these equations are either symmetric or related to one another under exchange of the two rapidities and recalling the original definitions in \eqref{original-f-functions}, one easily arrives at~(\ref{f12}). As a by-product we also derive that
\beq\label{f34}
f_{3}(u, v)  = f_{3}(v, u)\, , \qquad  f_{4}(u, v)  = f_{4}(v, u)\, .
\eeq
We stress that all these relations are valid at any coupling. This completes the proof of unitarity.

\subsection{Crossing}\label{Crossing}

The S-matrix for holes in $\mathcal{N}=4$ and $\mathcal{N}=6$ theories has a well-defined behaviour under the crossing transformation. These were given in \eqref{cross-rel} and \eqref{cross-rel-bis}. In this appendix we will provide their derivation.

Let us start with the $\mathcal{N}=4$ theory. As alluded to before, it is convenient to think of the crossing map as a sequence of two mirror transformations.  Hence, we shall first derive the expression for the scattering of two holes, of which one is continued to the mirror sheet while the other is kept physical. This is what we refer to as the mirror S-matrix, with a slight abuse of terminology. The mirror S-matrix is then nothing else than
\beq\label{defmirrS}
S^{\star}_{\mathcal{N}=4}(u, v) \equiv S_{\mathcal{N}=4}(u^{\gamma}, v)\, ,
\eeq
where $\gamma$ maps the rapidity $u$ to $u^{\gamma} = u+i$ after continuation through the cut $\mathcal{C}_{+}$ connecting the points $-2g+i/2$ and $2g+i/2$ in the upper-half $u$-plane, \textit{cf.} Figure \ref{fig:cpweak} and Figure \ref{fig:cpstrong}.

The main quantities that we need to continue are the functions $f_{1, 2}(u, v)$ entering the expression for the scattering phase \eqref{Sgeneral}. They have been defined explicitly in \eqref{f-functions}. Under a crossing transformation these functions map to each other. This is, however, not the case under a single mirror rotation. To make the algebra under mirror rotations complete, we should first enlarge this set of functions. This is done with the help of the functions $f_{3, 4}(u, v)$ previously defined in \eqref{original-f-functions}. They satisfy the relations \eqref{f34} together with
\beq
f_{3}(-u, -v)  = f_{3}(u, v)\, , \qquad f_{4}(-u, -v)  = f_{4}(u, v)\, ,
\eeq
which immediately follow from the parity property of the functions $\gamma^{v}, \tilde{\gamma}^{v}$.

We shall now prove that under a mirror rotation we have
\beq
\begin{aligned}\label{f-functions-App}
&f_{1}(u^{\gamma}, v) = -if_{4}(u, v) -i \int_{0}^{\infty}\frac{dt}{t}\frac{(e^{-iut+t/2}-J_{0}(2gt))(\cos{(vt)}e^{t/2}-J_{0}(2gt))}{e^{t}-1}\, ,\\
&f_{2}(u^{\gamma}, v) = -if_{3}(u, v)-\int_{0}^{\infty}\frac{dt}{t}\frac{(e^{-iut+t/2}-J_{0}(2gt))\sin{(vt)}e^{t/2}}{e^{t}-1}\, ,\\
&f_{3}(u^{\gamma}, v) = -if_{2}(u, v)-i\int_{0}^{\infty}\frac{dt}{t}\frac{(e^{iut-t/2}-J_{0}(2gt))\sin{(vt)}e^{t/2}}{e^{t}-1}\, ,\\
&f_{4}(u^{\gamma}, v) = -if_{1}(u, v)-\int_{0}^{\infty}\frac{dt}{t}\frac{(e^{iut-t/2}-J_{0}(2gt))(\cos{(vt)}e^{t/2}-J_{0}(2gt))}{e^{t}-1}\, .\\
\end{aligned}
\eeq
We illustrate the underlying technique by deriving the first and the fourth relation. The $\gamma$-transformation will be implemented in two steps. The first step is to shift the rapidity $u$ by $i/2-i0^+$. This way we move right under the first cut in the upper-half $u$-plane. The second step is to cross this cut and shift again by $i/2$ in order to reach $u^\gamma$. This will allow us to relate $f_1(u^\gamma,v)$ and $f_4(u,v)$. In the same vein one can perform the inverse rotation $u^{-\gamma} = u-i$ by shifting in two steps into the lower half-plane. This down shift, on the other hand, will allow us to derive the last relation in~(\ref{f-functions-App}). We start by performing the first shift
\beq\label{fupm1}
f_{1}(u^{\pm}, v) =  \int_{0}^{\infty}\frac{dt}{t}\frac{\sin{(ut)}\gamma^{v}(2gt)}{e^{t}-1} \pm \frac{i}{2}\int_0^\infty\frac{dt}{t}e^{\mp iut}\gamma^{v}(2gt)\, ,
\eeq
where $u^{\pm} = u\mp i0^+ \pm \ft{i}{2}$. Here, we plugged the identity
\beq
\sin(u^{\pm}t) = \sin{(ut)}e^{-t/2}\pm ie^{\mp iut}\textrm{sinh}{(t/2)}
\eeq
into the defining integral representation~(\ref{f-functions-App1}). The next step is to cross the cut at $-2g< \textrm{Re} (u) < 2g$ in the shifted rapidity. It comes from the second integral in~(\ref{fupm1}), which converges for $\textrm{Im}(u)>0$ (or $\textrm{Im}(u)<0$) if we perform the $+$ (or $-$) shift. We move through this cut by using the relation
\beq
\begin{aligned}
\pm \frac{i}{2}\int_0^\infty\frac{dt}{t}e^{\mp iut}\gamma^{v}(2gt) = &\mp \frac{i}{2}\int_0^\infty\frac{dt}{t}e^{\pm iut}\gamma^{v}(-2gt) \\
&+\int_{0}^{\infty}\frac{dt}{t}\sin{(ut)}\gamma^{v}_{-}(2gt) \pm i\int_{0}^{\infty}\frac{dt}{t}\cos{(ut)}\gamma^{v}_{+}(2gt)\, ,
\end{aligned}
\eeq
that leads to
\beq
\begin{aligned}\label{fupm2}
f_{1}(u^{\pm}, v) =& \mp \frac{i}{2}\int_0^\infty\frac{dt}{t}e^{\pm iut}\gamma^{v}(-2gt) +\int_{0}^{\infty}\frac{dt}{t}\sin{(ut)}\bigg[\frac{\gamma^{v}_{-}(2gt)}{1-e^{-t}} + \frac{\gamma^{v}_{+}(2gt)}{e^{t}-1}\bigg] \\
&\pm i\int_{0}^{\infty}\frac{dt}{t}(\cos{(ut)}-J_{0}(2gt))\gamma^{v}_{+}(2gt)\, .
\end{aligned}
\eeq
Note that, for the sake of convenience, we added to the right-hand side of~(\ref{fupm2}) the term
\beq
\mp i\int_{0}^{\infty}\frac{dt}{t}J_{0}(2gt)\gamma^{v}_{+}(2gt) = 0\,.
\eeq
It vanishes due to orthogonality of the Bessel functions~(\ref{orthoBessel}). Next, we utilise the first two equations in~(\ref{Eqs-uform}) to the two last terms in~(\ref{fupm2}). This way we find
\beq
\begin{aligned}\label{fupm3}
f_{1}(u^{\pm}, v) =& \mp i\int_{0}^{\infty}\frac{dt}{t}(\cos{((u\mp \ft{i}{2})t)}e^{t/2}-J_{0}(2gt))\frac{\gamma^{v}(-2gt)}{e^{t}-1} \\
&\mp i\int_{0}^{\infty}\frac{dt}{t}\frac{(e^{\mp iut}-J_{0}(2gt))(\cos{(vt)}e^{t/2}-J_{0}(2gt))}{e^{t}-1}\, .
\end{aligned}
\eeq
With this form at hand, it is straightforward to perform the second half of the mirror rotation by shifting $u$ by $\pm i/2$ once more
\beq\label{fupm4}
f_{1}(u^{\pm \gamma}, v) = \mp if_{4}(u, v) \mp i\int_{0}^{\infty}\frac{dt}{t}\frac{(e^{\mp iut +t/2}-J_{0}(2gt))(\cos{(vt)}e^{t/2}-J_{0}(2gt))}{e^{t}-1}\,.
\eeq
The derivation is now complete since the equalities in~(\ref{fupm4}) are equivalent to the first and last relations in~(\ref{f-functions-App}). In a similar manner we can prove the two  remaining relations in~(\ref{f-functions-App}).

Equipped with formulae~(\ref{f-functions-App}) we can derive the general expression for the mirror S-matrix~(\ref{defmirrS}). Starting from the real S-matrix~(\ref{Sgeneral})-(\ref{Shat}) we arrive at
\beq\label{mirrorSgen}
S^{\star}_{\mathcal{N}=4}(u, v) = \frac{u-v}{u-v+i}\hat{S}^{\star}_{\mathcal{N}=4}(u, v)\exp{(2f_{3}(u, v)-2f_{4}(u, v))}\, ,
\eeq
with
\beqa \nn
&&\log{\hat{S}^{\star}_{\mathcal{N}=4}(u, v)} = -2\int_{0}^{\infty}\frac{dt}{t}\frac{\cos{(u-v)t}-e^{t/2}J_{0}(2gt)(\cos{(ut)}+\cos{(vt)})+e^{t}J_{0}(2gt)^2}{e^t-1} \,. \label{Sstarhat}
\eeqa
By performing one more mirror rotation one easily arrives at the crossing relation~(\ref{cross-rel}). This follows again from~(\ref{f-functions-App}) together with some simple algebra required to continue the factor $\hat{S}^{\star}_{\mathcal{N}=4}(u, v)$ through the cut at $\textrm{Im}(u) = i/2$.

Finally, it is interesting to note that the mirror S-matrix~(\ref{mirrorSgen}) is almost symmetric under the exchange of $u$ and $v$. This is so because both $\hat{S}^{\star}_{\mathcal{N}=4}(u, v)$ and $f_{3,4}(u, v)$ have this property, see Eqs.~(\ref{f34}, \ref{Sstarhat}). The only term that breaks this symmetry is the rational prefactor in~(\ref{mirrorSgen}). We shall now demonstrate that these features lead to the mirror symmetry of the S-matrix mentioned in Section~\ref{O6Smat}, see~(\ref{mirrS}) therein. To start with we write
\beq\label{wdef}
S_{\mathcal{N}=4}(u^{\gamma}, v) = \frac{u-v}{u-v+i}w(u, v)\, ,
\eeq
where $w(u, v) = w(v, u)$ is the aforementioned symmetric factor. From the crossing relation of the S-matrix~(\ref{cross-rel}), we deduce that
\beq
w(u^{\gamma}, v) = \frac{(u^{\gamma}-v+i)(u-v)}{(u^{\gamma}-v)(u-v+2i)}S_{\mathcal{N}=4}(v, u) = \frac{u-v}{u-v+i}S_{\mathcal{N}=4}(v, u)\, ,
\eeq
where in the last equality we used that $u^{\gamma} = u+i$. Taking into account~(\ref{wdef}) and the permutation invariance of $w(u, v)$ we find the sought-after relation
\beq
S_{\mathcal{N}=4}(u^{\gamma}, v^{\gamma}) =  \frac{u-v^{\gamma}}{u-v^{\gamma}+i}w(v^{\gamma}, u) = S_{\mathcal{N}=4}(u, v)\, .
\eeq

\section{The real and mirror S-matrices at weak coupling}\label{wce}

In this appendix we explain how to perturbatively compute the scalar S-matrix of the $\mathcal{N}=4$ theory at weak coupling.

When looking at the general expression~(\ref{Sgeneral})-(\ref{f-functions}) for the S-matrix we see that it involves the relatively simple-to-expand factor $\hat{S}_{\mathcal{N}=4}(u, v)$ and the functions $f_{1,2}(u, v)$, the expansion of which requires expanding $\gamma^v(2gt)$ and $\tilde{\gamma}^v(2gt)$ first. Both can be computed at weak coupling by following the same procedure, which in essence boils down to Taylor expanding and applying the integral
\beq\label{usefulint}
\int_{0}^{\infty}\frac{dt}{t}\frac{e^{iwt}-1-iwt}{e^{t}-1} = \log{\Gamma(1-iw)}+iw\,\psi(1)\, ,
\eeq
or one of its derivatives. Let us first describe how to expand $\hat{S}_{\mathcal{N}=4}(u, v)$. We first notice that it can be written as
\beq
\begin{aligned}\label{expnShat}
\log{(-\hat{S}_{\mathcal{N}=4})} &= 2i\int_{0}^{\infty}\frac{dt}{t}\frac{\sin{(ut)}e^{t/2} - \sin{(vt)}e^{t/2}- \sin{((u-v)t)}}{e^{t}-1} \\
&\, \, +2i\int_{0}^{\infty}\frac{dt}{t}(J_{0}(2gt)-1)\frac{\sin{(ut)}e^{t/2} - \sin{(vt)}e^{t/2}}{e^{t}-1}\, .
\end{aligned}
\eeq
The first term is coupling independent and straightforward to evaluate using~(\ref{usefulint}). It gives the expression~(\ref{loSmat}) for the one-loop scalar S-matrix. The second term in~(\ref{expnShat}) is subleading at weak coupling, since $J_{0}(2gt) = 1-g^2t^2 + O(g^4)$. Computing the resulting integral by means of the generating formula~(\ref{usefulint}) we see that the second term in~(\ref{expnShat}) will produce derivatives of ascending order of the $\psi$ function with arguments $\ft{1}{2}\pm iu$ or $\ft{1}{2}\pm iv$.

We can proceed similarly with the functions $f_{1,2}(u, v)$. Recalling that the $\gamma^v(2gt)$ and $\tilde{\gamma}^v(2gt)$ functions admit an expansion over the Bessel functions~(\ref{Neumann}) whose coefficients $\gamma^{v}_{n},\tilde{\gamma}^v_n$ may be found by solving the system of linear equations (\ref{Eqs-system}) with source terms~(\ref{source-terms}), we find to leading order
\beq
\gamma_{1}^{v} = -\frac{g}{2}(\psi(\ft{1}{2}+iv)-\psi(\ft{1}{2}-iv)-2\psi(1)) + O(g^3) 
\eeq
and
\beq
\tilde{\gamma}_{1}^{v} = -\frac{\pi g}{2}\textrm{tanh}(\pi v) + O(g^3)\, .
\eeq
All higher coefficients are more suppressed with the coupling and thus appear irrelevant for the discussion here. Now to evaluate $f_{1,2}(u, v)$ we plug the functions
\beq
\gamma^{v}(2gt) = 2g\gamma_1 t + O(g^2)\, , \qquad \tilde{\gamma}^{v}(2gt) = 2g\tilde{\gamma}_1 t + O(g^2)\, ,
\eeq
in the integrals~(\ref{f-functions}) and differentiate formula~(\ref{usefulint}) to find
\beq\label{SN4pert}
S_{\mathcal{N}=4 }(u, v) = \frac{\Gamma(\ft{1}{2}-iu)\Gamma(\ft{1}{2}+iv)\Gamma(iu -iv)}{\Gamma(\ft{1}{2}+iu)\Gamma(\ft{1}{2}-iv)\Gamma(iv -iu)}\bigg[1+ \alpha(u, v) g^2 + O(g^4)\bigg]\, ,
\eeq
where
\beq
\alpha(u,v)=-2 H_{-\frac{1}{2}+i u}
   H_{-\frac{1}{2}-i v}+2
   H_{-\frac{1}{2}-i u}
   H_{-\frac{1}{2}+i
   v}+H_{-\frac{1}{2}-i
   u}^{(2)}-H_{-\frac{1}{2}+i
   u}^{(2)}-H_{-\frac{1}{2}-i
   v}^{(2)}+H_{-\frac{1}{2}+i
   v}^{(2)}\,.
\eeq
The function $H(z)=H_z$ and $H^{(2)}(z) = H^{(2)}_z = \partial_{z}^2H(z)$ are the harmonic sum of the first and second order, respectively. Subleading corrections to~(\ref{SN4pert}) are equally easily to construct and may be computed using a simple algorithm presented in~\cite{BSV}.

Let us also calculate the first few terms in the weak-coupling expansion of the mirror S-matrix~(\ref{mirrorSgen}). We have $f_{3, 4}(u,v)\sim g^2$ and thus
\beq
S^{\star}_{\mathcal{N}=4}(u, v) \simeq \frac{u-v}{u-v+i}\hat{S}^{\star}_{\mathcal{N}=4}(u, v)\, .
\eeq
Now to evaluate $\hat{S}^{\star}_{\mathcal{N}=4}(u, v)$ we arrange the expression~\eqref{Sstarhat} in three groups 
\beq
\begin{aligned}\label{wcsmirr}
\log{\hat{S}^{\star}_{\mathcal{N}=4}(u, v)} = &-2\int_{0}^{\infty}\frac{dt}{t}\bigg[J_{0}(2gt)^2-\frac{t}{e^{t}-1}\bigg]\\
&-2\int_{0}^{\infty}\frac{dt}{t}\frac{\cos{(u-v)t}-e^{t/2}(\cos{(ut)}+\cos{(vt)})+t+1}{e^{t}-1}\\
&+2\int_{0}^{\infty}\frac{dt}{t}(J_{0}(2gt)-1)\frac{e^{t/2}(\cos{(ut)}+\cos{(vt)})-J_{0}(2gt)-1}{e^{t}-1}\, .
\end{aligned}
\eeq
The last line is clearly suppressed at leading order. The second integral may be done explicitly using~(\ref{usefulint})
\beq
\begin{aligned}
&\int_{0}^{\infty}\frac{dt}{t}\frac{\cos{(u-v)t}-e^{t/2}(\cos{(ut)}+\cos{(vt)})+t+1}{e^{t}-1}\\
&\qquad \qquad \qquad  \qquad \qquad = \frac{1}{2}\log{\frac{(u-v)\cosh{(\pi u)}\cosh{(\pi v)}}{\pi \sinh{(\pi(u-v))}}}-\psi(1)\,.
\end{aligned}
\eeq
It thus remains to compute the first term. It is not allowed to Taylor expand this expression. Instead we have the exact result
\beq
\int_{0}^{\infty}\frac{dt}{t}\left[J_{0}(2gt)^2-\frac{t}{e^{t}-1}\right] = \int_{0}^{\infty}\frac{dt}{t}\left[J_{0}(2gt)-\frac{t}{e^{t}-1}\right]= -\log{g}+\psi(1)\, .
\eeq
Combined together these formula yield
\beq
\hat{S}^{\star}_{\mathcal{N}=4}(u, v) = \frac{\pi g^2 \sinh{(\pi(u-v))}}{(u-v)\cosh{(\pi u)}\cosh{(\pi v)}}+ O(g^4)\,,
\eeq
and therefore we have
\beq
S_{\mathcal{N}=4}(u^{\gamma}, v) = \frac{\pi g^2}{u-v+i}\frac{\sinh{(\pi(u-v))}}{\cosh{(\pi u)}\cosh{(\pi v)}} + O(g^4)\, .
\eeq
It is not difficult to compute the subleading corrections. The expressions for $f_{3,4}(u,v)$ may also be found using~(\ref{usefulint}) after some algebra
\beq
S_{\mathcal{N}=4}(u^{\gamma}, v) =  \frac{\pi g^2}{u-v+i}\frac{\sinh{\pi(u-v)}}{\cosh{(\pi u)}\cosh{(\pi v)}}\bigg[1+\beta(u, v)g^2+O(g^4)\bigg]\, ,
\eeq
where the function $\beta(u,v)$ is given by
\beq
\beta(u,v) = \frac{2\pi^2}{3}-\frac{\pi^2}{\cosh^2{(\pi u)}}-\frac{\pi^2}{\cosh^2{(\pi v)}}-2H_{iu-\ft{1}{2}}H_{iv-\ft{1}{2}}-2H_{-iu-\ft{1}{2}}H_{-iv-\ft{1}{2}}\,.
\eeq

\section{Subleading density and the spin-chain momentum}\label{sec:sub}

In this section we will first show how to compute the subleading all-loop density in the semiclassical regime from the Baxter equation and later use this result to evaluate the momentum. 
\subsection{Subleading density from Baxter equation}
We propose the following all-loop Baxter system corresponding to the Bethe equations for single-polarisation excitations, \textit{cf.} \eqref{BE1}-\eqref{BE2},
\beqa \label{t1}
&&\Delta_{+}(u+\tfrac{i}{2}) \bar{Q}(u+i)-\Delta_-(u-\tfrac{i}{2}) \bar{Q}(u-i)=t_1(u) Q(u)\,,\\ \label{t2}
&&\Delta_{+}(u+\tfrac{i}{2}) Q(u+i)-\Delta_-(u-\tfrac{i}{2})  Q(u-i)=t_2(u) \bar{Q}(u)\,.
\eeqa
The functions $\Delta_{\pm}(u)$ are defined by
\beqa \label{Dp}
&&\log \Delta_+(u)=L \log x+\sum^{\infty}_{n=1} \Gamma_n \left(\frac{ig}{x}\right)^n\,,\\ \label{Dm}
&&\log \Delta_-(u)=L \log x+\sum^{\infty}_{n=1} \Gamma^*_n \left(\frac{g}{ix}\right)^n\,.
\eeqa
The coefficients $\Gamma_n, \Gamma^*_n$ are moments of the distribution of magnons and anti-magnons. For real distribution of roots they are complex conjugate to each other. They can be related to the similar coefficients appearing in the Baxter equation for the $\mathfrak{sl}(2)$ sector of $\N=4$ SYM written in~\cite{Basso:2011rc}. In the current analysis only the first coefficients, $\Gamma_1$ and $\Gamma^*_1$, are relevant. They have a simple interpretation since they relate to the anomalous part of the total energy and momentum
\beq\label{App-ptot}
\delta\Delta = g \left(\Gamma_1 + \Gamma^*_1 \right) \, ,  \qquad p_{\textrm{tot}} = \sum^{K_h+\bar{K}_h}_{j=1}u_{h, j}+ig\left(\Gamma_1 - \Gamma^*_1\right)\, ,
\eeq
At large spin, the momenta are additive $p_{\textrm{tot}} = \sum^{K_h+\bar{K}_h}_{j=1}p(u_{h, j})$. The equivalent of \eqref{pholeN4} in $\mathcal{N}=6$ theory is $p(u) = u+2g\tilde{\gamma}^u$.

We will only consider polynomial solutions to the above system of equations with degrees $K$ and $\bar{K}$, respectively. The transfer matrices $t_{1,2}(u)$ have polynomial and non-polynomial parts. Consistency requires that the polynomial parts have degrees $\deg p_1 \equiv K_h = L-1+\bar{K}-K$ and $\deg p_2 \equiv \bar{K}_h= L-1+K-\bar{K}$. The following representation may be derived for the polynomial parts of the transfer matrices
\beqa
p_1(u)=q_1 \prod^{K_h}_{j=1} (u-u^{h}_j)\,,\qquad p_2(u)=\bar{q}_1 \prod^{\bar{K}_h}_{j=1} (u-\bar{u}^{h}_j)\,,
\eeqa
with
\beq
q_1=i(L+2\bar{K})+ig(\Gamma_1+\Gamma^*_1)\,, \qquad \bar{q}_1=i(L+2K)+ig(\Gamma_1+\Gamma^*_1)\,.
\eeq
We would like to point out a slight abuse of the notation in the above formulae due to the fact that the roots of transfer matrices may be identified with holes and anti-holes \textit{only when} the non-polynomial part does not contribute. Fortunately, this is the case in what follows. We notice that at large spin $q_1\sim \bar{q}_1\sim K\sim \bar{K} \sim S$. The polynomial parts of the transfer matrices are large $p_{1, 2} \sim S$ and dominate at large spin. This is similar to what happens in the  $\mathcal{N}=4$ theory except that there the scaling is stronger $\sim S^2$ and caused by the presence of the so-called large holes~\cite{Belitsky:2006en}. There are no large holes in the present case, but the analysis of~\cite{Belitsky:2006en} would still apply thanks to the scaling of the transfer matrices. It would lead to the same results as the analysis based on the density approach presented in Section~\ref{sec:wca}.

Taking the large $K, \bar{K}$ limit is straightforward if we use \eqref{t1}-\eqref{t2} to decouple $Q(u)$ and $\bar{Q}(u)$ from each other. For instance, the equation for $Q(u)$ reads
\beqa\label{BQonly}
&&\frac{\Delta_+(u+\tfrac{i}{2}) \Delta_+(u+\tfrac{3i}{2})}{t_2(u+i)}Q(u+2i)+\frac{\Delta_-(u-\tfrac{i}{2}) \Delta_-(u-\tfrac{3i}{2})}{t_2(u-i)}Q(u-2i)=\\
&&\qquad \qquad \left(t_1(u)+\frac{\Delta_+(u+\tfrac{i}{2})\Delta_-(u+\tfrac{i}{2})}{t_2(u+i)} +\frac{\Delta_+(u-\tfrac{i}{2})\Delta_-(u-\tfrac{i}{2})}{t_2(u-i)}\right) Q(u)\,.
\eeqa
Written in this form the above equation may be subjected to the semiclassical analysis of~\cite{Korchemsky:1995be}. We look for the WKB-like solution
\beq
Q(u) = e^{K\Phi(\tilde{u})}\, ,
\eeq
with $1/K \simeq 1/S$ playing the role of a Planck constant. For the sake of convenience, we also introduced the rescaled rapidity $\ub= u/K \simeq u/S$. The function $\Phi(\tilde{u})$ is defined up to an irrelevant constant and assumed to admit the following expansion at large spin
\beq\label{Phiexpand}
\Phi(\tilde{u}) = \Phi_0(\tilde{u}) + \Phi_1(\tilde{u}) /K + \ldots\,.
\eeq
We would like to stress that the hole rapidities are kept to be of order $O(1)$ when we send $S \to \infty$. This is different from the analysis in~\cite{Korchemsky:1995be},  where the main focus was on higher trajectories in the spectrum of large spin operators. Accordingly, we do not rescale the hole rapidities and they enter the analysis at subleading orders. Bearing that in mind, it is straightforward to derive the equations that $\Phi_0(\tilde{u})$ should satisfy. After plugging~(\ref{Phiexpand}) into~(\ref{BQonly}), rescaling the rapidity and expanding at large spin we find
\beq
\tilde{u}^2\sin^2{\Phi'_{0}(\tilde{u})} = 1\, ,
\eeq
with $\Phi'_{0}(\tilde{u}) = \partial_{\tilde{u}}\Phi_{0}(\tilde{u})$. This equation is closely related to the one considered in~\cite{Korchemsky:1995be}. It is straightforward to work out the relevant solution
\beq
\Phi'_{0}(\tilde{u}) = -i\log{\left(\sqrt{1-1/\tilde{u}^2}+i/\tilde{u}\right)}\, .
\eeq
We observe that the function $\Phi_0(\tilde{u})$ is analytic in the complex $\tilde{u}$ plane except for the cut $\tilde{u}\in (-1, 1)$. This is where the roots have condensed at large spin. The discontinuity across the cut gives the density $\rho$ in the semiclassical regime~\cite{Korchemsky:1995be}%
\footnote{Note that the density used in~\cite{Korchemsky:1995be} and the one defined in this paper are not exactly the same. They agree however at the  accuracy considered.}
\beq \label{density}
\rho(\ub) = \frac{i}{2\pi} \left(\Phi'(\ub+i0)-\Phi'(\ub-i0)\right)\,.
\eeq
We therefore have an expansion for the density similar to~(\ref{Phiexpand})
\beq
\rho(\ub) = \rho_0(\ub) +\rho_1 (\ub)/K+\dots\, ,
\eeq
with
\beq \label{LOdensity}
\rho_0(\ub)=\frac{1}{2\pi} \log \frac{1+\sqrt{1-\ub^2}}{1-\sqrt{1-\ub^2}}\,.
\eeq
This is the expression that is vital to determining the constant $C$ in Section \ref{sec:densofB}. As a side comment, we observe that it is half of the corresponding result for the $\mathfrak{sl}(2)$ spin chain~\cite{Korchemsky:1995be}.

The expression for $\Phi_1$ is more bulky, but most of the terms \textit{do not} contribute to the subleading density because of \eqref{density}. We find
\beq \label{NLOdensity}
\rho_1(\ub)= \frac{L+\bar{K}-K+g(\Gamma_1+\Gamma^*_1)}{2\pi \sqrt{1-\ub^2}}-\frac{\sum^{K_h+\bar{K}_h}_{j=1} u^h_j+ig(\Gamma_1-\Gamma^*_1)}{2\pi \ub \sqrt{1-\ub^2}}\, .
\eeq
Note that in the course of derivation we used $\bar{K}-K \sim O(1)$. In the following we will only need the second term in the RHS above,  which is odd in $\tilde{u}$ and controlled by the total momentum $p_{\textrm{tot}}$, see~(\ref{App-ptot}).

The expressions corresponding to~(\ref{NLOdensity}) for the density $\bar{\rho}$ of anti-magnons may be immediately found with the aid of the following substitutions
\beq
\bar{K} \leftrightarrow K\,, \qquad\ub = \frac{u}{K} \to \ub = \frac{u}{\bar{K}}\,.
\eeq
The equation \eqref{NLOdensity} and its conjugate may be used in conjunction with the densities found in Section \ref{sec:densofB} to compute the $Y(u)$ and $\bar{Y}(u)$ functions in Section \ref{sec:BYol}. It is however simpler to introduce the twists $q$ and $\bar{q}$ to regularise the expressions. The problem of dealing with divergent integrals is then shifted to the computation of $e^{iP}$. We will determine this quantity in the following subsection.
\subsection{The momentum operator}
In this section we will evaluate the exponential of the momentum operator,
\beq 
e^{iP} = \prod^K_{j=1} \frac{u_j+\tfrac{i}{2}}{u_j-\tfrac{i}{2}} \prod^{\bar{K}}_{j=1} \frac{\bar{u}_j+\tfrac{i}{2}}{\bar{u}_j-\tfrac{i}{2}} \prod^K_{j=1} \frac{1+g^2/(x^-_j)^2}{1+g^2/(x^+_j)^2} \prod^{\bar{K}}_{j=1} \frac{1+g^2/(\bar{x}^-_j)^2}{1+g^2/(\bar{x}^+_j)^2},
\eeq
at arbitrary value of the coupling constant. The crucial prerequisite to  this computation is \eqref{NLOdensity}. We first observe that
\beqa \nn
&&\log{\left[(-1)^{K+\bar{K}} e^{iP}\right]}= -2i \lim_{K\to \infty} \int^{K}_{-K} dv \arctan(2v) \rho_{+}(v)+2i\sum^{K_h+\bar{K}_h}_{j=1} \arctan \left({2 u_{h,j}}\right)\\ \label{eiPallloop}
&&+\int^{\infty}_{-\infty} dv \log{\frac{1+g^2/(x^-(v))^2}{1+g^2/(x^+(v))^2}} \rho_{+}(v)- \sum^{K_h+\bar{K}_h}_{j=1} \log{\frac{1+g^2/(x^-(u_{h,j}))^2}{1+g^2/(x^+(u_{h,j}))^2}} \,.
\eeqa
The second integral is finite and does not require regularisation. We are left with the computation of the integral\footnote{We have assumed that to the order considered the effect of unequal supports of $\rho(u)$ and $\bar{\rho}(u)$ may be ignored.} 
\beq
I=\lim_{K \to \infty} \int^{K}_{-K} \arctan(2v) \rho_+(v)\,.
\eeq
To separate the $v \sim 1$ and $v \sim K$ domains we introduce a new boundary $M$,
\beq
\lim_{K \to \infty} \int^{K}_{-K}dv \arctan(2v) \rho_+(v)=\lim_{M,K \to \infty} \left(\int^{-M}_{-K}+ \int^{M}_{-M}+\int^{K}_{M} \right) dv \arctan(2v) \rho_+(v) \,,
\eeq
and require that $\lim_{K,M \to \infty} M/K = 0$. Please note that we expect the end result \textit{not} to depend on $M$. If true, this will substantiate this ad hoc regularisation method. Using \eqref{NLOdensity} we easily compute
\beqa
I_1&=&\left(\int^{-M}_{-K}+\int^{K}_{M} \right) dv \arctan(2v) \rho_+(v) \\ \nn
&\simeq& \pi \left( \int^{-M/K}_{-1}+\int^{1}_{M/K}\right) d\tilde{v}\, \textrm{sign}(\tilde{v})\rho_1( \tilde{v}) = p_{\textrm{tot}}\log{\frac{M}{2K}}\,.
\eeqa
We still need to compute
\beq
I_2=\int^{M}_{-M}du \arctan(2u) \rho_{+}(u)=-\int^{\infty}_0 dt \left(\frac{1}{\pi}\int^{M}_{-M}du \arctan(2u) \sin(ut)\right) e^{t/2} \tilde{\Omega}^v(t)\,.
\eeq
The integral kernel in the bracket has the following large $M$ expansion
\beq \nn
\K(t,M)=\frac{1}{\pi} \int^{M}_{-M}du \arctan{\left(2u\right)} \sin{\left(ut\right)} = \frac{e^{-t/2}-\cos (M t)}{t}+ O\left(M^{-1} \right)\,.
\eeq
For any smooth enough function $h(t)$, an integral involving kernel $\K(t,M)$ is logarithmically divergent when $M \to \infty$
\beq
\int^{\infty}_0 dt\, \K(t,M) h(t) = h(0) \log M + \int^{\infty}_0 dt \,e^{-t/2} \, \frac{h(t)-h(0)e^{-t/2}}{t}+ O(M^{-1})\,.
\eeq
Putting $h(t) = e^{t/2} \tilde{\Omega}^v(t)$ and using~(\ref{FTomegat}) we immediately find
\beqa \nn
I_2&=&-\int^{\infty}_0 dt\, \K(t,M) e^{t/2} \tilde{\Omega}^v(t) =-p_{\textrm{tot}} \log(M) -\sum^{K_h+\bar{K}_h}_{j=1} \frac{i}{2} \log{\frac{\Gamma(\ft{3}{2}+i u_{h,j})}{\Gamma(\ft{3}{2}-iu_{h,j})}} \\
&& \qquad \qquad -\sum^{K_h+\bar{K}_h}_{j=1}\int^{\infty}_0 dt\, \frac{\tilde{\gamma}^{u_{h,j}}(2gt)-2g \tilde{\gamma}^{u_{h,j}}_1(1-e^{-t})}{t (e^t-1)}+ O(M^{-1})\,.
\eeqa
We check by combining $I_1$ and $I_2$ that the dependence on the arbitrary cut-off $M$ drops out as expected. Finally, the higher-loop correction in \eqref{eiPallloop} yields after performing Fourier transformation
\beq
\begin{aligned}
\int^{\infty}_{-\infty} dv &\log{\frac{1+g^2/(x^-(v))^2}{1+g^2/(x^+(v))^2}} \rho_{+}(v) - \sum^{K_h+\bar{K}_h}_{j=1} \log{\frac{1+g^2/(x^-(u_{h,j}))^2}{1+g^2/(x^+(u_{h,j}))^2}} \\
&\qquad = 2i \sum^{K_h+\bar{K}_h}_{j=1} \int_{0}^{\infty}\frac{dt}{t}(J_{0}(2gt)-1)\frac{\sin{(u_{h, j}t)}e^{t/2} + \tilde{\gamma}^{u_{h, j}}(2gt)}{e^{t}-1}\,.
\end{aligned}
\eeq
After combining all the pieces together and using that $K\sim \bar{K}\sim S$, we get the simple result
\beq
e^{iP} = (-1)^{K+\bar{K}}(2S)^{2ip_{\textrm{tot}}} \prod^{K_h}_{j=1} T_{\mathcal{N}=6}(u_{h, j})\prod^{\bar{K}_h}_{j=1} T_{\mathcal{N}=6}(\bar{u}_{h, j})\, ,
\eeq
with  $T_{\mathcal{N}=6}(u) = T_{\mathcal{N}=4}(u)^{1/2}$ the transmission amplitude of the $\mathcal{N}=6$ theory, and $T_{\mathcal{N}=4}(u)$ defined in~(\ref{defTv}). For comparison, in the $\mathfrak{sl}(2)$ subsector of the $\mathcal{N}$ theory one would find
\beq
e^{iP} = (-1)^{K}S^{ip_{\textrm{tot}}} \prod^{K_h}_{j=1} T_{\mathcal{N}=4}(u_{h, j})^{1/2}\, .
\eeq
\newpage
\bibliography{aba}
\bibliographystyle{nb}
\end{document}